\documentclass[12pt,twoside,floatfix]{mbook}
\usepackage{graphicx}
\usepackage{amssymb}
\usepackage{amsmath}
\usepackage[T1]{fontenc}
\usepackage[cp 1250]{inputenc}
\usepackage{textcomp}
\usepackage{amsfonts}
\usepackage{epsfig}
\usepackage{here}
\usepackage{cite}
\usepackage{esint}
\usepackage{fancyhdr}
\usepackage{color}
\title{{\bf Examination of the measurement possibility of the 
$\eta^{\prime}\to\pi^{+}\pi^{-}\pi^{0} $ decay using WASA-at-COSY apparatuss }}
\author{Marcin Zieli\'{n}ski}
\date{18 maj 2008 r.}
\begin{document}

\thispagestyle{empty}
\newpage
\thispagestyle{empty}
\begin{center}
{\Large INSTITUTE OF PHYSICS  }\\
{\Large FACULTY OF PHYSICS, ASTRONOMY  }\\
{\Large AND APPLIED COMPUTER SCIENCE}\\
{\Large JAGIELLONIAN UNIVERSITY}\\
\end{center}
\begin{center}\end{center}
\begin{center}\end{center}
\begin{center}\end{center}
\begin{center}\end{center}
\begin{center}
{\LARGE{\bf Feasibility study}}
\end{center}
\begin{center}
{\LARGE{\bf of the  $\eta^{\prime}\to\pi^{+}\pi^{-}\pi^{0} $ decay }}
\end{center}
\begin{center}
{\LARGE{\bf using WASA-at-COSY apparatus}}
\end{center}
\begin{center}
\end{center}
\begin{center}\end{center}
\begin{center}\end{center}
\begin{center}
{\Large Marcin Zieli\'{n}ski}
\end{center}
\begin{center}\end{center}
\begin{center}\end{center}
\begin{center}
{\normalsize Master Thesis  }\\
{\normalsize prepared in the Nuclear Physics Division\\}
{\normalsize of the Jagiellonian University\\}
{\normalsize supervised by } 
\end{center}
\begin{center}
{\Large Prof. dr hab. Pawe\l{} Moskal}
\end{center}
\begin{center}
\end{center}
\begin{center}\end{center}
\begin{center}\end{center}
\begin{center}
\begin{figure}[h]
\hspace{6.6cm}
\vspace{-3.5cm}
\parbox{0.10\textwidth}{\centerline{\epsfig{file=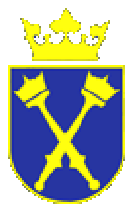,width=0.13\textwidth}}}
\end{figure}
\end{center}
\begin{center}
{\vspace{-.2cm}\normalsize }
\end{center}
\begin{picture}(0,0)
\put(0,580){\line(1,0){440}} 
\put(0,-35){\line(1,0){440}} 
\put(175,-50){CRACOW 2008}
\end{picture}

\newpage
\thispagestyle{empty}
\begin{center}\end{center}
\newpage
\thispagestyle{empty}
\begin{center} 
\vspace{2.cm}
\Large{\bf Abstract} 
\end{center}

\vspace{2.cm}
One of the objectives of the 
vast physics programme of the recently
comissioned WASA-at-COSY facility is the study of 
fundamental 
symmetries via the measurements 
of the  $\eta$ and $\eta^{\prime}$ mesons decays. 
Especially interesting are isospin violating 
hadronic precesses of these mesons
into 3$\pi$ systems  driven by the term of QCD Lagrangian 
which depends on the mass difference 
of the $u$ and $d$ quarks.

When an $\eta$ or an $\eta^{\prime}$ meson is created
in the hadronic reaction signals 
from such decays may be significantly obscured by the prompt production of $\pi$ mesons.
In this thesis we present the estimation of the  
upper limit of the background 
due to prompt pion production for
the $\eta'\to 3\pi^0$ and $\eta'\to\pi^+\pi^-\pi^0$ decays. 
Using the data from proton-proton collisions measured 
by the COSY-11 group 
we have extracted differential cross sections for the multimeson
production with the invariant 
mass corresponding to the mass of the $\eta^{\prime}$ meson.
Based on these results and on parametrizations of
the total cross sections for the $\eta^{\prime}$ meson
as well as  parametrization of the upper limit for the prompt $\pi^+\pi^-\pi^0$ 
production in the collisions of protons
we discuss in details the feasibility 
of a measurement of the $\eta'$ meson decay into $3\pi$ channels
with the WASA-at-COSY facility. 
Based on the chiral unitary approach the value of the branching ratio BR($\eta^{\prime}\to \pi^+\pi^-\pi^0$)
was recently predicted to be about 1\%. 
We show that the WASA-at-COSY has a potential to verify this result empirically.

Furthermore, we discuss the possible usage of the time signals and of the energy loss measurement in the forward part of 
the WASA-at-COSY detector for the determination of the energy of the forward scattered charged particles.
We briefly describe the detectors to be used for this 	
purpose and introduce the computational algorithm which was developed to be applied with this technique.     

\newpage
\thispagestyle{empty}
\begin{center}\end{center}
\newpage
\thispagestyle{empty}
\begin{center}\end{center}
\begin{center}\end{center}
\begin{center}\end{center}
\begin{center}\end{center}
\begin{center}\end{center}
\begin{center}\end{center}
\begin{center}\end{center}
\begin{center}\end{center}
\begin{center}\end{center}
\begin{center}\end{center}
\begin{center}\end{center}
\begin{center}\end{center}
\begin{center}\end{center}
\begin{center}\end{center}
\begin{center}\end{center}
\begin{center}\end{center}
\begin{center}\end{center}\begin{center}\end{center}
\begin{center}\end{center}
\begin{center}\end{center}
\begin{tabbing}
\=\bf''There is no expedient $~~~~~~~~~~~~~~~~~~~~~~~~~~~$ \=\bf''Nie ma takiego fortelu,\\ 
\>\bf to which a man will not go                \>\bf do którego nie odwołałby się człowiek,\\
\>\bf to avoid the labor of thinking.''         \>\bf aby uniknąć pracy zwanej myśleniem.''
\end{tabbing}
\begin{center}\end{center}\begin{center}\end{center}
\begin{flushright}
{\it Thomas Alva Edison (1847 -- 1931)}
\end{flushright}

\newpage
\thispagestyle{empty}
\begin{center}\end{center}

\newpage
\pagestyle{fancy}
\fancyhead{}
\fancyfoot{}
\renewcommand{\headrulewidth}{0.8pt}
\fancyhead[RO]{\textbf{\sffamily{{{\thepage}}~}}}
\fancyhead[RE]{\bf\footnotesize{\nouppercase{\leftmark~}}}
\fancyhead[LE]{\textbf{\sffamily{{{\thepage}}}}}
\fancyhead[LO]{\bf\footnotesize{{\nouppercase{\rightmark~}}}}
\advance\headheight by 5.3mm
\advance\headsep by 0mm
\tableofcontents
\markboth{Contents}{Contents}

\chapter{Introduction}
\hspace{\parindent}
The Standard Model is nowadays a well known and established 
theory describing the particles and theirs strong, electromagnetic and 
weak interactions. Therefore, tests of the applicability of this model are 
very important and are carried out in many particle physics laboratories.
In particular particle physicists endeavour to discover phenomena  
which the Standard Model does not describe. 
One of such laboratory is WASA-at-COSY~\cite{proposal} where we investigate the limits of
applicability of the Standard Model by studying production and decays of pseudoscalar mesons 
like $\eta$ and $\eta^{\prime}$. The examination of these meson production and their decays give us
a chance to probe fundamental symmetries like C (charge conjugation), P (parity), T (time reversal) and their combinations
and to determine the parameters of the Standard Model like for example the quark masses. 

The masses of the light quarks (up -$u$, down - $d$, strange - $s$) are known only 
approximately due to the fact that quarks are 
confined inside hadrons and cannot be observed directly as physical particles.
But there are several indirect possibilities of establishing these masses 
by measuring the mass differences and mass ratios.   
The mass difference of light quarks may be derived from the partial widths of the decay of the  
$\eta$ and $\eta^{\prime}$ mesons into $\pi^{+}\pi^{-}\pi^{0}$ and theirs total widths.
The possibility of determining of the difference between the mass of the light quarks motivates 
us for studing these decays. 

Detailed studies of $\eta\to\pi^{+}\pi^{-}\pi^{0}$ decay was conducted by the  
WASA/CELSIUS collaboration \cite{pauly06-celcius}, 
but so far the decay $\eta^{\prime}\to\pi^{+}\pi^{-}\pi^{0}$ has never been observed.
In both of these processes the isospin conservation is violated \cite{kupsc-aip}. 
But despite this fact in case of the $\eta$ meson 
the branching ratio for the $\eta\to\pi\pi\pi$ decay is in the order of 50\%.
However, for the $\eta^{\prime}$ meson the situation is quite different the branching ratio for the 
$\pi^{0}\pi^{0}\pi^{0}$ is at the permil level and for the $\pi^{+}\pi^{-}\pi^{0}$ system so far
only an upper limit of 5\% was established \cite{rittenberg69}.  

For the decay of both $\eta$ and $\eta^{\prime}$ mesons there exists a physical background from the direct production 
of three pions via the $pp\to pp\pi^{+}\pi^{-}\pi^{0}$ reaction channel.  
In case of the $\eta$ meson the signal to background ratio is large, amounting to about 10 
for tagging by means of the missing mass technique with a resolution of a few MeV, and 
permits a clear identification of the $\eta\to\pi^{+}\pi^{-}\pi^{0}$ decay. 
However, this ratio is expected to be worse by more than three orders of magnitude for the $\eta^{\prime}$
meson making the investigations much more challenging experimentally, especially since the hadronic 
production cross section is by about a factor of thirty smaller for the $\eta^{\prime}$ meson in 
comparison to the $\eta$ ($\sigma_{\eta} \approx 30 \sigma_{\eta^{\prime}}$) meson  at the same excess energy. 

The main aim of this thesis is the determination of the optimum beam momentum for the measurement of
the branching ratio for the $\eta^{\prime}\to\pi^{+}\pi^{-}\pi^{0}$ decay channel via  the 
$pp\to pp\eta^{\prime}\to pp\pi^{+}\pi^{-}\pi^{0} \to pp\pi^{+}\pi^{-}\gamma\gamma$
reaction chain.  The measurement is planned to be carried out in the near future and its result will
be used for the derivation of the light quarks mass difference ($m_d - m_d$).  
The experiment will take place at the Research Centre J\"{u}lich in 
Germany and will be conducted by the WASA-at-COSY collaboration. 
The identification of the reaction is based on selecting events where the 
$\pi^{+}\pi^{-}\pi^{0}$ system was produced.
For this subsample of data the fractions corresponding to the direct $\pi^{+}\pi^{-}\pi^{0}$ production 
and to the decay of the $\eta^{\prime}$ meson will be extracted  based on the missing mass of the two 
outgoing protons registered in the Forward Detector of the WASA-at-COSY system. 

To calculate the number of  expected $\eta^{\prime}$ events we have parametrized the total cross 
section for the $\eta^{\prime}$ meson production and have etablished a parametrization 
for an upper limit of the background production. Further on we have parametrized a 
missing mass resolution taking into account effects which are related with the energy resolution of the forward detector and  
the beam and target spread. In the 
case of the $\eta^{\prime}$ meson the missing mass resolution is at present 
not satisfactory for studies of rare
decays due to the high energies of the protons. Therefore, we have  proposed the 
usage of the time-of-flight method (TOF) for the  
identification of charged particles emitted in forward direction based on
time signals from  the scintillator detectors. 

In the next chapter of this work we will outline difficulties of the mass determination
for light quarks. 
We will quote the 
idea of an indirect determination of quark masses proposed by D.~Gross~, B.~Treiman and
F.~Wilczek \cite{gross79}, and by H.~Leutwyler \cite{leutwyler96}.

The measurment and identification method of the decay 
$\eta^{\prime}\to\pi^{+}\pi^{-}\pi^{0}$ as well as the WASA-at-COSY detector 
setup will be presented in Chapter 3. 

The first section of Chapter 4 describes how the total cross section for
the production of the $\eta^{\prime}$ meson depends on the excess energy near the kinematical threshold. 
Further on in this chapter the method of estimating an upper limit of the background under 
the $\eta^{\prime}$ peak in the missing mass distribution will be given~\cite{kupsc-menu}. 
Next, applying the parametrization of the energy resolution of the
detector, the accuracy of the branching ratio determination as a function
of excess energy and time will be shown. 
For the calculation we consider a range of values for the upper limit of~5\% down to a value lower  
by one order of magniude~(0.5\%).  

In chapter 5 we will propose a  method of particle energy reconstruction by measuring the 
time in the Forward Detector.
For the purpose of this thesis we will consider only thin scintillators, but in the future the 
reconstruction will be based on time signals form all detectors. 
First we will show results 
of simulations for an energy resolution based on TOF measurements and describe
how passive material of the detector influences the precision of the TOF determination.
As a result of the simulations the missing mass distributions calculated 
based on the TOF measurment will be shown. In the last section of this chapter we will
describe the computational algorithm which was developed to reconstruct the 
energy of particles based on the time signals from scintillators and energy losses in the 
five independent layers of the range hodoscope.   

Chapter 6 summarises the whole thesis and brings the conclusions and remarks
regarding the beam momentum and the duration of the planned measurement of the 
$\eta^{\prime}\to \pi^{+}\pi^{-}\pi^{0}$ decay.

The work is supplemented with Appendics where section~\ref{app:mezony} introduces basic informations about
the $\eta^{\prime}$ meson and the SU(3) symmetry. 
Appendix~\ref{app:mm} describes
the missing mass technique used for tagging the $\eta^{\prime}$ meson production.
Section~\ref{app:tof} provides
an analytical calculation of the fractional momentum resolution as a function of the fractional time resolution.  
In appendix~\ref{app:3pi} the parametrization of the total cross section for the $\pi^{+}\pi^{-}\pi^{0}$
production will be shown.
Section~\ref{app:fsi} introduces the parametrization of the Final State 
Interaction (FSI) and production dynamic of the multipion system. 
The next section presents
simulations of the $3\pi$ production in proton-proton collisions 
using models described in section~\ref{app:fsi}. The last appendix shows 
schemes of programs used for the simulations.  

\chapter{Relation between the partial width $\Gamma_{\eta^{\prime}\to\pi^{+}\pi^{-}\pi^{0}}$\\ 
          and the $u$ and $d$ quark mass }

\hspace{\parindent}
The determination of the light quark masses is one of the important  goals of hadron physics experiments,
and we intend to contribute to their estimations by determining the  
quark mass difference $m_d - m_u$ which induces an isospin breaking. 
Studing the isospin-violating decays $\eta(\eta^{\prime})\to\pi^{+}\pi^{-}\pi^{0}$ and 
$\eta(\eta^{\prime})\to\pi^{0}\pi^{0}\pi^{0}$ was pointed out as an accurate way of extracting the quark 
mass difference \cite{gross79,leutwyler96}.

The quark masses are one of the Standard Model parameters and their values depend
on how they are defined. The field theory which describes the strong interaction
between gluons and quarks is the Quantum Chromodynamics (QCD).
A general QCD Lagrangian for $N$ flavors reads:
\begin{equation}
\mathbb{L}_{QCD} = \sum_{k=1}^{N}\bar{q_k}(iD - m_k)q_k - \frac{1}{4}G_{\mu\nu}G^{\mu\nu},
\label{qcd_lag}
\end{equation} 
where $m_k$ denotes the quark masses, $D$ indicates the gauge convariant derivative and 
$G_{\mu\nu}$ represents the gluon field strength. 

In the low energy regime where the strong coupling constant $\alpha_s$ is large, the perturbative approach of QCD 
cannot be applied in the same way as for high energies. 
Thus other methods  must be applied in low energy hadron physics. 
One of these approaches is an effective field theory of QCD at low energies - the Chiral Perturbation Theory (ChPT).
It is based on the observation that, in the low energy region, the
relevant and  effective degress of freedom of strong interactions are hadrons composed of confined 
quarks and gluons.
This leads to the effective Lagrangian which is formulated in terms of the effective degrees of freedom: 
\begin{equation}
\mathbb{L}_{QCD}^{eff} = \mathbb{L}_0 + \mathbb{L}_m.
\label{eff_lag}
\end{equation}
We can use this effective Lagrangian instead of the formula (\ref{qcd_lag}). 
The $\mathbb{L}_0$ term is the part
of flavor symmetry of QCD and $\mathbb{L}_m$ contains a contribution:
\begin{equation}
-\frac{1}{2}(m_d - m_u)(\bar{u}u - \bar{d}d)
\label{lm_lag}
\end{equation}
responsible for the isospin changing in QCD ($\Delta I = 1$).
One can identify the degrees of freedom as eight Goldstone bosons 
which are $\pi$, $K$ and $\eta$ mesons from the pseudoscalar meson nonet.
This effective Lagrangian breakes down spontaneously the chiral symmetry 
$SU(3)_L\times SU(3)_R$ to $SU(3)_V$. It is believed that
this violation of the chiral symmetry constitutes the source of the mass generation 
in QCD. The standard ChPT provides an accurate description of the strong and electroweak 
interactions of the pseudoscalar mesons at low energies. 

The $\eta$ and $\eta^{\prime}$ mesons are members of the pseudoscalar nonet, and play an  
important role in the understanding of the low energy QCD. Due to the mixing of the $\eta$ and $\eta^{\prime}$
fields the $\eta$ treatment in the standard ChPT is complicated. The $\eta^{\prime}$ meson is related 
with the axial U(1) anomaly. This anomaly in QCD prevents the $\eta^{\prime}$ meson from being a 
Goldston boson which is manifested in its large mass ($m_{\eta^{\prime}} = 958$ MeV), a mass which is much 
larger than the masses of other members of pseudoscalar nonet (see. App. \ref{app:mezony}). 
Hence the $\eta^{\prime}$ meson is not included explicitly 
in the conventional SU(3) ChPT, albeit its effects are hidden in coupling constants.
A recent extension of ChPT methods has provided tools which enable to include $\eta^{\prime}$ 
in a consistent way and perform reliable calculations. One of these tools is a chiral unitary approach,
which is based on the chiral perturbation theory and the unitarization using the Bethe-Salpeter equation.  

Concernig hadronic decays of the $\eta$ and $\eta^{\prime}$ mesons into three pion systems, 
we can certify that 3$\pi$ can be in isospin 0 state only if a two pion subsystem is in I = 1 state.
In case of the $\pi^{0}\pi^{0}\pi^{0}$ system the two pion can have I$_{2\pi}$ = 0, 1, 2 but coupling with 
the remaining pion to I$_{3\pi}$ = 0 is  only possible if I$_{2\pi}$ = 1. However,  the $(\pi^{0}\pi^{0})_{I=1}$ 
dose not exist (a corresponding Clebsch-Gordan coefficient is equal to zero~~\cite{pdg})
and as a conseqence the decay $\eta(\eta^{\prime})\to\pi^{0}\pi^{0}\pi^{0}$ has to violate isospin.
 
In the case of the $\eta(\eta^{\prime})\to\pi^{+}\pi^{-}\pi^{0}$ decay taking into account the 
Clebsch-Gordan coefficients~\cite{pdg} one can write~\cite{andrzej-priv}:
\begin{equation}
(3\pi)_{I=0} = \sqrt{\frac{1}{3}}\left[ (\pi^{+}\pi^{0})_{I=1}\vert\pi^{-}\rangle - 
                                        (\pi^{+}\pi^{-})_{I=1}\vert\pi^{0}\rangle +
                                        (\pi^{-}\pi^{0})_{I=1}\vert\pi^{+}\rangle \right],
\end{equation} 
where:
\begin{equation}
\nonumber
(\pi^{+}\pi^{0})_{I=1} = \sqrt{\frac{1}{2}}\left[ 
                  \vert\pi^{+}\rangle\vert\pi^{0}\rangle - \vert\pi^{0}\rangle\vert\pi^{+}\rangle\right]  ,
\end{equation}
\begin{equation}
\nonumber
(\pi^{+}\pi^{-})_{I=1} = \sqrt{\frac{1}{2}}\left[ 
                  \vert\pi^{+}\rangle\vert\pi^{-}\rangle - \vert\pi^{-}\rangle\vert\pi^{+}\rangle\right] ,
\end{equation}
\begin{equation}
\nonumber
(\pi^{-}\pi^{0})_{I=1} = \sqrt{\frac{1}{2}}\left[ 
                  -\vert\pi^{-}\rangle\vert\pi^{0}\rangle + \vert\pi^{0}\rangle\vert\pi^{-}\rangle\right]  .
\end{equation}
Thus the full wave function for the 3$\pi$ system reads:
\begin{eqnarray}
\nonumber
(3\pi)_{I=0} = \sqrt{\frac{1}{6}}\left[ \vert\pi^{+}\rangle\vert\pi^{0}\rangle\vert\pi^{-}\rangle -
                                        \vert\pi^{0}\rangle\vert\pi^{+}\rangle\vert\pi^{-}\rangle -
                                        \vert\pi^{+}\rangle\vert\pi^{-}\rangle\vert\pi^{0}\rangle +\right.\\
                                        \left.\vert\pi^{-}\rangle\vert\pi^{+}\rangle\vert\pi^{0}\rangle -
                                        \vert\pi^{-}\rangle\vert\pi^{0}\rangle\vert\pi^{+}\rangle + 
                                        \vert\pi^{0}\rangle\vert\pi^{-}\rangle\vert\pi^{+}\rangle \right].
\end{eqnarray}
This wave function is antisymmetric against any exchange of pions: $\pi^{0}\leftrightarrow\pi^{+}$, $\pi^{-}\leftrightarrow\pi^{+}$ and $\pi^{0}\leftrightarrow\pi^{-}$. In particular, by applying charge conjugation we have:
\begin{equation}
C(3\pi)_{I=0} = - (3\pi)_{I=0}.
\end{equation}
This is in contradiction with C = +1 for $\eta(\eta^{\prime})$. Therefore the decay
$\eta(\eta^{\prime})\to\pi^{+}\pi^{-}\pi^{0}$ should violate C or I.
 
On the other hand there exist a $G$ operator which is constructed from the $C$ parity  and isospin  $I_2$ operators in the 
following way:
\begin{equation}
G = C e^{i\pi I_2}.
\end{equation}
The eigenvalue of this operator is given by $\lambda_G = (-1)^I\lambda_C$, thus  $\lambda_G = -1$ for pions 
and $\lambda_G = +1$ for the $\eta$ mesons. Therefore the decay $\eta(\eta^{\prime}) \to \pi\pi\pi$ does not conserve $G$.   
 
Historically these decays were considered as electromagnetic processes with partial widths
smaller than the second order electromagnetic decay $\eta(\eta^{\prime})\to\gamma\gamma$. But it appears that 
the electromagnetic contribution is small \cite{sutherland,bell68} and 
instead the process is expected to be dominated by the isospin-violating term in the strong interaction.
Due to this fact we can neglect the electromagnetic terms, and thus the decay 
amplitudes of the $\eta(\eta^{\prime})\to3\pi$ decay becomes directly proportional to the quark mass 
difference ($m_d - m_u$) \cite{leutwyler96}. 
\begin{figure}[h]
\parbox{0.32\textwidth}{\centerline{\epsfig{file=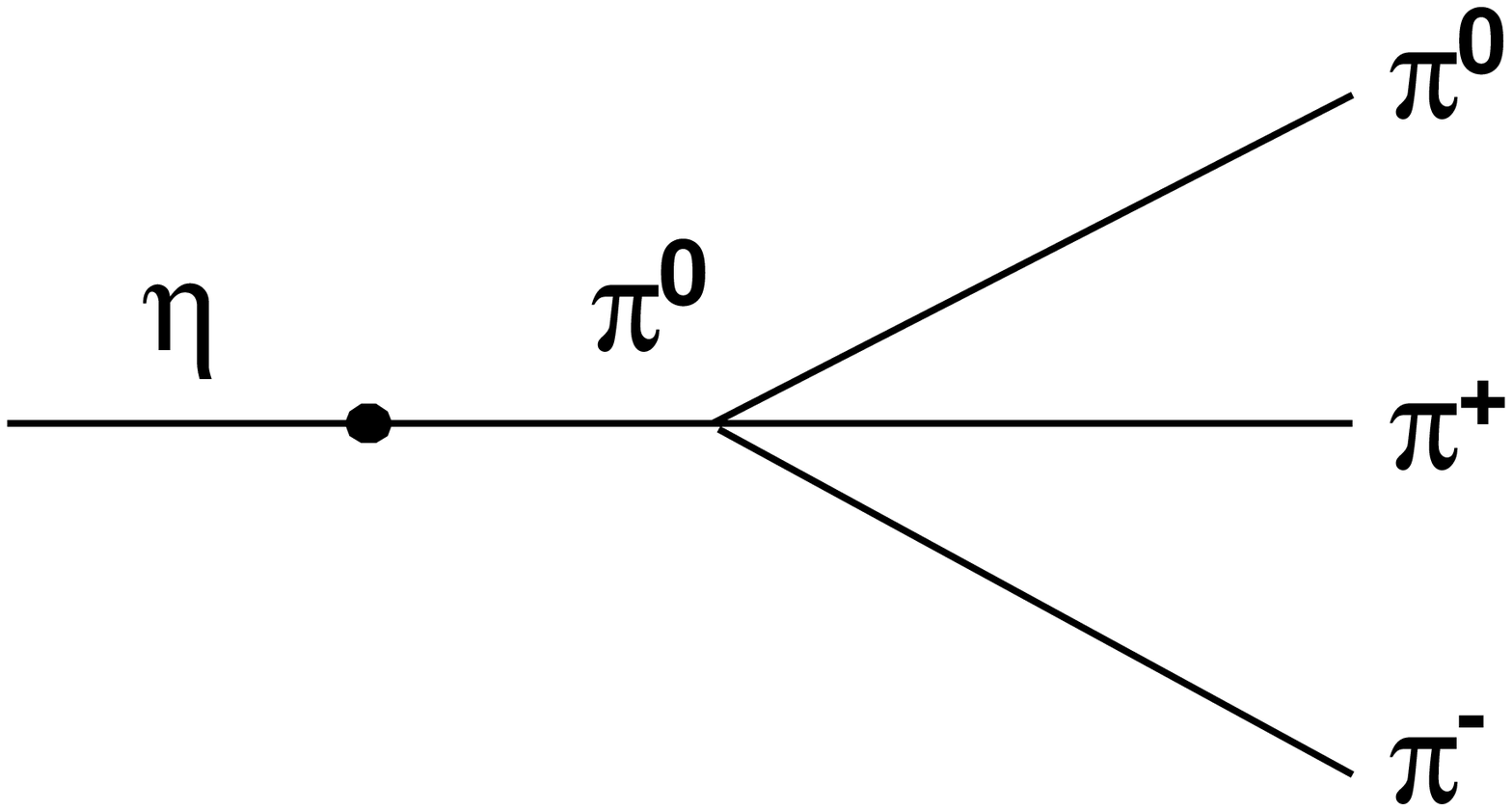,width=0.36\textwidth}}}
\parbox{0.32\textwidth}{\centerline{\epsfig{file=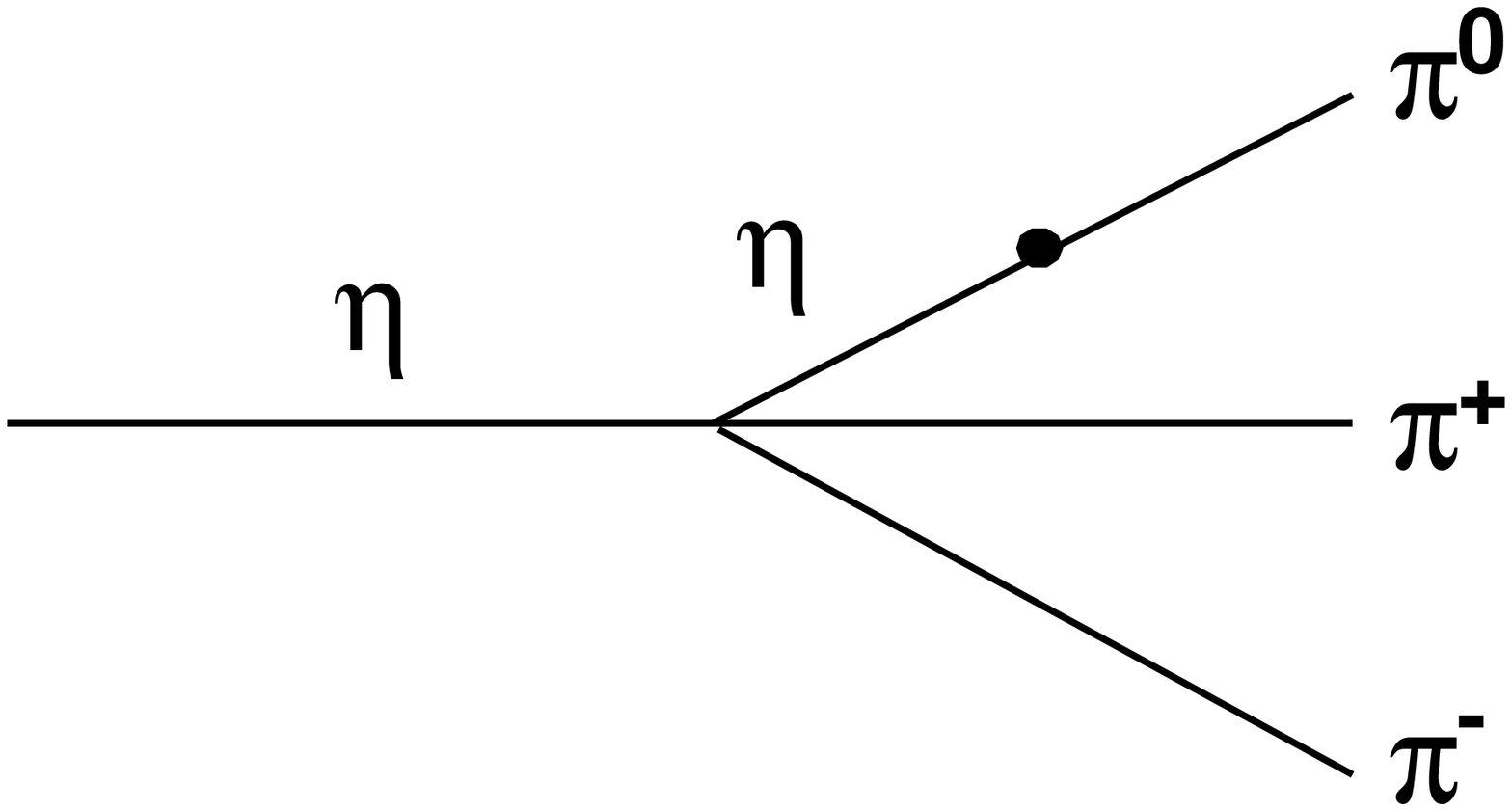,width=0.36\textwidth}}}
\parbox{0.32\textwidth}{\centerline{\epsfig{file=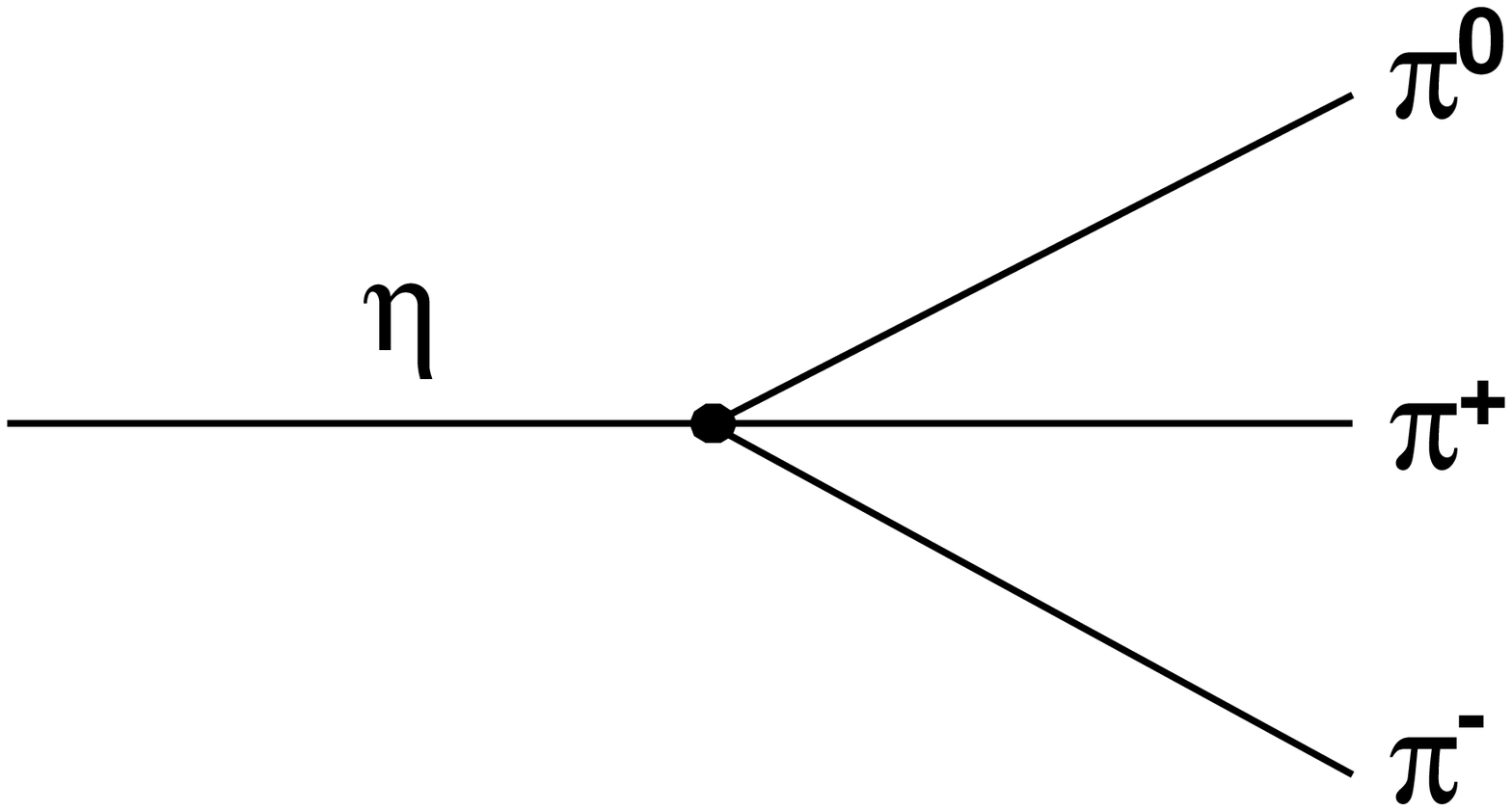,width=0.36\textwidth}}}
\begin{picture}(0,0)
\put(-400,10){($\eta^{\prime}$)}
\put(-240,10){($\eta^{\prime}$)}
\put(-100,10){($\eta^{\prime}$)}
\put(-400,-30){(a)}
\put(-240,-30){(b)}
\put(-100,-30){(c)}
\end{picture}\caption{
Isospin violation in $\eta(\eta^{\prime})\to 3\pi$ decay. 
Lowest order effective lagrangian contribution (courtesy of A.~Kup\'{s}\'{c}~\cite{andrzej-priv}).
}
\label{grafy}
\end{figure}
The lowest order contribution to the decay mechanism is given by the Current Algebra (CA).
Figure~\ref{grafy} shows graphs consisting of a combination of the $\eta-\pi^{0}$ mixing (a,b) 
and crossed graph (c) of the elementary low energy QCD processes - scattering of two pseudoscalar mesons.
The partial width of the  $\eta\to\pi^{+}\pi^{-}\pi^{0}$ decay calculated using Current Algebra is 66 MeV \cite{bardeen},
which is much below the experimental value of 294$\pm$16 MeV \cite{pdg}. Gasser and Leutwyler had corrected this value 
to 160 eV \cite{gasser85} using the second order in the low energy expansion of the effective QCD Lagrangian. 
This change implies the importance of the $\pi\pi$ interaction in the final state and involving  
higher loop calculations should improve this value since they give a better description of the $\pi\pi$ final state
interaction. An other approach which includes a pion-pion interaction up to higher orders uses the dispersion relation,
which connects the imaginary part of the decay amplitude with the amplitude itself. There are two estimations 
based on this method \cite{kambor,anisovich} but using different formalisms. They lead consistently to 
an enhancement of the decay rate by about 14\%. 

\begin{figure}[h]
\hspace{3.5cm}
\parbox{0.55\textwidth}{\centerline{\epsfig{file=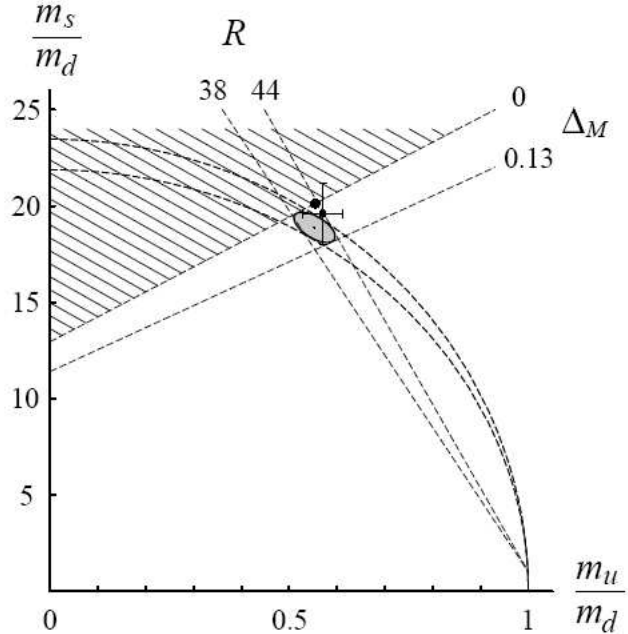,width=0.59\textwidth}}}
\caption{
Ellipse which characterises the quark mass ratios (for the description see text). 
The figure is adapted from~\cite{leutwyler96}. 
}
\label{elipsa}
\end{figure}

It was suggested by H.~Leutwyler that the decay width of the $\eta(\eta^{\prime})\to \pi^{+}\pi^{-}\pi^{0}$ 
is sensitive to the mass difference of the light quarks following the relation \cite{leutwyler96}:
\begin{equation}
\Gamma_{\eta^{\prime}\to \pi^{+}\pi^{-}\pi^{0}} \propto \Gamma_{0} (m_d - m_u)^2 .
\label{wzor_na_gamma}
\end{equation}
More explicitly the decay width can be written in a convenient form as:
\begin{equation}
\Gamma_{\eta(\eta^{\prime})\to \pi^{+}\pi^{-}\pi^{0}} = \Gamma_{0} \left(\frac{Q_D}{Q}\right)^4
\end{equation}
where the dependence of the $d$ and $u$ quarks mass difference is contained in the $Q$ term:
\begin{equation}
\frac{1}{Q^2}  = \frac{m^2_d - m_u^2}{m^2_s - \frac{1}{4}(m_d + m_u)^2},
\end{equation}
and $Q_D$ is given by a relation:
\begin{equation}
Q^2_D = \frac{m^2_K}{m^2_{\pi}}\frac{m^2_K - m^2_{\pi}}{m^2_{K^0} - m^2_{K^+} m^2_{\pi^+} - m^2_{\pi^0} }.
\end{equation}
$Q_D$ and $\Gamma_0$ correspond to $Q$ and $\Gamma$ calculated in the Dashen limit \cite{dashen} 
where the quark masses are constrained by the assumption that the electromagnetic mass difference 
for kaons and pions are equal. Using the leading order expression of masses of the pseudoscalar mesons and 
applying the Dashem theorem, it was numerically calculated that $Q_D = 24.1$. 
The decay width $\Gamma$ is very 
sensitive to the exact value of $Q$ and thus provides the precise constraints for the light quark mass ratios.  
Q determines the major axis of the 
ellipse of the light quark mass ratios which is given by the formula:
\begin{equation}
\left(\frac{m_u}{m_d} \right)^2 + \frac{1}{Q^2} \left(\frac{m_s}{m_d} \right)^2 = 1.
\end{equation} 
The ellipse is shown in Fig. \ref{elipsa}.

For the determination of Q we can use $\Gamma_0$ together with the experimantal value of $\Gamma$.
Recently Q value of 22.8 was derived by means of the dispersion relation 
approach from preliminary KLOE data on the $\eta\to\pi^{+}\pi^{-}\pi^{0}$ decay~\cite{martemyanov}. 
In Fig.~\ref{elipsa}, the quark mass ratio is indicated by the shaded uncertainity ellipse. 
The upper and lower sides of the ellipse is 
bound by two dashed lines corresponding to Q=22.7~$\pm$~0.8. From the left it touches the hatched
region, excluded by the bound $\Delta_M >0$, calculated  in the leading term in 
the expansion of powers  $1/N_c$ and extended to first non-leading order.
Other bounds are given by the limits of the branchning ratios 
$R = \Gamma_{\Psi^{\prime}\to\Psi\pi^0}/\Gamma_{\Psi^{\prime}\to\Psi\eta}$.  
The dot corresponds to Weinberg's value, and the cross represents the estimations described in the reference~\cite{gasser82}.
The Weinberg ratio corresponds to $\Delta_M = 0$, and is located at the boundary of this region. 
In particular this implies $m_u/m_d > \frac{1}{2}$ and excludes a massless $u$ quark.  

The $\Gamma_{0}$ factor can be calculated in the chiral limit where the mass of 
quarks tends toward zero $m_u = m_d = 0$.
Estimations of the decay width by the Chiral Perturbation Theory are based on the leading term of the 
expansion in the quark masses and the precise calculations of $\Gamma_0$ in the isospin limit.
In case of this method the normalization for the decay width must be obtained from other 
experminents with the electroproduction of the $\eta^{\prime}$ meson.

Gross, Treiman and Wilczek~\cite{gross79} have proposed another method of establishing the 
light quark mass difference by finding the ratio of partial widths for the isospin violating
$\eta^{\prime}\to \pi^{+}\pi^{-}\pi^{0}$ decay to the isospin conserving process of $\eta^{\prime}\to\eta\pi^{+}\pi^{-}$:
\begin{equation}
r = \frac{\Gamma_{\eta^{\prime}\to \pi^{+}\pi^{-}\pi^{0}}}{\Gamma_{\eta^{\prime}\to \eta\pi^{+}\pi^{-}}} 
 \approx (16.8)\frac{3}{16}\left(\frac{m_d - m_u}{m_s}\right)^2 ,
\label{br_gross}
\end{equation}
where the factor 16.8 denotes the ratio of phase-space volumes. Formula~(\ref{br_gross}) 
was claimed to be true 
under two assumptions: that the amplitudes for both decays are constant over the phase space, and the 
amplitudes are related via \cite{gross79}:
\begin{equation}
A(\eta^{\prime}\to \pi^{+}\pi^{-}\pi^{0}) = \sin\theta~A(\eta^{\prime}\to \eta\pi^{+}\pi^{-}) , 
\end{equation}
where $\sin\theta$ indicates the $\pi^{0}-\eta$ mixing angle:
\begin{equation}
\sin\theta = \frac{\sqrt{3}}{4}\frac{m_d - m_u}{m_s} .
\end{equation}
The second assumption directly implies that the decay $\eta^{\prime}\to \pi^{+}\pi^{-}\pi^{0}$
proceeds entirely through the channel $\eta^{\prime}\to \eta\pi^{+}\pi^{-}$ followed by $\pi^{0}-\eta$ mixing. 

From equation (\ref{br_gross}) one can see that the measurement of the ratio does not require the 
information from other experiments for normalization of the partial decay width. Additionally  with the 
simultaneous measurement of these decays, due to the similar final state, many systematical uncertainties 
will cancel. 

But recently Borasoy et al. \cite{borasoy06} claimed that the light quark 
masses can not be extracted from the ratio (\ref{br_gross}). They studied the two assumptions which were 
mentioned above using the U(3) chiral unitary framework which is in good agreement with $\eta^{\prime}$
data as regarding widths and spectral shapes. They showed that results from the chiral unitary approach  are in 
disagreement with these two assumptions. 
Concluding that on the theoretical side a still more sophisticated treatment including the final state 
interaction is required for univocal  statements about the quark masses~\cite{barasoy_aip}.  

In our experiment we intend to establish the ratio 
$\Gamma_{\eta^{\prime}\to\pi^{+}\pi^{-}\pi^{0}}/\Gamma_{\eta^{\prime}\to\pi^{+}\pi^{-}\eta }$
hoping that the near future progress on the theory side will permit us to determine univocally from 
this ratio the light quark mass difference $m_d - m_u$.
Further more, in combination with experiments determining values of quark mass ratios 
$m_s/m_d$ and $m_u/m_d$, it would be possible to obtain the absolut values of the $u$ and $d$ quark masses.

\chapter{Measurement method of the $\Gamma_{\eta^{\prime}\to\pi^{+}\pi^{-}\pi^{0}}$ }
\hspace{\parindent}
The partial width $\Gamma_{\eta^{\prime}\to\pi^{+}\pi^{-}\pi^{0}}$ may be determined from the 
branching ratio of the $\eta^{\prime}\to\pi^{+}\pi^{-}\pi^{0}$ decay and the total width of the $\eta^{\prime}$ meson.
In the WASA-at-COSY experiment the $\eta^{\prime}$ meson will be created in 
two proton collisions via the $pp\to pp\eta^{\prime}$ reaction. The measurement have to be performed
at best possible conditions, because the $\eta^{\prime}\to\pi^{+}\pi^{-}\pi^{0}$ decay channel is very rare.  
So far the $\eta^{\prime}\to \pi^{+}\pi^{-}\pi^{0}$ decay was never observed
and only an upper limit of 5\% for the branching ratio has been established \cite{rittenberg69}.
Recent theoretical calculations, based on a chiral unitary approach, predict the branching ratio
of about 1\% \cite{borasoy06}. 

\section{Branching ratio $BR(\eta^{\prime}\to\pi^{+}\pi^{-}\pi^{0})$}
\hspace{\parindent}

The branching ratio is a quantity which informs about the probability of the particle 
decay into a specific channel. It is defined as: 
\begin{equation}
BR_i = \frac{\Gamma_i}{\Gamma_{tot}},
\label{br}
\end{equation} 
where $\Gamma_i$ and $\Gamma_{tot}$ denote the partial and total (natural)
width of the particle, respectively. Index $i$ indicates a decay channel.   
From (\ref{br}) we can see that the determination of  partial 
widths will rely on the precise determination of branching ratio and of the total width.

For the $\eta^{\prime}\to\pi^{+}\pi^{-}\pi^{0}$ decay the branching ratio can be 
explicitly expressed as:
\begin{equation}
BR(\eta^{\prime}\to \pi^{+}\pi^{-}\pi^{0}) = \frac{\Gamma_{\eta^{\prime}\to \pi^{+}\pi^{-}\pi^{0}} } { \Gamma_{\eta^{\prime}}^{tot} },
\end{equation}
where $\Gamma_{\eta^{\prime}\to \pi^{+}\pi^{-}\pi^{0}}$ is the searched observable. 
The $\Gamma_{\eta^{\prime}}^{tot}$ is known from the Particle Data Group (PDG) estimations to be  
$\Gamma_{\eta^{\prime}}^{tot} = 0.202$~$\pm$~$0.016$ MeV/c$^2$ \cite{pdg}\footnote{
The COSY-11 collaboration is working on a more precise estimation of 
this value with a dedicated measurement of the $\eta^{\prime}$ 
mass distributions by using the missing mass techniques applied to the $pp\to ppX$ 
reaction very close to the kinematical threshold \cite{eryk_aip07}.}, and  
BR($\eta^{\prime}\to \pi^{+}\pi^{-}\pi^{0}$) still remains to be established experimentally.
With the WASA-at-COSY facility we plan to determine
this branching ratio as the ratio of the number of events where $\eta^{\prime}$ decays into the $\pi^{+}\pi^{-}\pi^{0}$ 
system ($N_{\eta^{\prime}\to \pi^{+}\pi^{-}\pi^{0}}$) to the number of all produced 
$\eta^{\prime}$ mesons ($N_{\eta^{\prime}}^{tot}$):
\begin{equation}
BR(\eta^{\prime}\to \pi^{+}\pi^{-}\pi^{0}) = \frac{N_{\eta^{\prime}\to \pi^{+}\pi^{-}\pi^{0}} } { N_{\eta^{\prime}}^{tot} }.
\label{br_exp}
\end{equation}
For this purpose 
the $\eta^{\prime}$ mesons will be produced in collisions of the proton beam with 
the hydrogen pellet target via the $pp\to pp\eta^{\prime}$ reaction.
The outgoing protons and the products of the decay of the $\eta^{\prime}$ meson will be 
detected and identified using the WASA-at-COSY detection system. More details of the 
measurement techniques will be given in the following sections.   

\section{Description of the WASA-at-COSY detector facility} 
\begin{figure}[H]
\hspace{2.2cm}
\parbox{0.72\textwidth}{\centerline{\epsfig{file=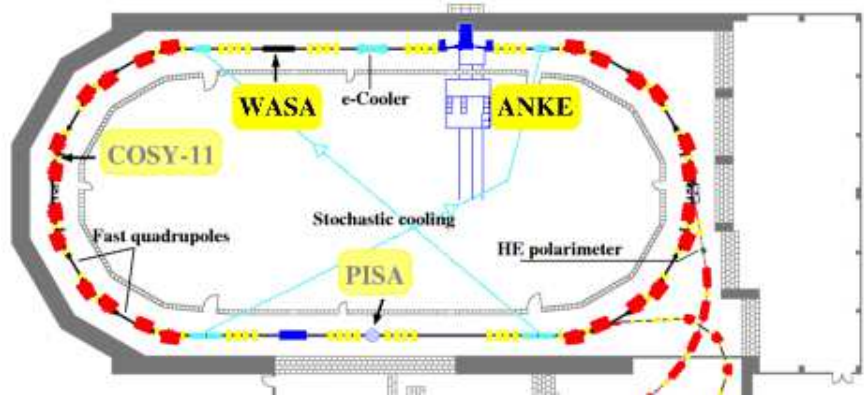,width=0.77\textwidth}}}
\caption{
Schematic view of the COoler SYnchrotron ring in the Research Centre J\"{u}lich.
The presently used detector systems: WASA~\cite{proposal} and ANKE~\cite{barasov} and completed  experiments 
COSY-11~\cite{brauksiepe96} and PISA~\cite{barna04} are shown.  
}
\label{cosyring}
\end{figure}

\hspace{\parindent}
The COoler SYnchrotron (COSY) is a storage ring operating at the Research Centre 
J\"{u}lich (Germany) since 1993 \cite{maier}. It delivers unpolarized and polarized 
proton and deuteron \cite{raport_ikp_2005} beams in the momentum range  
between 300 and 3700 MeV/c.
The first step of the particle acceleration  takes place in the
isochronous cyclotron (JULIC).
Next the beam is injected into the 184 m long COSY ring (see Fig. \ref{cosyring}) where the particles 
experience further acceleration and at present may be used for experiments with 
the WASA \cite{proposal}, ANKE \cite{barasov} and TOF (external beam) \cite{tof-proposal} 
detection setups where the energy of the beam allows for production 
of all basic pseudoscalar and vector mesons. 

The COSY accelerator is equipped with two types of beam cooling systems: an electron and stochastic cooling 
used for low and high energies, respectively~\cite{stockhorst2001}. 
Both cooling systems allow to decrease the momentum and geometrical spread of the beam.
The whole process for the  beam preparation from injection till final state of acceleration
takes a few seconds. The ring can be filled with up to $10^{11}$ particles, 
and the life time of the circulating beam ranges from minutes to hours depending 
on the thickness of the used target.     

\begin{figure}[ht]
\hspace{1.1cm}
\parbox{0.90\textwidth}{\centerline{\epsfig{file=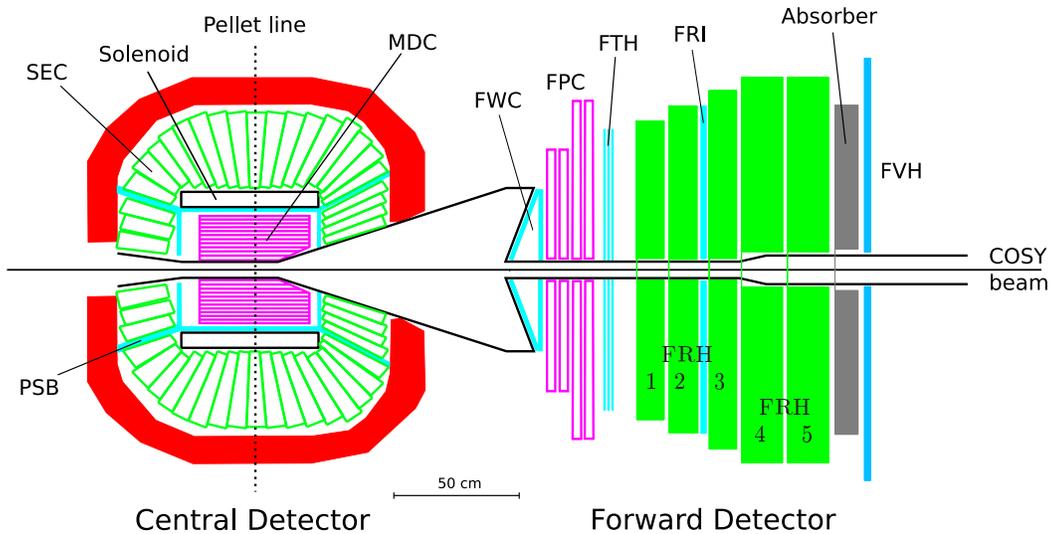,width=0.95\textwidth}}}
\begin{picture}(0,0)
\put(-113,-45){\footnotesize{FRH}}
\put(-115,-55){\footnotesize{4}}
\put(-97,-55){\footnotesize{5}}
\put(-150,-25){\footnotesize{FRH}}
\put(-156,-35){\footnotesize{1}}
\put(-145,-35){\footnotesize{2}}
\put(-130,-35){\footnotesize{3}}
\end{picture}
\caption{
Cross section of the WASA-at-COSY detector with new FRH4 and FRH5 layers,
mounted in August 2006 \cite{calen_raport07} and newly assambled two 
layers of the FWC \cite{pricking07}.
Names of the detectors are explained in the text.  
}
\label{wasacross07}
\end{figure}

In 2006 the WASA detector \cite{bargholtz} has been transferred from the CELSIUS \cite{bilger} facility in Uppsala 
to the COSY ring \cite{maier} in J\"{u}lich where it was successfully installed and brought into operation.
The WASA-at-COSY, shown schematically in Fig. \ref{wasacross07}, is a large acceptance detector which consists of
three main parts: the Central Detector (CD), the Forward Detector (FD) and the 
pellet target system \cite{trostell95,ekstrom96,alexander}. 

The Central Detector (CD)~\cite{schuberth} is used for detection and identification of charged
and neutral particles ($\gamma, \pi^{+}, \pi^{-}, e^{+}, e^{-}$) which are the decay products 
of short lived mesons like $\pi^0$, $\eta$ and $\eta^{\prime}$. 
It covers scattering angles between $20^{o}$ and $169^{o}$. 
The most outer part of the CD constitutes an electromagnetic calorimeter (SEC) built out of 1012 CsI 
scintillating crystals positioned around the interaction point \cite{brjany}. 
The crystals are ordered
in 24 rings between the iron yoke and super-conducting solenoid which provides an axial magnetic
field enabling the determination of the momenta of charged particles by 
measuring their tracks with the Mini Drift Chamber (MDC) \cite{jacewicz}.
The MDC is mounted around the beam pipe inside of the solenoid. 
It is a cylinder which consists of 17 layers of straw tubes. Each straw is 
made out of thin mylar foil tube, with a gold plated sensing wire in its center.
The drift chamber is surrounded with the Plastic Scintillator Barrel (PSB) \cite{jacewicz} used for 
trigger purposes and for the determination of the energy loss for charged particles. 
PSB together with SEC and MDC permits the identification of charged particles by means of the
energy loss method ($\Delta E$).

The forward part of the WASA detector is built out of thirteen scintillating layers 
and four layers of straw drift chambers. The forward detector was designed for the 
detection and identification of protons, deutrons and He nuclei. 
The detector closest to the scattering chamber - Forward Window Counter (FWC) - 
covers the conical exit window of the axially symmetric scattering chamber towards the 
Forward Detector assembly \cite{pricking07}. Presently it is used for trigger purposes, but in the future it will 
also serve as a start detector for the time-of-flight determination. 

For the track reconstruction the Forward Proportional Chamber (FPC)~\cite{dyring} is used. 
It provides information about particle scattering angles with a precision better than 0.2$^o$~\cite{janusz07}. 
The FPC planes are rotated by $45^o$ with respect to each other. 
Directly behind the straw detector the three layer Forward Trigger Hodoscope (FTH)~\cite{dahmen,waters} is placed. 
It consists out of 96 individual plastic scintillator elements arranged in three layers: two layers with modules 
in the form of Archimedean spiral and one layer with cake-piece shaped modules.  
The FTH provides information about charged particle multiplicities in the FD used 
for the first level trigger, as well as in the offline track reconstruction~\cite{pauly-an07}. Future trigger 
developments aim at a complete real time scattering angle reconstruction of individual tracks. 
This information, combined with the information of deposited energy in successive detector layers,
will allow to determine the missing mass of forward going particles on the trigger level, and thus will provide a 
very efficient meson tagging~\cite{pauly-an07}. The FTH detector can also be used to deliver the time information
and it is thus helpful for the application of the TOF technique.

\begin{figure}[h]
\hspace{4.1cm}
\parbox{0.50\textwidth}{\centerline{\epsfig{file=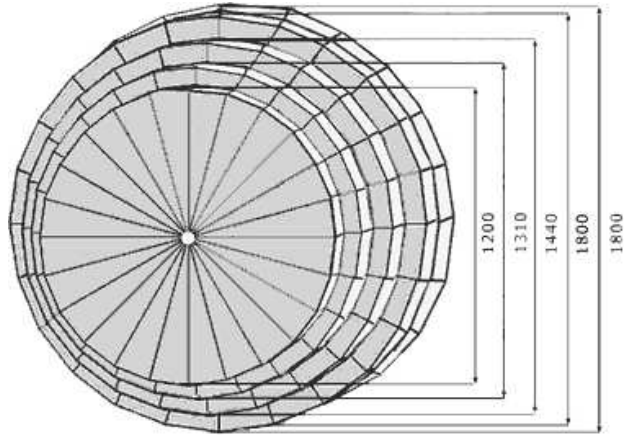,width=0.55\textwidth}}}
\caption{
View of the Forward Range Hodoscope build out of the cake-pieced scintillaiting 
modules arranged in five layers~\cite{calen_raport07}.
}
\label{frh_view}
\end{figure}

The next five thick planes are called Forword Range Hodoscope (FRH) and are made out of cake-piece 
shaped plastic scintillator modules (see Fig.~\ref{frh_view}). This detector enables  identification 
of charged particles from the absolute value and the pattern of energy deposited in 
different layers.
Among the second and third layer of the FRH
the Forward Range Interleaving hodoscope (FRI) \cite{pauly05A547} is mounted. It is composed of 
two interleaving planes of 32 plastic scintillators bars aligned horizontally and vertically.   
The main function of the FRI 
is the measurement of the hit position of the charged particles improving the resolution 
for the vertex reconstruction. In addition, the FRI delivers also time information which can be used in the 
future application of the time-of-flight method.

The Forward Veto Hodoscope (FVH) \cite{brodowski} is the last detection layer. It consists of
12 horizontally placed plastic scintillator modules equipped with the photomultiplayers 
on both sides. The FVH gives the position of particle hits, reconstructed
from time signals determined on two side of the module.
It is important to note that 
this detector can also be used to deliver time information ($t_{stop}$) 
for the TOF technique.
It is also worth mentioning that in the near future it is planned to extend the FVH 
by a second layer of vertically arranged scintillators~\cite{mikhail-priv}.  

Finally, depending on the studied reaction, a passive absorber (FRA) made of iron 
can be positioned in front of the FVH, enabling to disantangle between slow 
and fast particles. The thickness of the material can be choosen from 5~mm up to 100~mm.   

From the beginning of its operation at CELSIUS and COSY the WASA detector is equipped
with a pellet target system~\cite{trostell95,ekstrom96,alexander},
providing a stream of frozen  hydrogen droplets with a diameter of about 35~$\mu$m. 
Pellets are passing to the interaction region through a thin 2~m long pipe. 
Usage of this type of target enables to achieve densities of up to 10$^{15}$ atoms/cm$^{2}$ 
resulting  in luminosities of up to 10$^{32}$cm$^{-2}$s$^{-1}$ when combined with the 
COSY beam of 10$^{11}$ particles stored in the ring.

\section{Identification of the $pp\to pp\eta^{\prime}\to pp\pi^{+}\pi^{-}\pi^{0}\to pp\pi^{+}\pi^{-}\gamma\gamma$ 
          reaction chain}
          
\hspace{\parindent}
The proposed experiment aiming at the determination of the branching ratio for the 
$\eta^{\prime}\to\pi^{+}\pi^{-}\pi^{0}$ decay will be based on the production of the $\eta^{\prime}$
meson in proton-proton collisions and registration of its decay products. The complete 
reaction chain which needs to be identified  (production and decay) reads:
\begin{equation}
\nonumber
pp \to pp \eta^{\prime} \to  pp\pi^{+}\pi^{-}\pi^{0} \to pp\pi^{+}\pi^{-}\gamma\gamma .
\end{equation}

The measurement relay on the registration of all particles in the final state and on the determination 
of their four-momenta: 
\begin{equation}
\nonumber
\{\mathbb{P}_{p_1}, \mathbb{P}_{p_2}, \mathbb{P}_{\pi^{+}}, 
\mathbb{P}_{\pi^{-}}, \mathbb{P}_{\gamma_{1}}, \mathbb{P}_{\gamma_{2}}\}.
\end{equation}
In order to measure the energy of the two forward scattered protons the FRH1-5 planes in the Forward 
Detector are used which provide information of the energy losses. To determine the 
direction $\vec{r}$ of these two protons we use the straw chambers of the FPC and the FRI detector.
Charged pions scattered under angles larger then 18$^o$  
will be registered in the Central Detector.
The MDC will provide information about the direction $\vec{r}(\theta,\phi)$ 
and the SEC will allow to measure the energy losses. 
In the case when the charged pion will be scattered forward 
(angular range of 2.5$^o$ - 18$^o$) it will be detected by the Forward Detector 
in the same way as the protons. The two gamma quanta originating from the $\pi^0$ decay will be 
recorded in the SEC from which we will receive the information about their energy and direction. 

\section{Invariant and missing mass techniques}

\hspace{\parindent}

At the first stage of the data analysis we have to select from the full data sample 
only these events where $\pi^{+}\pi^{-}\pi^{0}$ were produced and in the second step using a
missing mass technique we will determine fractions of selected events corresponding
to the direct production and to the decays of $\eta^{\prime}$ meson.

In order to identify signals from the $\pi^{0}$ mesons, we will reconstruct the invariant mass of two 
gamma quanta:
\begin{equation}
m_{\gamma_1\gamma_2} = \sqrt{\mathbb{P}^{2}_{\gamma_1} + \mathbb{P}^{2}_{\gamma_2}} ,
\end{equation}
which should be equal to the mass of the neutral pion within the expected resolution.
Next we identify the charged pions on the basis of the energy measured in the
SEC ($E_{\pi^{\pm}}$) and their momenta ($\vec{p}_{\pi^{\pm}}$) derived from the curvature
of the tracks reconstructed from signals measured with the MDC. In the case of the $\pi^{\pm}$
the energy and momenta should fulfill (within the expected resolution) the relation:
\begin{equation}
m_{\pi^{\pm}} = \sqrt{E^{2}_{\pi^{\pm}} - \vec{p}^{2}_{\pi^{\pm}}},
\end{equation}
where $m_{\pi^{\pm}}$ denotes the mass of a charged pion. 

Further on, from measuring the energy loss in the five scintillator layers of the
FD and tracks in the FPC (in the future also
from the time-of-flight) we reconstruct the four-momentum vector of the forward emitted protons~\cite{jozef}.
Next in order to check whether the identified pions originate from the decay of 
the $\eta^{\prime}$ meson, we calculate a missing mass of the $pp \to pp X$ reaction
according to the equation (see Appendix~\ref{app:mm}):  
\begin{equation}
m_X = \sqrt{ (E_{beam} + m_{target} - E_{p_1} - E_{p_2})^2 - (\vec{p}_{beam} - 
              \vec{p}_{1} - \vec{p}_{2})^2 } ,
\label{masa_brakujaca}
\end{equation}
where $m_X$ corresponds to the mass of an unobserved particle, 
$E_{beam},\vec{p}_{beam}$ denote the energy and momentum of the beam respectively, and $E_{p_1},E_{p_2},\vec{p}_{1},\vec{p}_{2}$ represent energies and momenta of two registered protons.
\begin{figure}[ht]
\hspace{2.7cm}
\parbox{0.60\textwidth}{\centerline{\epsfig{file=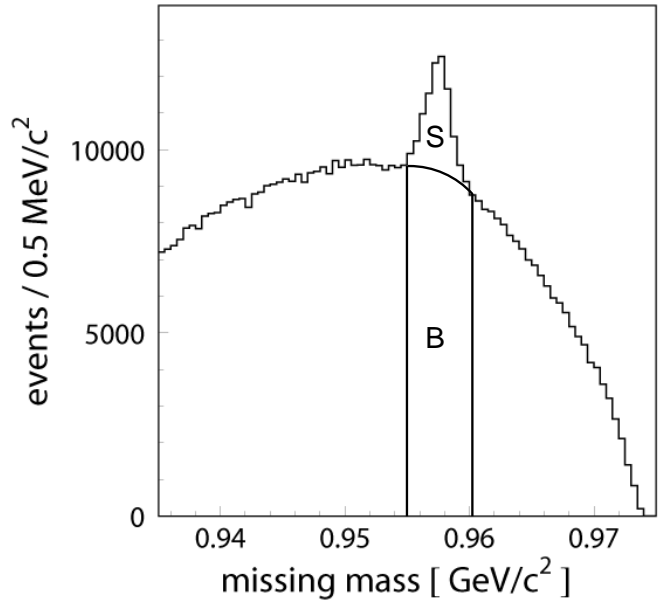,width=0.65\textwidth}}}
\caption{
Example of the missing mass distribution for the reaction $pp\to ppX$ from the
COSY-11 measurements at the excess energy Q = 15.5 MeV~\cite{moskal_meson06}. 
$S$ denotes the signal originating from the $\eta^{\prime}$ production and $B$ 
indicates the background under the peak.   
}
\label{wykres_masy}
\end{figure}
At present the experiment is still in preparation therefore   
as an example of the reconstructed missing mass distribution in Fig.~\ref{wykres_masy} 
we show data derived from the COSY-11 measurement  carried out at an excess energy of 
Q=15.5 MeV above the threshold for the $\eta^{\prime}$ meson production~\cite{moskal_meson06}.
The number of events which correspond to the production of the $\eta^{\prime}$ meson 
is marked by the letter S, 
and the field of the slice under the $\eta^{\prime}$ peak correspond to the number of the 
background events marked by B. 
The background originates from the direct two, three and more pions production.
It maybe treated as the upper limit for the expected background of the 
$pp\to pp\pi^{+}\pi^{-}\pi^{0}$ reaction. 
In the WASA-at-COSY experiment we expect a similar spectrum for the missing mass 
of the  $pp\to pp(X=\pi^{+}\pi^{-}\pi^{0})$ events.
However, the signal to background ratio could be different due to the different mass 
resolution of the WASA-at-COSY detector. Additionally it will vary with the 
excess energy\footnote{Excess energy Q is a kinetic energy available in the reaction exit channel
in the center-of-mass system.}. 
Thereforee, in order to find an optimum beam momentum for the BR($\eta^{\prime}\to \pi^{+}\pi^{-}\pi^{0}$)
determination the energy dependence 
(i)~of the signal ($pp\to pp\eta^{\prime}\to pp\pi^{+}\pi^{-}\pi^{0}$), 
(ii)~of the background ($pp\to pp\pi^{+}\pi^{-}\pi^{0}$), 
(iii)~of the missing mass resolution and 
(iv)~of the detection efficiency 
has to be established. 
The upper limit of number of background events may be obtained from results achieved by the COSY-11 experiment
(where one spectrum is shown in Fig.~\ref{wykres_masy}). A detailed description of the 
derivation of the dependence of S(Q) and B(Q) will be given in the next chapter.

\chapter{Conditions for the determination of the BR($\eta^{\prime}\to \pi^{+}\pi^{-}\pi^{0}$)\\
          via the $pp\to pp\eta^{\prime}\to pp\pi^{+}\pi^{-}\pi^{0}$ reaction chain } 
\hspace{\parindent}

The number for all produced $\eta^{\prime}$ mesons 
$N_{\eta_{\prime}}^{tot}$ (see eq.~\ref{br_exp}) can be determined from the known 
total cross section for the $pp\to pp\eta^{\prime}$ reaction~\cite{pawel,alfons2004,cosy11-98,cosy11-00,hibou,balestra} 
and the luminosity which will be established by the measurement of the reaction with the well known cross 
section e.g. from elastic scattering of protons ($pp\to pp$)~\cite{albers}.
  
When we assume that the relative statistical error of $N_{\eta_{\prime}}^{tot}$ can be neglected 
due to large number of measured $pp\to pp$ or $pp\to pp\eta^{\prime}$ events  the formula 
for the relative error of the branching ratio is reduced to:
\begin{equation}
\frac{\sigma(BR)}{BR} = \frac{\sigma(N_{\eta^{\prime}\to \pi^{+}\pi^{-}\pi^{0}})}{N_{\eta^{\prime}\to \pi^{+}\pi^{-}\pi^{0}}},
\end{equation}
where $\sigma(N_{\eta^{\prime}\to \pi^{+}\pi^{-}\pi^{0}})$ denotes the statistical uncertainty
of the signal.  
Assuming that the shape of the background is known the statistical error of the 
signal can be  approximated   as:
\begin{equation}
\sigma(N_{\eta^{\prime}\to \pi^{+}\pi^{-}\pi^{0}}) \approx \sqrt{N_{\eta^{\prime}\to \pi^{+}\pi^{-}\pi^{0}} + N_{B}},
\label{blad_22}
\end{equation}
where $N_{B}$ indicates the number of all background events under the signal.
Equation~\ref{blad_22} was obtained under the assumption that the statistical errors propagate according to the formula \cite{kamys}:
\begin{equation}
\sigma(Y(x_1,...,x_n)) = \sqrt{ \sum\limits_{i=1}^n \left(\frac{\partial Y}{\partial x_i}
                         \sigma(x_i)\right)^2 },
\end{equation} 
where the variables $x_1,...,x_n$ are independent and  $\sigma(x_i)$ denotes their uncertainty. 
Thus the relative accuracy of the branching ratio can be expressed as: 
\begin{equation}
\frac{\sigma(BR)}{BR} = \frac{\sqrt{N_{\eta^{\prime}\to \pi^{+}\pi^{-}\pi^{0}} + N_{B}} }{N_{\eta^{\prime}\to \pi^{+}\pi^{-}\pi^{0}}}.
\label{blad_wzgledny_br}
\end{equation}
The number of events where $\eta^{\prime}$ decayed into three pions 
which we expect to register with the WASA-at-COSY facility,
can be expressed by the formula:
\begin{equation}
N_S(Q) \equiv N_{\eta^{\prime}\to \pi^{+}\pi^{-}\pi^{0}}(Q) = \sigma^{tot}_{\eta^{\prime}}(Q)\cdot BR(\eta^{\prime}\to \pi^{+}\pi^{-}\pi^{0}) \cdot A(Q) \int\limits_{0}^{\Delta t} L\cdot dt,
\label{liczbaet}
\end{equation}
where $L$ is the luminosity which is expected to be around 
10$^{32}$ cm$^{-2}$s$^{-1}$,
$\sigma^{tot}_{\eta^{\prime}}$ denotes the total cross section for the $\eta^{\prime}$ meson production
which depends on the excess energy Q, $BR$ is the supposed  branching ratio for
that decay, $\Delta t$ indicates the measurement time and $A(Q)$ representes the acceptance  
of the WASA-at-COSY detector for the measured reaction.

In the following section we will estimate the uncertainty of the branching ratio
determination as a function of the excess energy and measurement time taking into 
account the energy resolution of the forward part of the WASA detector.
To this end, in the following we will parametrize the energy dependence of the total cross section
for the $\eta^{\prime}$ meson and for the multimeson production. 
Presented estimations of the multimeson production will allow to estimate an upper limit of the 
background for the $\eta^{\prime}\to\pi^{+}\pi^{-}\pi^{0}$ decay. 
This is because it will be based on the missing mass distributions which 
includes also other pion channels like e.g. $pp\to pp2\pi$, $pp\to pp4\pi$, $pp\to pp2\pi\eta$.
    
\section{Parametrization of the total cross section for the $\eta^{\prime}$ meson production}
\begin{figure}[H]
\hspace{2.7cm}
\parbox{0.63\textwidth}
{\vspace{-1.6cm}\centerline{\epsfig{file=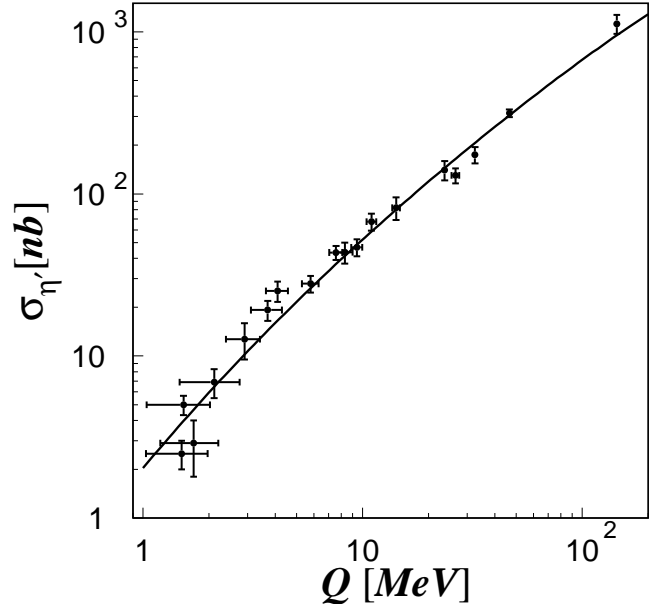,width=0.67\textwidth}}}
\caption{
Experimental data points of total cross section for the $pp\rightarrow pp\eta^{\prime}$ reaction 
derived from the COSY-11, DISTO and SPESIII measurements~\cite{alfons2004,cosy11-98,cosy11-00,hibou,balestra}.
The solid line denotes parametrization of the cross section using formula~(\ref{przekrojparam}).
The figure was adapted form references~\cite{moskal_meson06,eryk_mgr}.
}
\label{przekrojeta}
\end{figure}

\hspace{\parindent}
In order to estimate the expected production rate of the $\eta^{\prime}$ meson 
given by formula (\ref{liczbaet})
it is mandatory to know the total cross section for the production of this meson. 
From measurements of the COSY-11, DISTO and SPESIII~\cite{alfons2004,cosy11-98,cosy11-00,hibou,balestra} 
collaborations we have several experimental data 
points for the excess energy range from 1.3 MeV up to 150 MeV. 
The data are shown in Fig.~\ref{przekrojeta}, where it is clearly seen that near the threshold 
production of the $\eta^{\prime}$ meson depends strongly on the excess energy and hence this dependence
must be taken into account.
Generally the production cross section can be described 
as an integral of the square of the production amplitude $\vert$M$\vert^2$ 
over the phase space volume V$_{ps}$ using formula:     
\begin{equation}
\sigma^{tot}_{\eta^{\prime}}(Q) = \frac{1}{F}\int dV_{ps} \vert M\vert^2, 
\end{equation}
where F denotes the flux factor of the colliding particles. 
This energy dependence of the cross section can be written in a closed analitical form using  the 
F\"{a}ldt and Wilkin model \cite{faldt1,faldt2} which takes into account
the proton-proton Final State Interaction (FSI):
\begin{equation}
\sigma^{tot}_{\eta^{\prime}}(Q) = C_1 \frac{V_{ps}}{F} \frac{1}{1 + \sqrt{ 1 + \frac{Q}{\epsilon}}} = C_2 \frac{Q^2}{\sqrt{\lambda(s,m^2_1,m^2_2)}}\frac{1}{1 + \sqrt{ 1 + \frac{Q}{\epsilon}}},
\label{przekrojparam}
\end{equation}
where $C_1$ and $C_2$ denotes the normalization constant, $\epsilon$ stands for the binding energy \cite{faldt1,faldt2}, 
and the $\lambda(s,m^2_1,m^2_2)$ is the triangle function~\cite{particlekin}, defined as:
\begin{equation}
\lambda(x,y,z) = x^2 + y^2 + z^2 - 2xy - 2xz - 2yz.
\end{equation}
The free parameters $\epsilon$ and $C_2$ have to be established by fitting the 
function~(\ref{przekrojparam}) to the experimental data.
The estimation of these parameters and the fit was done in reference~\cite{eryk_mgr}, 
and the obtained values amount to~\cite{moskal_meson06}:
\begin{equation}
\nonumber
\epsilon = 0.62 \pm 0.13 \;\textrm{MeV}
\end{equation}
\begin{equation}
\nonumber
C_2 = 84 \pm 14 \;\textrm{mb}.
\end{equation} 
Knowing the parametrization from~(\ref{przekrojparam}) and having the values of the 
parameters we can compute the cross section for the $\eta^{\prime}$ 
meson production for  excess energies near threshold.    
The result of the fit together with the experimental data is shown 
in Fig.~\ref{przekrojeta}.

\section{Differential cross section $\left.   
         \frac{d\sigma}{dm}\right|_{m=m_{\eta^{\prime}}}$ for multimeson production }
\hspace{\parindent}
For the estimation of the background from the direct $\pi^{+}\pi^{-}\pi^{0}$ production 
in the missing mass spectrum of the $pp\to pp(X=\pi^{+}\pi^{-}\pi^{0}$) reaction, 
expected to be observed by the WASA-at-COSY detector, we 
need to calculate the $\left.\frac{d\sigma(pp\to pp\pi^{+}\pi^{-}\pi^{0})}{dm_x}\right|_{m_x=m_{\eta^{\prime}}}$ and the 
missing mass resolution ($\Delta_{w})$ of the WASA detector setup.

To our knowledge there are no data available on the invariant mass distributions or even 
on the total cross section for the $\pi^{+}\pi^{-}\pi^{0}$ production in the proton-proton 
reaction near the kinematical threshold of the $\eta^{\prime}$ meson production\footnote{
The total cross section of the $pp\to pp\pi^{+}\pi^{-}\pi^{0}$ reaction has been established experimentally 
only at three proton beam energis~\cite{pickup,hart,eisner}~(see App.~\ref{app:3pi}).}.
Therefore, in order to estimate at least an upper limit of the differential cross section for the $\pi^{+}\pi^{-}\pi^{0}$ production 
we have taken the missing mass spectra of the $pp\to ppX$ reaction determined by the COSY-11 
collaboration for several beam energies near threshold for the $\eta^{\prime}$
meson production~\cite{pawel, alfons2004, cosy11-98, cosy11-00} (see e.g. Fig. \ref{wykres_masy}).
The differential cross section for the background was calculated according to the formula:
\begin{equation}
 \frac{d\sigma_B}{dm} = \frac{N_B(Q)}{N_S(Q)} \frac{\sigma_{\eta^{\prime}}^{tot}(Q)}{\Delta m } ,
\label{diff_cross_section}
\end{equation} 
which was obtained by dividing N$_S$(Q) and N$_B$(Q) as given below. 

\begin{figure}[H]
\hspace{2.5cm}
\parbox{0.65\textwidth}{\vspace{-0.2cm}\centerline{\epsfig{file=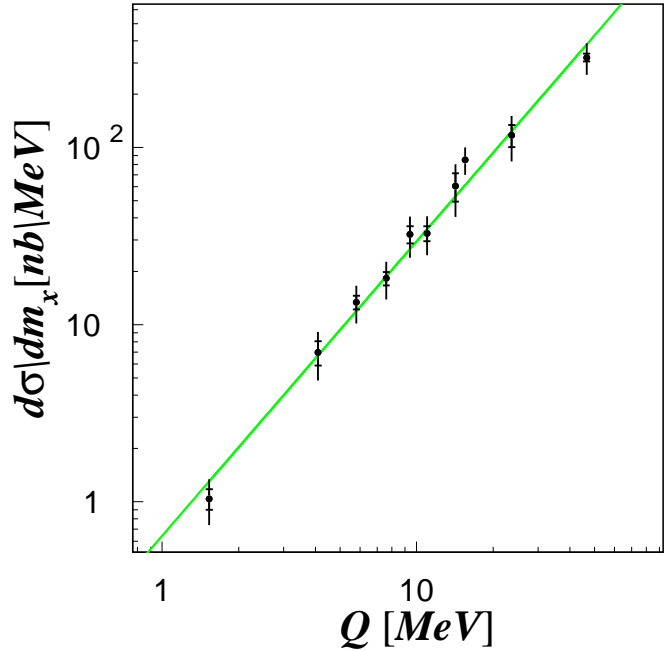,width=0.69\textwidth}}}
\caption{
Inclusive differential cross section for multimeson production derived from the COSY-11 data
\cite{alfons2004, cosy11-98, cosy11-00}.  
The superimposed line shows the function 
$\frac{d\sigma}{dm_x} = \alpha\cdot Q^{\beta}$ fitted to the COSY-11 data, treating $\alpha$ and $\beta$ as free parameters. }
\label{diffcross}
\end{figure}

The number of all measured $pp\to pp\eta^{\prime}$ events can be expressed as:
\begin{equation}
N_{S}(Q) = \sigma_{\eta^{\prime}}^{tot}(Q)   
\cdot A(Q)\int\limits_{0}^{\Delta t}L\cdot dt,
\end{equation} 
and the number of the measured background events can be approximated by:  
\begin{equation}
N_B(Q) \approx \frac{d\sigma_B}{dm}\cdot \Delta m\cdot A(Q) \int\limits_{0}^{\Delta t}L\cdot dt,
\end{equation}
where $A$ denotes the acceptance of the COSY-11 detector,
which in a very good approximation depends only on the mass of the produced system and on the excess 
energy~Q~\cite{moskal2003}, 
$L$ indicates the luminosity, $\Delta t$ is the time of the measurement and $\Delta m$ denotes the 
range of the missing mass values around the $\eta^{\prime}$ signal which is approximated by:    
\begin{equation}
\Delta m \approx 2\Gamma \approx 4.7\sigma_{std}.
\end{equation}
Figure~\ref{diffcross} and Tab.~\ref{tabela1} 
show values of the $\left.\frac{d\sigma}{dm_x}\right|_{m_x=m_{\eta^{\prime}}}$ extracted from the 
experimental data~\cite{alfons2004,cosy11-98,cosy11-00} by means of formula~\ref{diff_cross_section}.           
\begin{table}
\begin{center}
\begin{tabular}{||c|c|c|c||}\hline\hline
Q & $\left.\frac{d\sigma}{dm_x}\right\vert_{m_x = m_{\eta^{\prime}}}$ & $\Delta_{stat}$ & $\Delta_{syst}$   \\ 
$\left[MeV\right]$ & $\left[\frac{nb}{MeV}\right]$ & $\left[\frac{nb}{MeV}\right]$ & $\left[\frac{nb}{MeV}\right]$ \\ \hline\hline 
1.53  & 1.04   & 0.14   & 0.16 \\
4.10  & 7.0   & 1.1   & 1.1 \\
5.80  & 13.4  & 1.2   & 2.0 \\
7.60  & 18.2  & 1.6   & 2.8 \\
9.42  & 32.3  & 3.6   & 4.9 \\
10.98 & 32.7  & 3.2   & 4.9 \\
14.21 & 60  & 11  & 9.0 \\
15.50 & 85  & 2.4   & 13\\
23.64 & 117 & 17  & 17\\
46.60 & 322  & 16  & 48\\ \hline\hline
\end{tabular}
\end{center}
\caption{Differential cross section $\left. \frac{d\sigma}{dm_x}\right|_{m_x = m_{\eta^{\prime}}}$ for multimeson production 
in proton-proton collisions extracted from the COSY-11 data \cite{alfons2004,cosy11-98,cosy11-00}.
The systematical error amounts to 15\% as established in reference \cite{cosy11-00}.}
\label{tabela1}
\end{table}
A good description of the data was obtained by the function of the form:
\begin{equation}
\left. \frac{d\sigma}{dm_x}\right|_{m_x = m_{\eta^{\prime}}} \left(Q\right) = 
\alpha\cdot Q^{\beta},
\label{tloparam}
\end{equation}
where $\alpha$ and $\beta$ are free parameters which, 
for Q expressed in units of MeV, were estimated to be 
$\alpha = 0.64 \pm 0.14$~nb/MeV and $\beta = 1.662 \pm 0.081$~\cite{kupsc-menu,raport07-1}.
Therefore, conservatively the signal to background ratio for the WASA-at-COSY detector can 
be calculated using formula \ref{diff_cross_section} by replacing $\Delta m$ by 
the WASA missing mass resolution, and by replacing the cross section $\sigma_{\eta^{\prime}}^{tot}$ 
by the product $\sigma_{\eta^{\prime}}^{tot}\times BR(\eta^{\prime}\to\pi^{+}\pi^{-}\pi^{0})$.

In order to understand the production mechanism of the $\pi^{+}\pi^{-}\pi^{0}$ system 
more precise studies are needed, because till now it did not receive a proper 
attention neither experimentally nor theoretically \cite{kupsc-menu}. 
Therefore, we have conducted investigations by simulating the $\frac{d\sigma}{dm}$ distributions
for several production mechanisms and compared the results to the values of 
$\left.\frac{d\sigma}{dm_x}\right|_{m_x = m_{\eta^{\prime}}}$ extracted from the experimental data. 
A more detailed description of these studies is presented in appendix~\ref{app:sym}.      
   
\section{Energy resolution of the Forward Detector}
\hspace{\parindent}
The last parameter needed to be calculated for the estimation of the background expected  to be observed 
with WASA-at-COSY is the missing mass resolution $\Delta m_{w}$ which depends on the energy resolution 
of the Forward Detector. 

The Forward Detector was previously used for studies of the protons from
the $pp\to pp\eta$ reaction. 
It consists of plastic scintillator layers which measure 
the energy losses, and on their basis the kinetic energy of the 
particles can be reconstructed. The thickness of the detector was optimized for a measurement 
of protons with kinetic energies in the range from 100 to 550~MeV. 
\begin{figure}[ht]
\hspace{2.8cm}
\parbox{0.65\textwidth}{\centerline{\epsfig{file=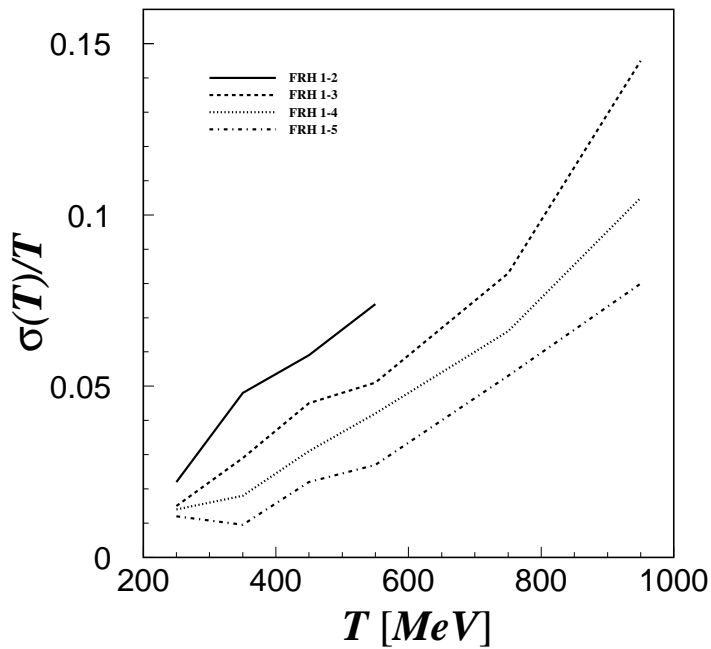,width=0.69\textwidth}}}
\caption{
Relative energy resolution as a function of the proton kinetic energy $T$
achieved using different numbers of FRH planes of the WASA-at-COSY detection setup (courtesy of H.~Cal\'{e}n~\cite{hans08}).
}
\label{FDres}
\end{figure}

In order to study the production and decays of the $\eta^{\prime}$ meson 
an upgrade of the Forward Range Hodoscope was necessary, due to the 
higher background to signal ratio and higher energies of outgoing 
protons from the $pp\to pp\eta^{\prime}$ reaction (kinetic energy ranges 
from 300 to 800~MeV). The extension was done by removing one 11~cm thick layer and instead 
adding two new layers of 15~cm thickness each built out of cake-piece shaped plastic 
scintillators with a photomultiplier tube attached at the end~\cite{calen_raport07}. 
At present the total thickness of FRH detector amounts to 63~cm, which
enables to stop protons with kinetic energies up to 360 MeV.          
Thanks to the new layers the accuracy of the determination of proton energies
was improved by about 25\%. Fig.~\ref{FDres} shows the relative 
\begin{figure}[h]
\hspace{0.5cm}
\parbox{0.95\textwidth}{\vspace{-0.1cm}\centerline{\epsfig{file=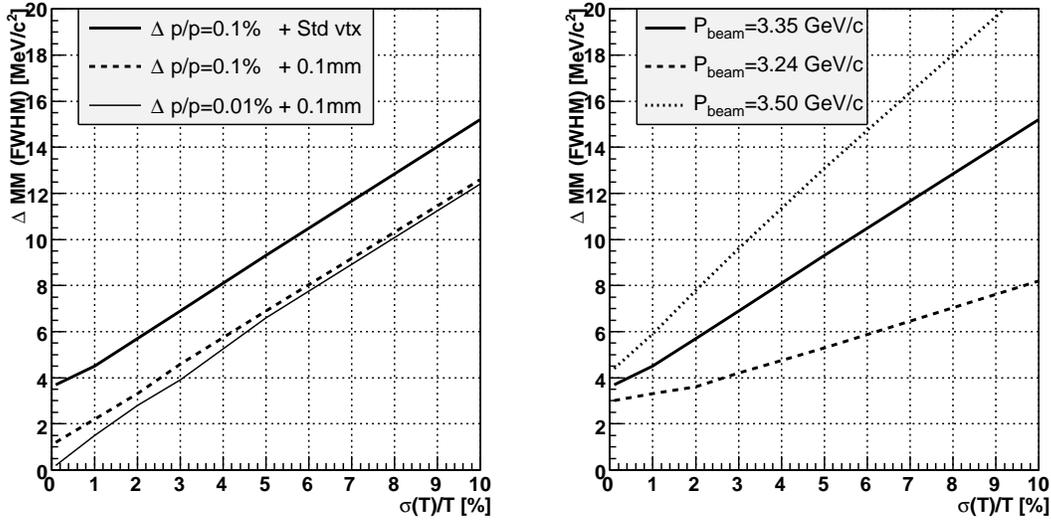,width=0.99\textwidth}}}
\caption{
({\bf left}) Accuracy of the missing mass reconstruction as a function of fractional kinetic energy resolution simulated for
a beam momentum of $P_{beam}$=3.35 GeV/c.
The $\frac{\Delta p}{p}$ denotes the beam momentum resolution and the 'Std vtx' indicates the standart vertex resolution.
({\bf right}) Simulation assuming $\frac{\Delta p}{p}$ = 0.1\% and standard vertex resolution, 
for three beam momenta as indicated in the figure (courtesy of A. Kup\'{s}\'{c} \cite{andrzej-priv}).    
}
\label{mmresolutionFRH}
\end{figure}
energy resolution of protons for different number of FRH planes.
The accuracy of the energy reconstruction is a crucial 
point in the analysis. The systematical and statistical errors in the 
evaluations of branching ratios for the studied decay channels will strongly 
depend on the precision of the missing mass reconstruction, which in turn 
will depend on the precision of the reconstruction of the momenta of the forward 
emitted protons. The effective energy resolution of the newly assembled Forward Range
Hodoscope is estimated to be about $\sigma=$ 3\% for protons from the 
$pp\to pp\eta^{\prime}$ reaction, at a beam momentum of 3.35~GeV/c.

In order to investigate the missing mass resolution for the $pp\to ppX$ reaction 
we assumed that the COSY beam momentum spread 
is $\Delta p/p \approx 10^{-3}$, perpendicular beam profiles: horizontaly $\sigma_X$=2~mm, vertical $\sigma_Y$=5~mm \cite{moskal-2001A466} and the pellets are passing through the interaction
point distributed homogeneously in a cylinder with a diameter of 2.5~mm. The result is presented in 
Fig.~\ref{mmresolutionFRH}, where the width of the missing mass peak 
is plotted as a function of the relative resolution of the kinetic energy.
It is seen that even in the case of a perfect energy resolution of the FRH 
the missing mass resolution amounts to about 4~MeV/c$^2$. If we further assume a perfect 
interaction point (vertex) determination the contribution from 
the beam momentum spread is seen to be at the order of 1~MeV/c$^2$.
In the right panel of Fig.~\ref{mmresolutionFRH} the dependencies were plotted for three different  
beam energies: 3.24, 3.35 and 3.5~GeV/c  
from the near threshold $\eta^{\prime}$ production region.  The broadening of the signal with increasing beam momentum is a 
kinematical effect due to the error propagation discussed in detail e.g. in reference~\cite{smyrski-2000B474}.
\begin{figure}[ht]
\parbox{0.5\textwidth}{
\centerline{\epsfig{file=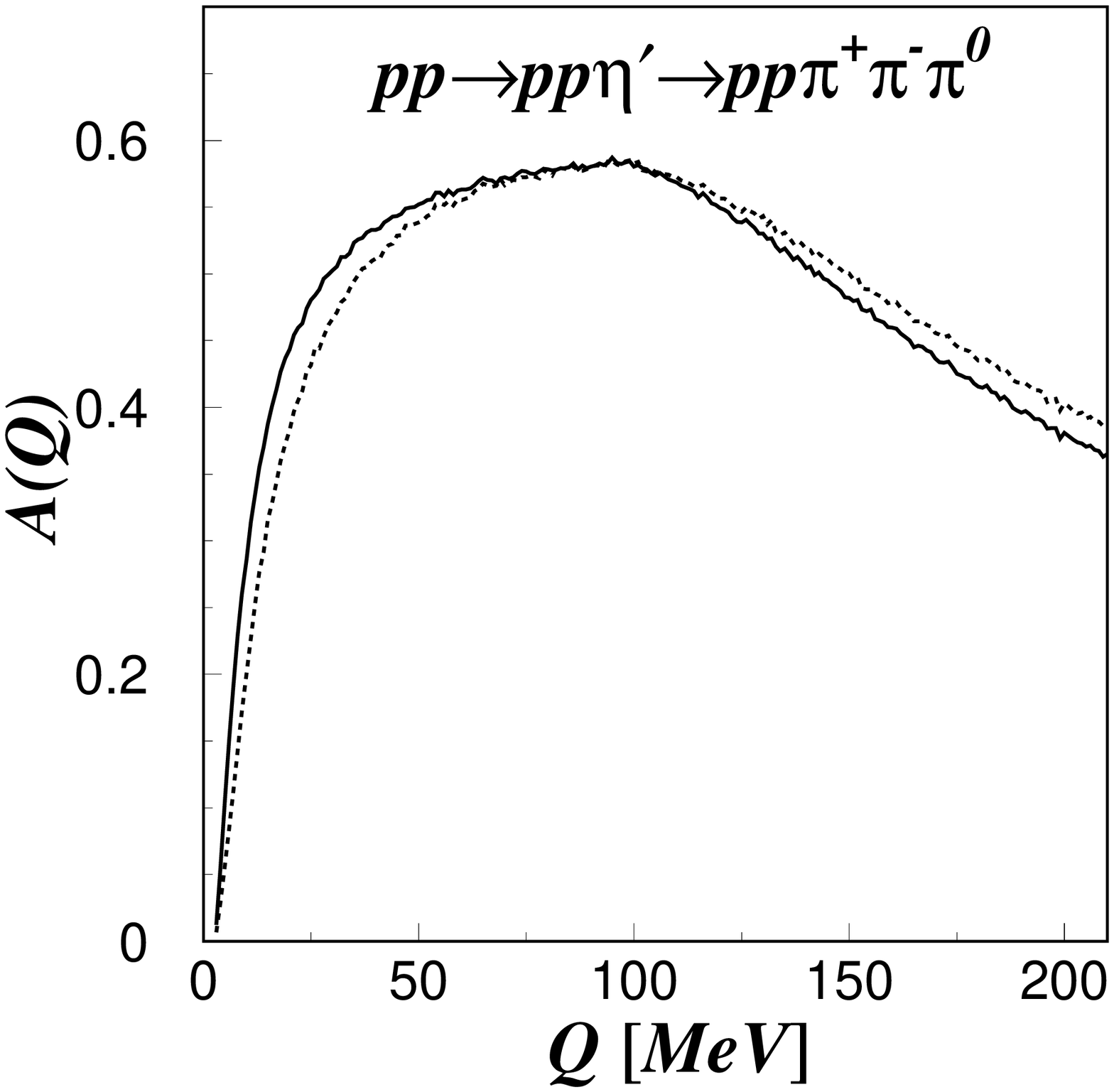,width=0.55\textwidth}}}
\parbox{0.5\textwidth}{
\centerline{\epsfig{file=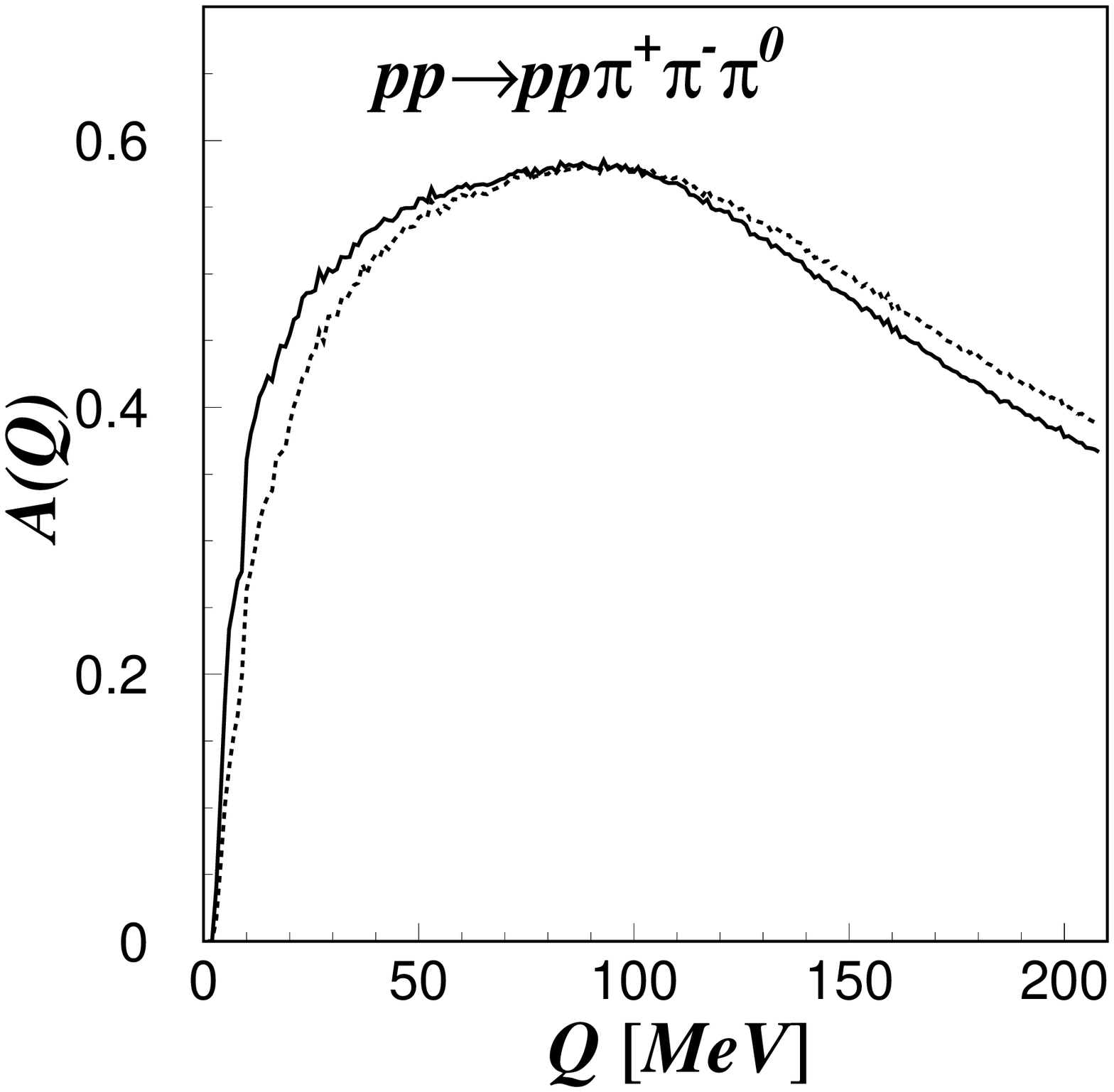,width=0.55\textwidth}}}
\caption{
Acceptance of the WASA-at-COSY detector as a function of the excess energy: 
(left) for the $pp\to pp\eta^{\prime}\to pp\pi^{+}\pi^{-}\pi^{0}$ and 
(right) for the $pp\to pp\pi^{+}\pi^{-}\pi^{0}$ reaction. Solid line indicates  
acceptance calculated assuming a homogeneous  phase space distribution, and the dashed line 
denotes the acceptance where additionally a proton-proton FSI was taken into account.   
}
\label{akceptancja}
\end{figure}

Taking into account all effects which we mentioned above, the missing mass 
resolution can be parametrized as a function of excess energy and the kinetic energy resolution
of forward detector by the following formula~\cite{pawel-priv1}:
\begin{equation}
\Gamma = (0.87\sqrt{Q} + 1.25)(0.48 + \frac{dT}{T}0.17).
\label{pawel_wzor}
\end{equation}
This relation will be useful in the next section where the accuracy of the branching ratio determination as a function of 
excess energy will be calculated. 

Additionally we have included the acceptance of the WASA-at-COSY  apparatus for detecting the 
$pp\to pp\eta^{\prime}\to pp\pi^{+}\pi^{-}\pi^{0}$ and $pp\to pp\pi^{+}\pi^{-}\pi^{0}$ channels.
We have taken into account the geometrical acceptance of the central and the forward detector which for detecting the 
protons and pions covers the following ranges of the polar angle: $2.5^o \leqslant \theta \leqslant 18^o$, and  
$20^o \leqslant \theta \leqslant 169^o$.
Furthermore the Final State Interaction (FSI) between outgoing protons was included using the square 
of the on-shell proton-proton scattering amplitude calculated according to the Cini-Fubini-Stanghellini 
formula including the Wong-Noyes Coulumb corrections~\cite{22swave,23swave,24swave} (for details see appendix~\ref{app:fsi}).
Fig.~\ref{akceptancja} indicates the excess energy dependence of the acceptance assuming that the phase space
is homogeneously  populated (solid line) and including the FSI between protons (dashed line).  

\section{Accuracy of the branching ratio determination}
\hspace{\parindent}
The   direct  three   pion   production  and   the
$\eta^{\prime}\to  \pi^{+}\pi^{-}\pi^{0}$ decay will  be disentangled
by using  the missing mass  of the  two outgoing
protons measured  in the forward  detector.
It is worth noting that in order to distinguish
(on the base of a statistically significant sample of events)
the direct and resonant multi-pion production
it is mandatory first to
select a sample of events with the studied final state channel (e.g. $\pi^+\pi^-\pi^0$)
and then only for this selected sample  to construct a distribution
of the missing mass  to the proton-proton system.

To estimate the accuracy of the BR($\eta^{\prime}\to\pi^{+}\pi^{-}\pi^{0}$)  determination we have parametrized the total cross 
section for the $\eta^{\prime}$ meson production (eq.~\ref{przekrojparam}) and have etablished a parametrization 
for an upper limit of the background production (eq.~\ref{tloparam}). Further on, the 
missing mass resolution taking into account effects which are related to the energy resolution of the forward detector and  
\begin{figure}[ht]
\hspace{3.cm}
\parbox{0.60\textwidth}{
\centerline{\epsfig{file=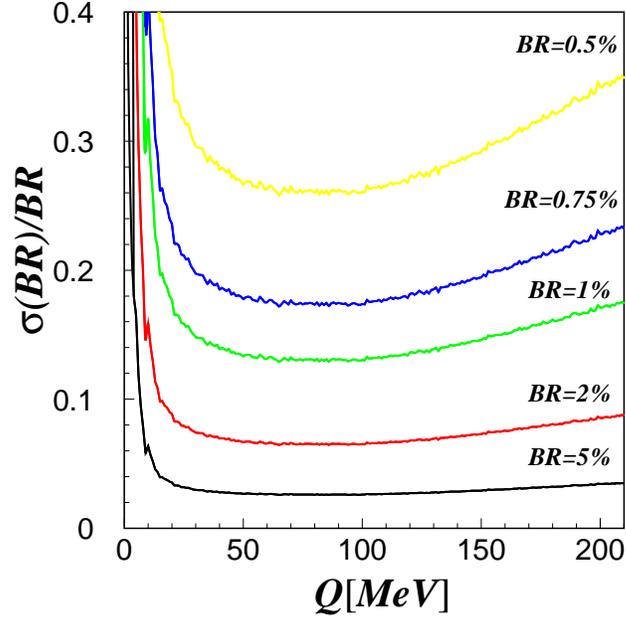,width=0.65\textwidth}}}
\caption{
The relative accuracy of the determination of the BR($\eta^{\prime}\to\pi^{+}\pi^{-}\pi^{0}$) 
as a function of the excess energy Q for the 
$pp\to pp\eta^{\prime}$ reaction. 
One week of data taking with WASA-at-COSY with luminosity L~=~10$^{32}$~cm$^{-2}$~s$^{-1}$ is assumed for the calculations.
}
\label{br_accuracy}
\end{figure}
the beam and target spread (eq.~\ref{pawel_wzor}) has been parametrized. 

\begin{figure}[H]
\parbox{0.5\textwidth}{\centerline{\epsfig{file=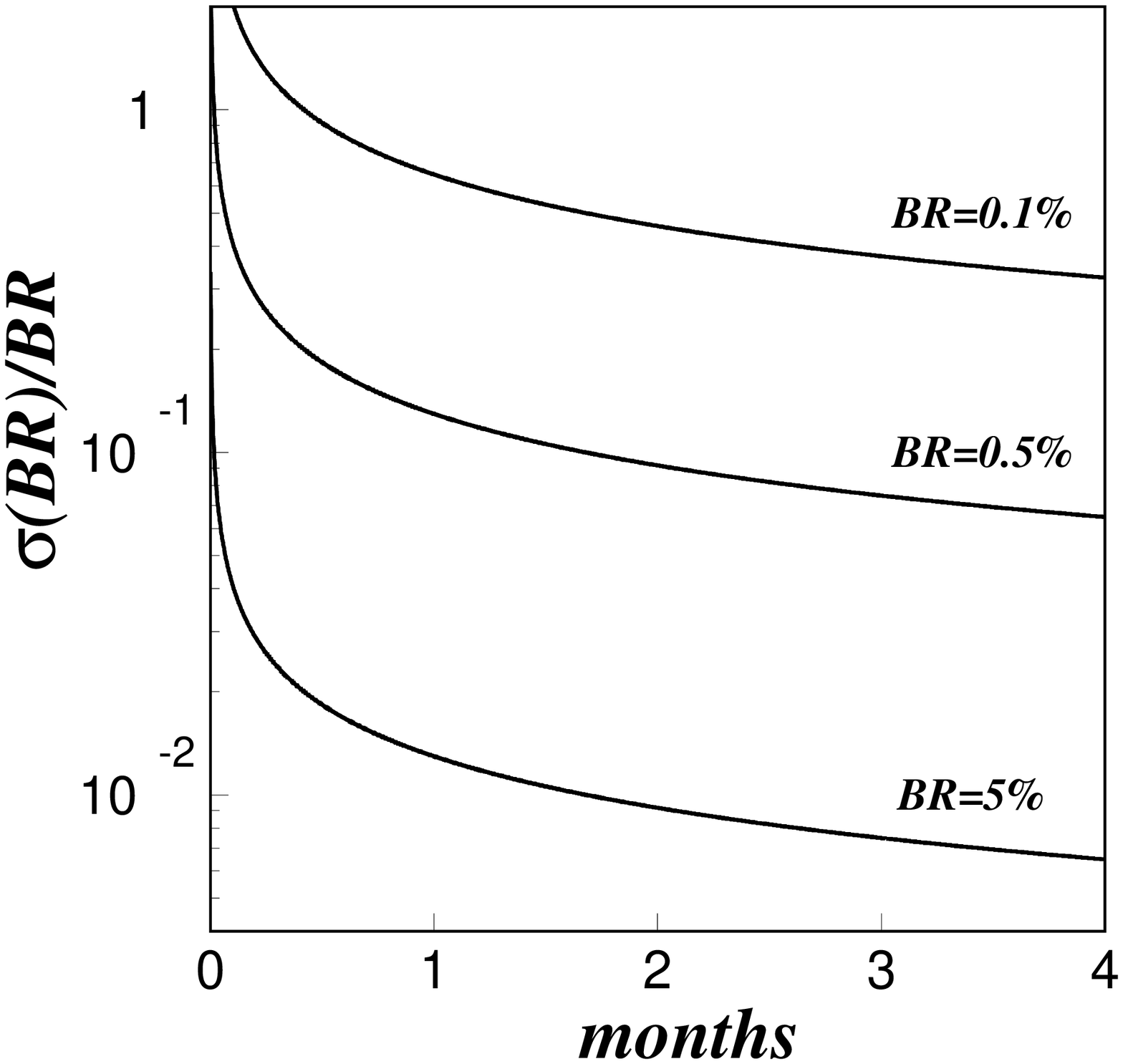,width=0.55\textwidth}}}
\parbox{0.5\textwidth}{\centerline{\epsfig{file=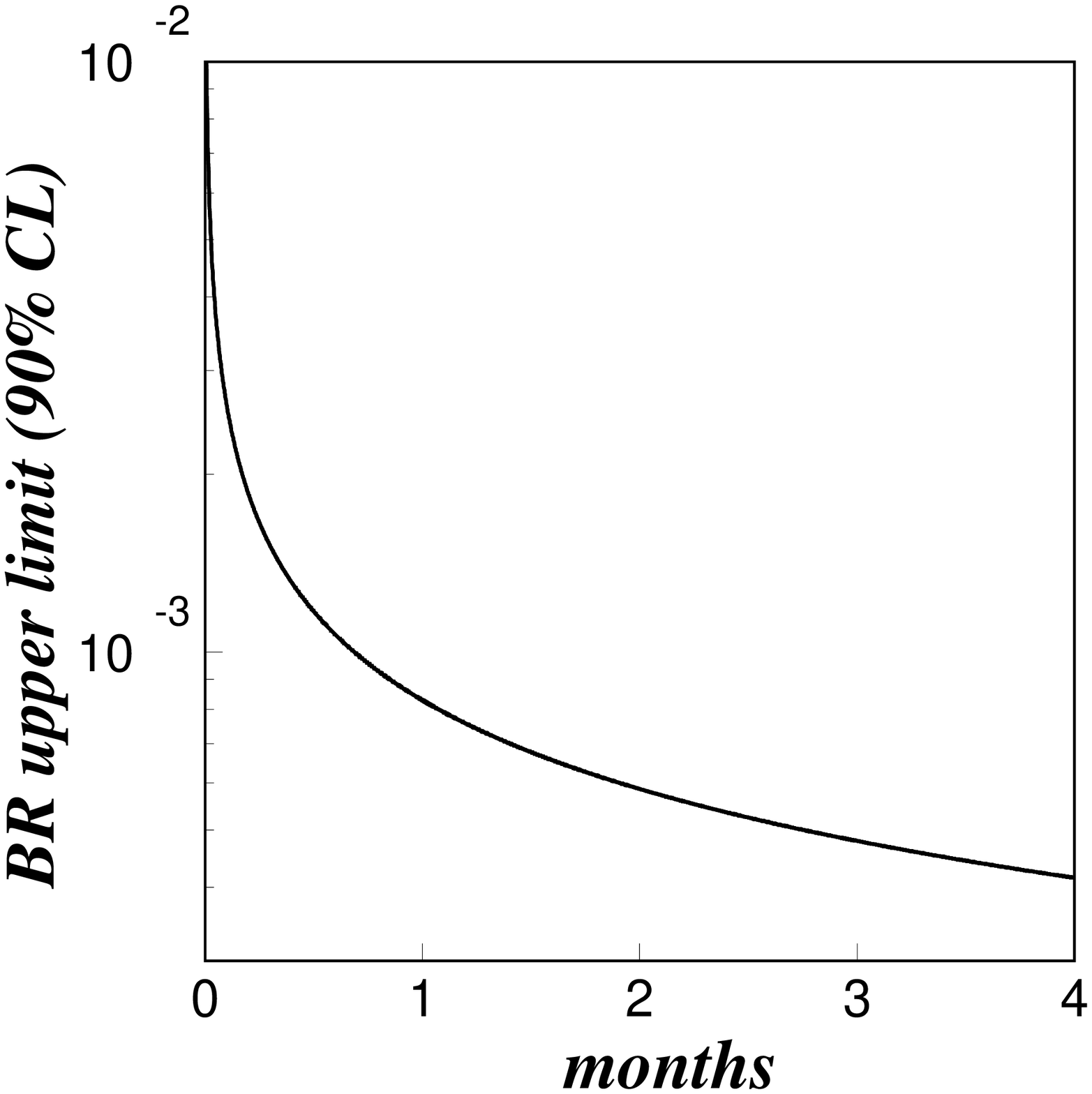,width=0.55\textwidth}}}
\caption{
({\bf left}) The relative accuracy of the BR($\eta^{\prime}\to\pi^{+}\pi^{-}\pi^{0}$) determination as 
a function of measurement time for three assumed values of the branching ratio and a beam momentum of $p_b$ = 3.45 GeV/c.
({\bf right}) Upper limit of determination of the BR($\eta^{\prime}\to\pi^{+}\pi^{-}\pi^{0}$) for 
a confidence level of 90\% versus the time of the experiment. Estimations are performed assuming a luminosity of L~=~10$^{32}$~cm$^{-2}$~s$^{-1}$ and beam momentum of $P_b$~=~3.45~GeV/c.}
\label{br_time}
\end{figure}

Applying these parametrizations and assuming five values of the BR($\eta^{\prime}\to\pi^{+}\pi^{-}\pi^{0}$): 5\%, 2\%, 1\%, 0.75\% and 0.5\% 
we have made calculations for the relative accuracy of the branching ratio determination. 
For the calculations we considered the range of values from the established upper limit of 5\% 
down to 0.5\%.
Assuming for example that the measurement will last one week with a luminosity of L=10$^{32}$cm$^{-2}$s$^{-1}$, we obtained 
the result shown in Fig.~\ref{br_accuracy} where
the relative error of the branching ratio is plotted as a function of the excess energy Q. 
We can see that the relative statistical error scales nearly  with the value of the assumed branching ratio, and that 
the optimum accuracy is achieved for excess energies between 60 and 90 MeV independently of the BR magnitude.
Relation (\ref{blad_wzgledny_br}) implies that the relative statistical error of the branching ratio determination
will improve with time as $1/\sqrt{t}$, as it is shown in Fig.~\ref{br_time} (left) for a beam 
momentum of $P_{beam}$~=~3.45 GeV/c corresponding to an excess energy of Q = 75~MeV. 
The plot shows that if the branching ratio was equal to 0.5\% a relative accuracy of 10\% 
would require two months of data taking~\cite{raport07-2}. 

Additionaly, for the case if the BR($\eta^{\prime}\to\pi^{+}\pi^{-}\eta$) is to small to be observed we have 
estimated the upper limit at a confidence level of 90\% as a function of measurement time. 
The estimated upper limit of the branching ratio as a function of measurement time is shown in Fig.~\ref{br_time} (right). 
We can see that e.g. a sensitivity of 0.001 can be reached after few weeks of data taking with the WASA-at-COSY detector. 
As it was pointed out the theoretical calculation based on the chiral unitary approach predicts the branching ratio 
af about 1\%~\cite{borasoy06}. From the right panel of Fig.~\ref{br_time} we may infer that this prediction can be falsified or confirmed 
within a few days of beam time using the WASA-at-COSY detection setup.  

It is worth to mention that an additional source of background, not discussed here, comes form 
other decays of the $\eta^{\prime}$ meson involving similar particles: $\eta^{\prime}\to\pi^{+}\pi^{-}\eta$ and 
$\eta^{\prime}\to\omega\gamma$. This background cannot be suppressed using the missing mass method,
yet it can be identified using reconstructed invariant masses of the decay products.

\chapter{The Time-of-Flight method}
\hspace{\parindent}
In order to study rare decays of the $\eta$ and $\eta^{\prime}$ mesons produced in the reaction $pp\to pp\eta(\eta^{\prime})$
using the WASA-at-COSY detection setup, one has to determine the
four-momentum vectors of the decay products and of the forward scattered protons. 
At present the protons four-momentum vectors are derived only taking into
account the energy loss  measured  in the Forward Detector.
In this chapter we will present a possible future improvement of the momentum reconstruction  
using the time information from the scintillators of the Forward Detector~\cite{wasa-note,zielinski}. 
We will also describe the alghorithm which can be implemented to reconstruct the 
momentum of particles crossing the Forward Detector using time information obtained from 
the crossed detection layers.     
 
\section{Proposal of using the Forward Detector for a TOF method}
\hspace{\parindent}
In order to measure time-of-flight (TOF) of charged particles emitted in forward direction we can use all
scintillator detectors which are placed in the forward part of the WASA setup.
But for the 
purpose of this thesis we will consider only the thin plastic scintillator
detectors: Forward Window Counter, FWC (2$\times$3~mm), Forward Trigger Hodoscope, FTH
(3$\times$5~mm), Forward Range Intermediate Hodoscope, FRI (2$\times$5~mm), and the Forward
Veto Hodoscope, FVH (20~mm).
The five layer Forward Range Hodoscope can also be used to determine time, but it is made of
scintillators which thickness ranges from 11 to 15 cm, and more precise studies of the light signal generation and 
propagation would be necessary for the usage of the time information. 

Using only thin detectors, enables us to measure time in eight points on a distance of about 2.0 meters. 
But in practice one can combine the individual times measured in layers of each
detector to obtain one time value per detector.
Additionally the information about the trajectory of particles is needed 
to recontruct the time in the FWC, FTH and FRI counters, due to one side readout of the 
scintillating modules (only one photomultiplier).
The direction of particles will be obtained from  measuring tracks in the straw chambers (FPC). 
The FPC chambers are placed between the first two scintillator detectors - FWC and FTH. 
The material which they are made of is very thin, therefore  
we can neglect the particle energy losses in these layers. 
\begin{figure}[h]
\hspace{2cm}
\parbox{0.70\textwidth}{\centerline{\epsfig{file=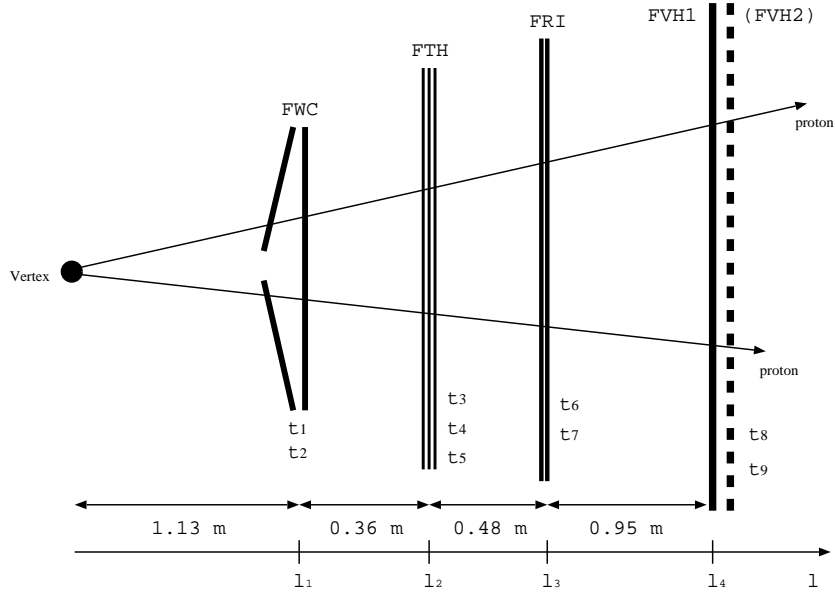,width=0.74\textwidth}}}
\caption{
Schematic view of detectors used for future TOF determination with the distance to each other given in meters. 
The FVH2 is the second veto layer which is planned to be build in the near future \cite{annette-priv}. 
}
\label{tof_schemat}
\end{figure}

The five layers of the Forward Range Hodoscope, FRH, are placed between FTH
and FVH. Their total thickness amounts to 63~cm, therefore particles which
travels through the FRH will be significantly slowed down.
However, the deposited energy in the FRH layers is measured with good accuracy
in the order of few per cent, and by combining the TOF information and the
deposited energy 
one can reconstruct the initial velocity of the particle at the interaction point.
Additionally the time signals will be measured by the FRI detector which is placed between the 
second and third layer of the range hodoscope. 

Generally, taking into account the time information from thin scintillators and the energy loss from thick 
scintillators we can write that all information which will be measured is collected 
in eight independent values of time and five energy losses:
\begin{equation}
(t,\Delta E)_{exp} \equiv t_1^{exp},t_2^{exp},t_3^{exp},t_4^{exp},t_5^{exp},t_6^{exp},t_7^{exp},t_8^{exp},\Delta E_1^{exp},\Delta E_2^{exp},\Delta E_3^{exp},\Delta E_4^{exp},\Delta E_5^{exp}.
\end{equation}
Using Monte-Carlo methods we can simulate the same quantities as from the experiment:
\begin{equation}
(t,\Delta E)_{mc} \equiv t_1^{mc},t_2^{mc},t_3^{mc},t_4^{mc},t_5^{mc},t_6^{mc},t_7^{mc},t_8^{mc},\Delta E_1^{mc},\Delta E_2^{mc},\Delta E_3^{mc},\Delta E_4^{mc},\Delta E_5^{mc}.
\end{equation}
Therefore, in order to reconstruct the correct particle energy (which the
particle had at the interaction point), we have to assume different true
particle energies, and compare the Monte-Carlo predicted times and energy losses
for each assumption with the measured qunatities until we find the optimum
(most probable) value for the kinetic energy of the particle.
A figure of merit in comparing this two samples could be a $\chi^2$ criterion. 
A more detailed description of the algorithm is given in section~\ref{sec:algorytm}.     

\section{Description of the detectors used for the time measurement}
\hspace{\parindent}
In this section we will briefly describe the detectors relevant for the time measurement.
The first detector which delivers time information on the way of the forward scattered particles 
is the Forward Window Counter (FWC)~\cite{pricking07}. The light collection in the detector is optimized to keep the 
detection efficiency as homogeneous as possible over the full hodoscope area.
The detector is 48-fold segmented and is composed of two layers \'{a} 24 elements made out of 3~mm plastic scintillator.
The first layer is of conical shape whereas the elements of the second layer are assembled in a plane.
The elements of the second plane are rotated by one half of a module (7.5$^o$) 
with respect to the first layer. 
This geometry provides a complete coverage  of the forward area without holes. 
In addition the 48-fold granularity coincides with the 48-fold granularity of the FTH.  
\begin{figure}[h]
\hspace{2.cm}
\parbox{0.30\textwidth}{\centerline{\epsfig{file=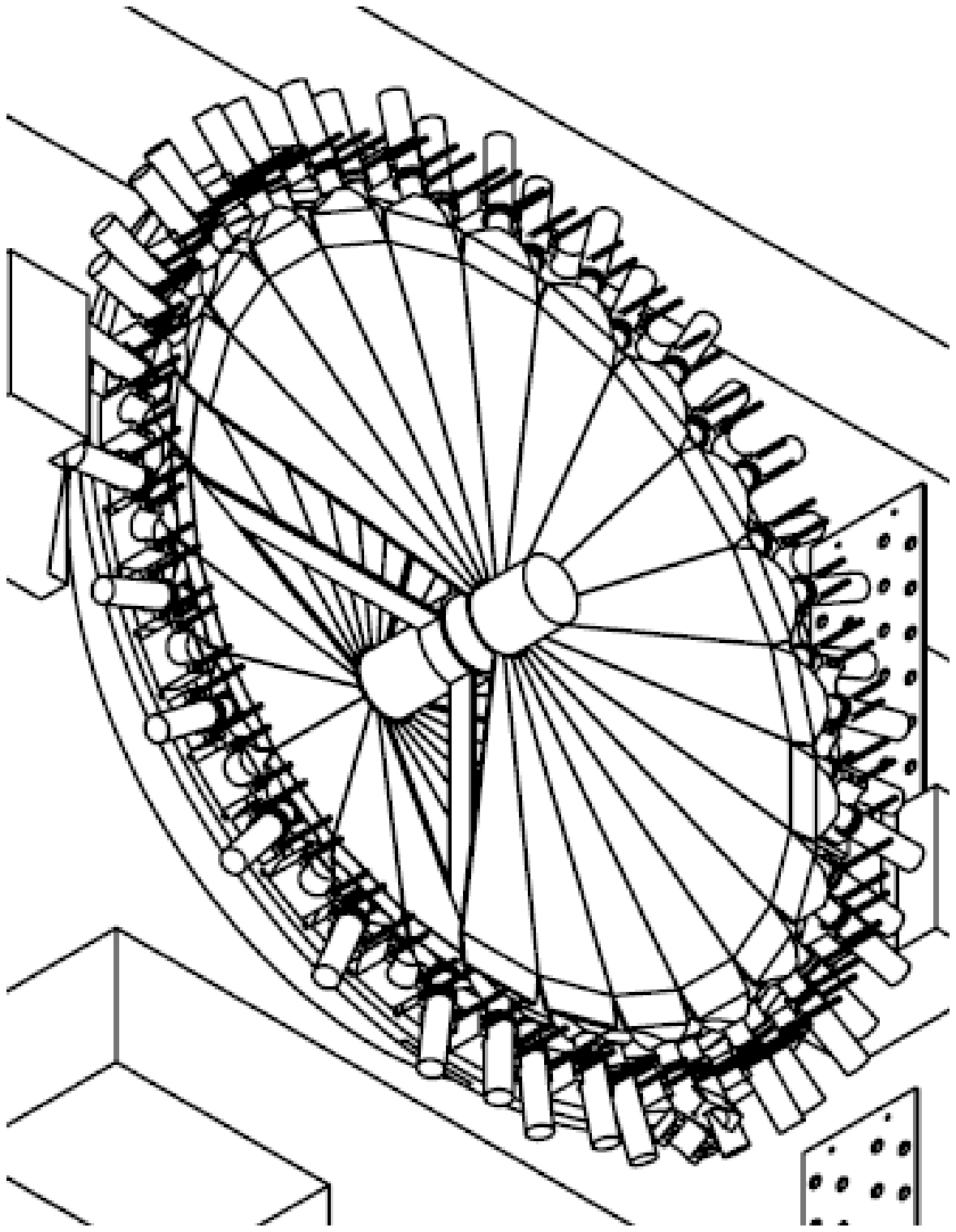,width=0.35\textwidth}}}
\hspace{2.cm}
\parbox{0.30\textwidth}{\centerline{\epsfig{file=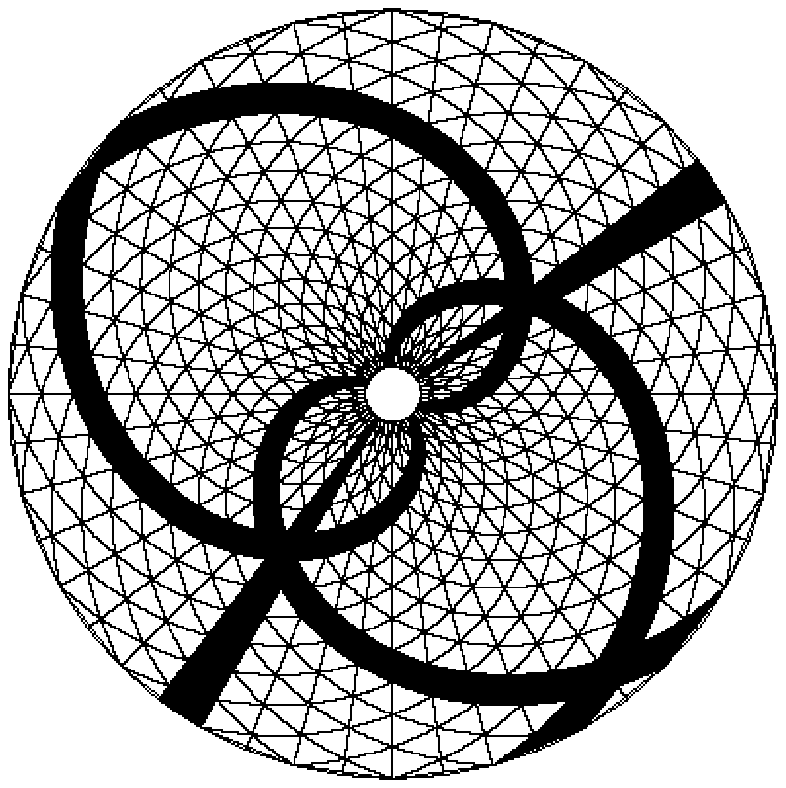,width=0.35\textwidth}}}
\caption{
Schematic view ({left}) of the Forward Window Counter (FWC) \cite{annette-priv} which is the ''start'' 
counter for TOF measurement ($t_1,t_2$), and ({\bf right}) of the Forward Trigger Hodoscope (FTH) ($t_3,t_4,t_5$).   
}
\label{tof_detektory}
\end{figure}
We expect to achieve a time resolution for each layer in the order of $\sigma$~(FWC)~$\approx$~200~ps. 
A schematic view of FWC is shown in the left panel of Fig.~\ref{tof_detektory}.

The next detector yielding time information on the way of the forward flying particles 
is the Forward Trigger Hodoscope (FTH). It consists of three scintillating layers which were renewed in 
the last year~\cite{pauly-an07,christoph}.
Setup has a highly homogeneous detection efficiency and shows a fairly uniform behavior ~\cite{pauly-an07}.
In case of using the FTH detector to measure time it is expected to achieve a resolution of about  
$\sigma$(FTH)~$\approx$~200~ps  for each layer. The front view of the FTH detector is shown in the 
right panel of Fig.~\ref{tof_detektory}.  

\begin{figure}[ht]
\hspace{0.3cm}
\parbox{0.60\textwidth}{\centerline{\epsfig{file=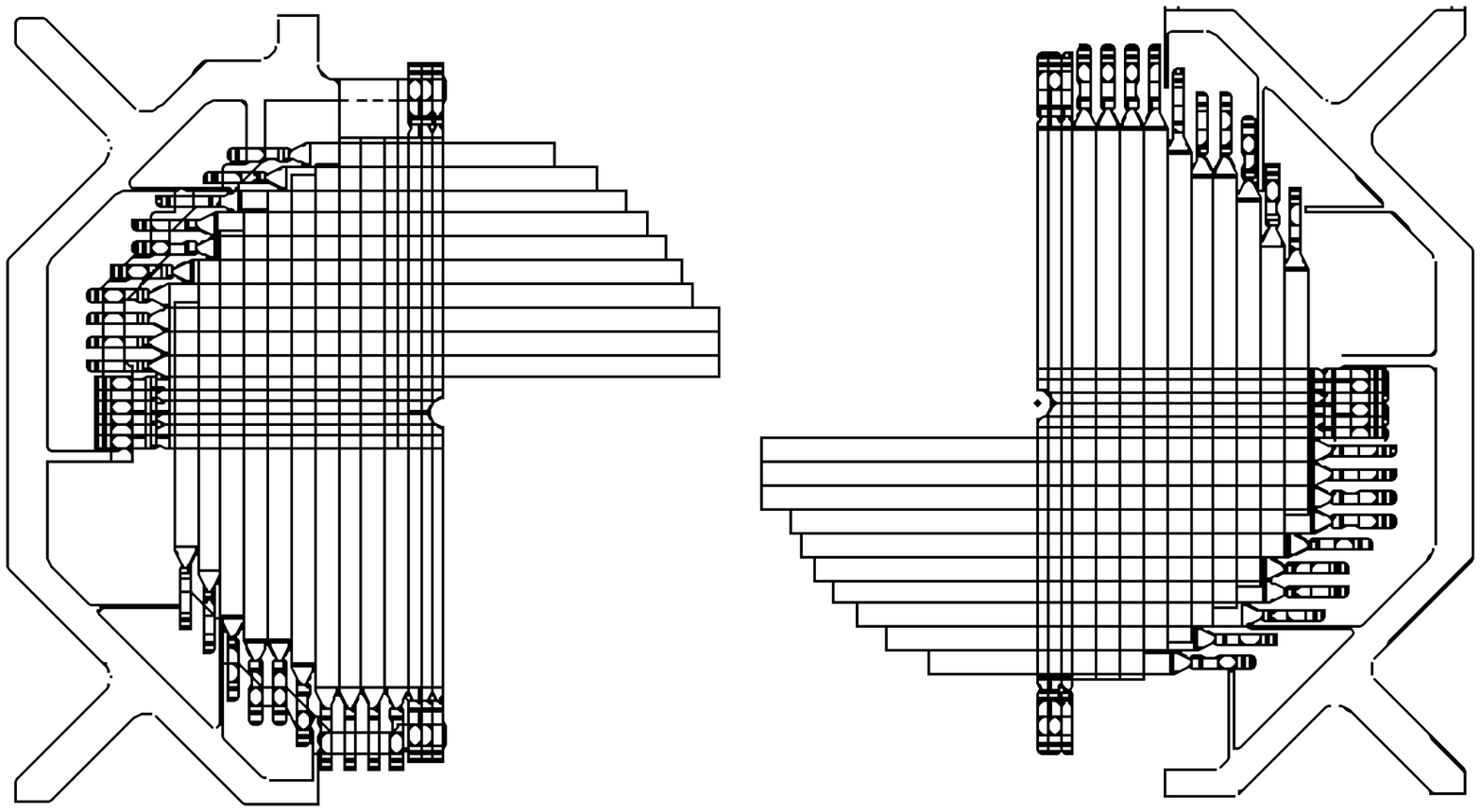,width=0.64\textwidth}}}
\hspace{.5cm}
\parbox{0.30\textwidth}{\centerline{\epsfig{file=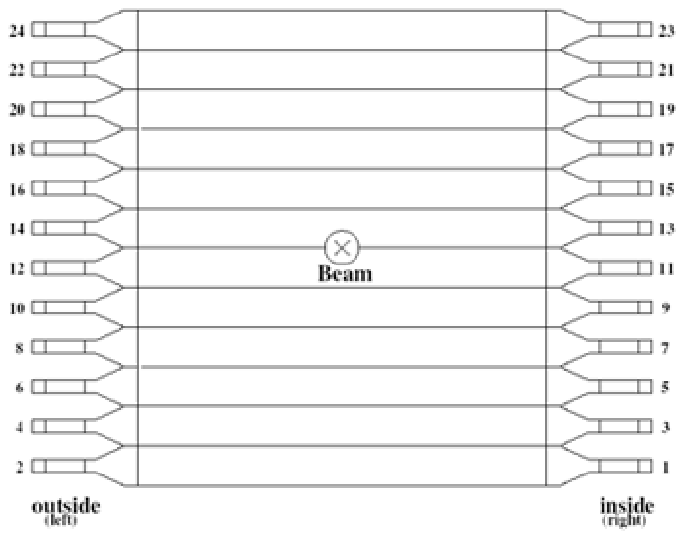,width=0.34\textwidth}}}
\caption{
Schematic view of the: ({left}) Forward Interleaving Hodoscope(FRI) ($t_6,t_7$), 
and ({\bf right}) the Forward Veto Hodoscope (FVH) which will be the ''stop'' counter ($t_8,t_9$).  
}
\label{tof_detektory2}
\end{figure}

FRI is a thin scintillator hodoscope, designed to provide fast spatial and
time information from inbetween the FRH.
It consists of two layers of horizontal and vertical scintillator bars of 5~mm
thickness each, and a maximum length of 1405~mm.
Each bar is read out on one side via small and fast photomultiplier with light guides of fishtail type.
The expected time resolution of the FRI is about $\sigma$(FTH)~$\approx$~300~ps.
Schematic view of the FRI is shown on left panel of Fig.~\ref{tof_detektory2}.

The last thin detector in the pathway of a forward scattered particle is the Forward Veto Hodoscope (FVH).
The FVH is a veto detector build out of 12 horizontal scintillator bars which
are read out on both sides (right panel of Fig.~\ref{tof_detektory2}). 
The accuracy of a time measurement amounts to about $\sigma\approx$~150~ps. 
In future it is planned to build a second veto detector with 20 modules placed vertically 
and readout by photomultipliyers on both sides.
This detector will be optimized for the TOF measurment, and it will be placed about 80~cm down stream 
with respect to the first FVH~\cite{mikhail-priv}.   

By combining the time-of-flight method and presently used energy loss technique we expect to increase the 
momentum determination by a factor of about 1.5 and increase correspondingly the accuracy of particle identification.
In order to estimate an expected improvement we have conducted Monte Carlo simulations of  
the $pp\to pp\eta^{\prime}$ reaction in the range of few tens of MeV above the threshold where the velocity 
of outgoing protons is around $\beta\approx 0.75$. The determination of the kinetic energy of protons 
from energy loss will at best be possible with a fractional resolution of $\frac{\sigma(T)}{T}$=3\%, which is equivalent 
to a relative momentum resolution of (see App. C):
\begin{equation}
\frac{\sigma(p)}{p} = \frac{T + m}{T + 2m}\frac{\sigma(T)}{T} \approx 2.4 \%
\end{equation} 
The determination of the time-of-flight in vacuum would lead to a fractional resolution of the momentum reconstruction expressed
by the following formula:
\begin{equation}
\frac{\sigma(p)}{p} = \frac{1}{1 - \beta^2} \frac{\sigma(tof)}{tof}.
\end{equation}
Thus, from the above anticipated resolution of about 150~ps for FWC and FVH counters even with this two detectors, 
one could obtain a fractional momentum resolution of $\frac{\sigma(p)}{p}$ = 3\%. But this is 
only a conservative limit of the expected improvement since the particles will be slowed down in the FRH and the 
time will be measured by more detectors.  

\section{Fractional energy resolution from the TOF measurement}
\hspace{\parindent}
To check the posibility of energy reconstruction from the time-of-flight method we 
have performed studies of fractional energy resolution as a function of the kinetic energy and 
kinetic energy as a function of time-of-flight between start and stop detectors.
We have simulated homogeneously in the phase space the $pp\to pp\eta^{\prime}$ reaction using the GENBOD~\cite{genbod}
procedure with a nominal beam momentum of $p_{beam} = 3.35$~GeV/c without taking 
into account any beam spread. 
\begin{figure}[H]
\hspace{3.cm}
\parbox{0.65\textwidth}{
\vspace{-1.cm}
\centerline{\epsfig{file=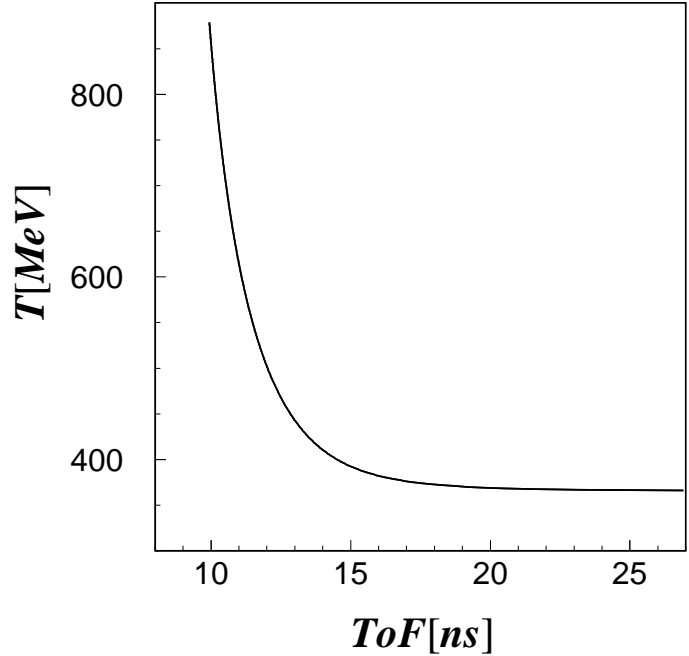,width=0.69\textwidth}}}
\caption{
The time-of-flight between the FWC and FVH detectors as a function of the proton kinetic energy.
For the calculations the energy losses in the scintillators of the Forward Detector were taken into account.   
}
\label{T_vs_tof}
\end{figure}

To calculuate the kinetic energy as a function of the time-of-flight we used only two detectors here: the FWC as
start counter and the FVH as stop counter, which are about 2~m apart (see Fig~\ref{tof_schemat}).
The energy loss in the detector 
material of FD was calculated using the Bethe-Bloch formula \cite{leo}, which for plastic scintillators 
and protons in the kinetic energy range from 200 to 830 can be approximated by~\cite{pdg}:
\begin{equation}
-\frac{dE}{dx} = \text{const}\cdot\beta^{-\frac{5}{3}},
\end{equation}
where the factor $\text{const} = 1.76$ was calculated~\cite{pawel-priv1} from the experimental data~\cite{jani}.

The time-of-flight between the start and stop detectors was calculated iteratively using numerical Monte-Carlo
techniques.
Technically 
for each proton flying through the detector the time-of-flight was derived by summing very small time intervals for passing
a distance on which the changes of the ionization power ($dE/dx$) can be neglected: 
\begin{equation}
\textrm{TOF} = \int\limits_0^{\textrm{d}}\frac{dx}{\beta(x)} 
 = \int\limits_E^{\textrm{E(d)}}\frac{dE}{\beta(E)\frac{dE}{dx}(E)} 
\label{calka-tof}
\end{equation}  

The plot in Fig.~\ref{T_vs_tof} shows the time-of-flight calculated as a function of the proton kinetic energy.
We can see that with decreasing energy the time-of-flight between the start and stop 
couters increases. And for particles which have an energy smaller then 360 MeV the 
time-of-flight cannot be established because they do not reach the stop counter. 
\begin{figure}[H]
\parbox{0.5\textwidth}{\vspace{-1.cm}\centerline{\epsfig{file=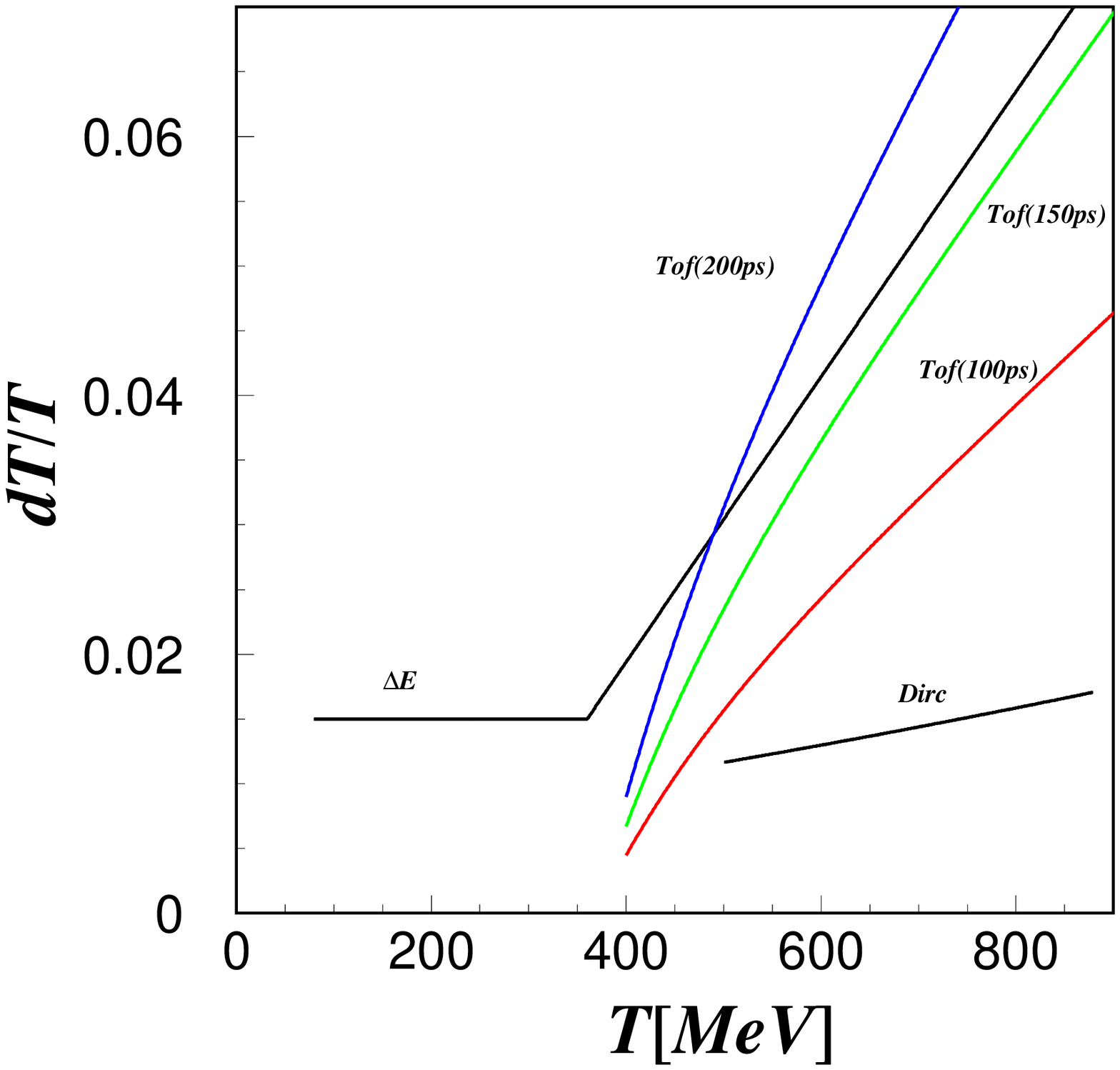,width=0.55\textwidth}}}
\parbox{0.5\textwidth}{\vspace{-1.cm}\centerline{\epsfig{file=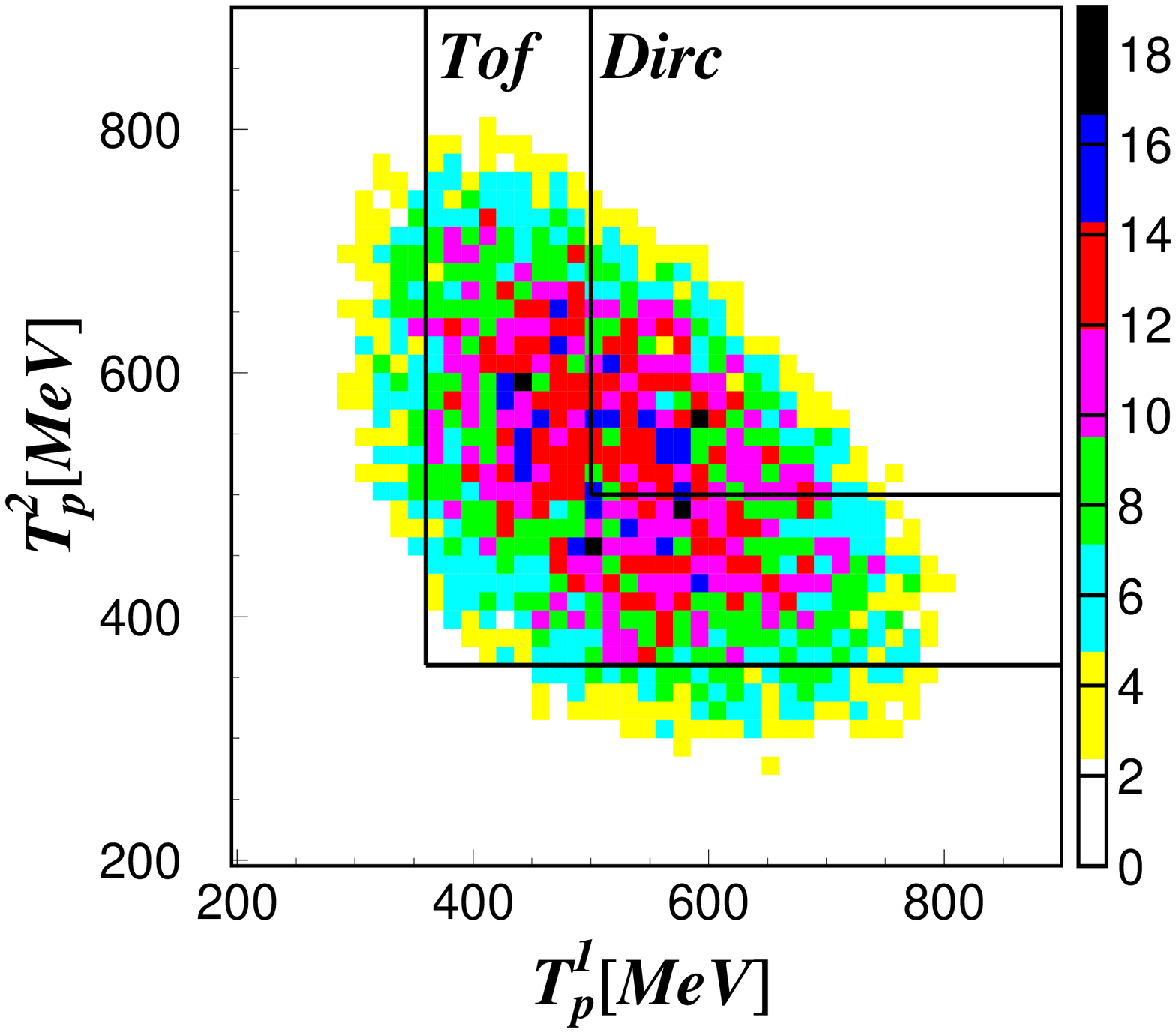,width=0.55\textwidth}}}
\caption{
({\bf{left}}) Fractional kinetic energy resolution estimated for three considered techniques - energy loss, DIRC and TOF.  
({\bf{right}}) Distribution of kinetic energy of protons emitted from $pp\to pp\eta^{\prime}$ reaction 
simulated for a  beam momentum of 3.35 GeV/c.  
}
\label{rozdz_tof_dirc}
\end{figure} 
Another possible improvement of the energy measurement of the forward emitted particles assumes the
installation of a DIRC detector which would enable to determine the velocity of protons~\cite{peter,teufel}. 
Therefore, in the following we will compare the missing mass resolution obtained using  the 
energy loss method with the resolution expected when using the time-of-flight method and also  
with the resolution achievable when using a DIRC counter. 

For the comparison we expressed the properties of time-of-flight technique, DIRC and energy loss method in terms
of fractional kinetic energy resolution. The result is shown in the left panel of Fig.~\ref{rozdz_tof_dirc}, 
it indicates that both DIRC and time-of-flight can yield better resolution than that obtained from energy loss only. 
However, for the estimation of the missing mass resolution it must be taken into account that the TOF method can be 
used  for protons which passed the whole detector with an energy T~$\geqslant$~360 MeV, whereas the DIRC can
deliver signals only for protons above the Cerenkov threshold which corresponds to 500~MeV~\cite{peter}.    
This implies that only a fraction of protons (see right panel of Fig.~\ref{rozdz_tof_dirc})
can be reconstructed by means of these two methods.
The right panel of Fig.~\ref{rozdz_tof_dirc} illustrates that the energy of both protons 
can be reconstructed for only about 20\% of 
events using the DIRC detector, and for about 70\% of events using time-of-flight method. In other
cases the energy of either one or both protons must be determineted using energy loss technique. 

\section{Accuracy of the missing mass reconstruction}
\hspace{\parindent}
For a rough estimation of a missing mass accuracy we have assumed that we will measure the time only with the 
FWC and FVH but to account for the fact that in reality we will use much more detectors we assumed the precision of
$\sigma(tof)=100$~ps.
Further more we have included the geometrical acceptance of the WASA Forward Detector for particles emitted in the 
forward direction, which is: $2.5^o\leqslant\theta\leqslant 18^o$. 
The simulation did not include contributions due to a finite interaction
region and the momentum spread of the COSY beam,
since we focused on the comparison of the resolution resulting from the
reconstruction of the proton four-momenta by the discussed methods.   

In order to calculate the missing mass distributions 
we have simulated proton four-momenta using an event generator based on
GENBOD~\cite{genbod}, which produces particle four-momenta in the CM-frame homogeneously
distriuted in phase space.
These particle four-momenta were then boosted into the lab-system by means of a
Lorentz transformation.
To reproduce the missing mass as measured in the experiment we then smeared
the kinetic energy of the simulated protons applying the known energy
resolutions for each of the studies reconstruction methods.
Next using the calculated 
four-momenta of protons the mass of an unobserved particle was calculated according to the equation:
\begin{equation}
m_X = \sqrt{ (E_{beam} + m_{target} - E_{p_1} - E_{p_2})^2 - (\vec{p}_{beam} - 
              \vec{p}_{1} - \vec{p}_{2})^2 } ,
\end{equation}    

\begin{figure}[H]
\parbox{0.30\textwidth}{\centerline{\epsfig{file=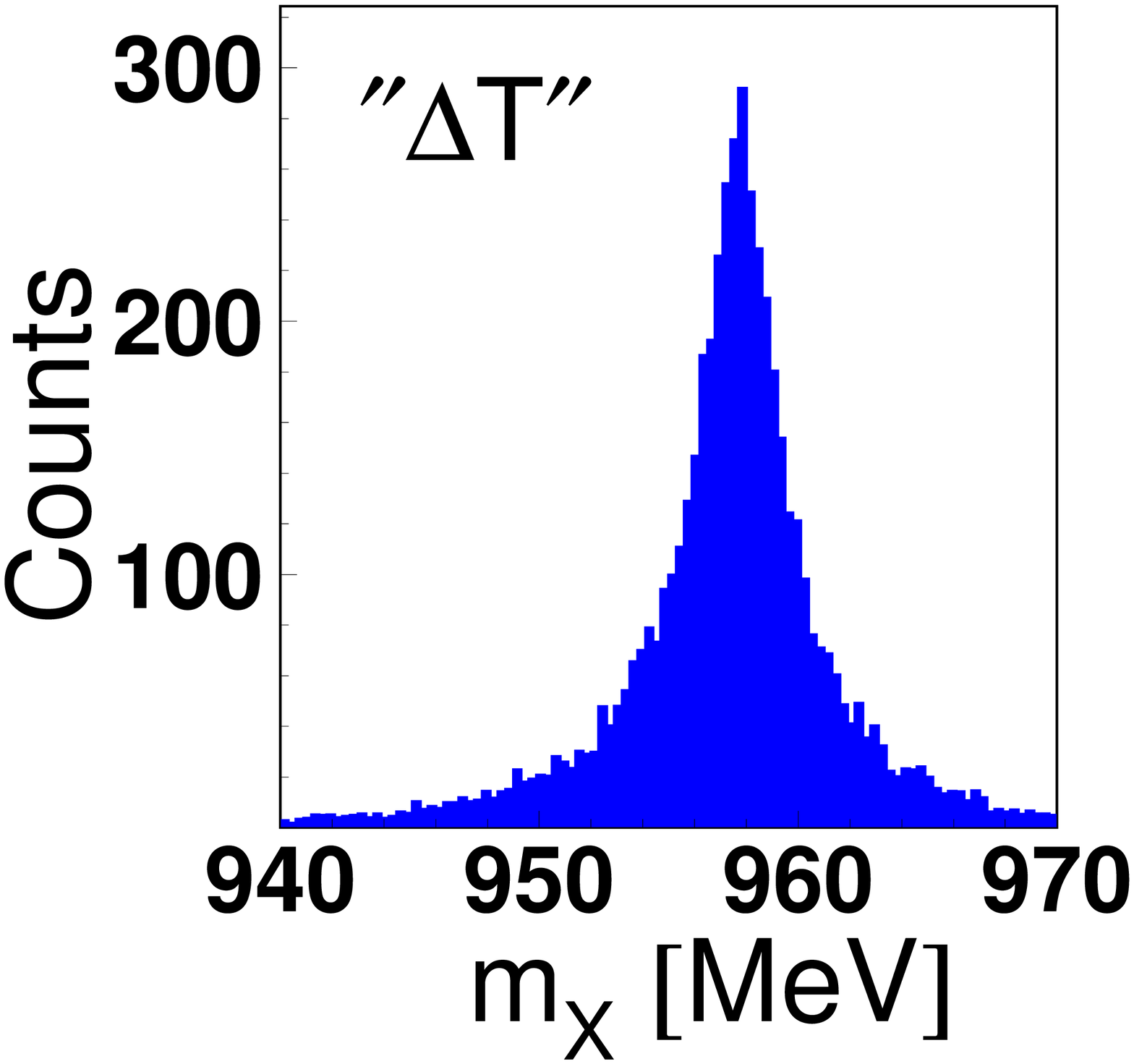,width=0.35\textwidth}}}
\parbox{0.30\textwidth}{\centerline{\hspace{0.6cm}\epsfig{file=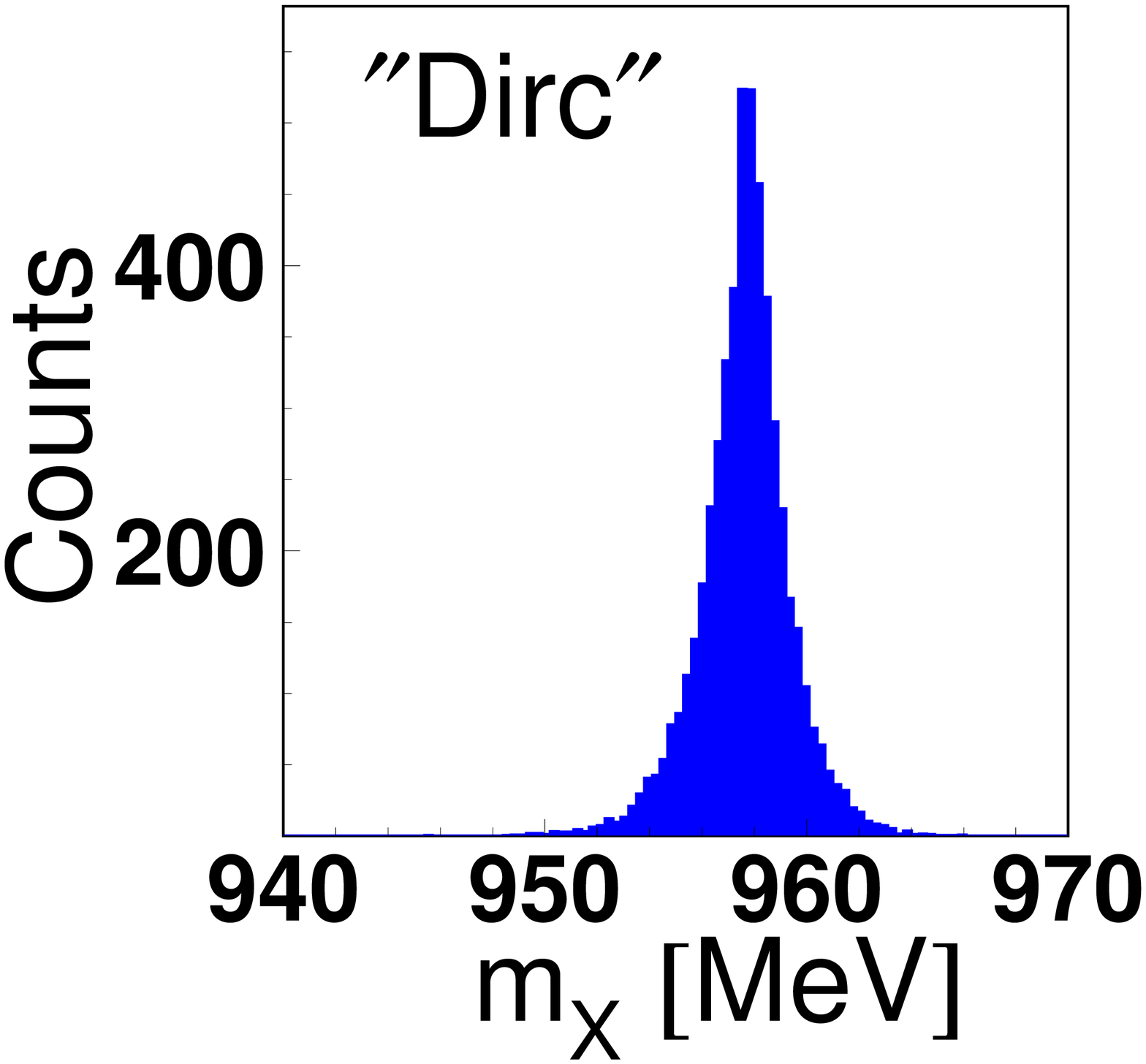,width=0.35\textwidth}}}
\parbox{0.30\textwidth}{\centerline{\hspace{1.2cm}\epsfig{file=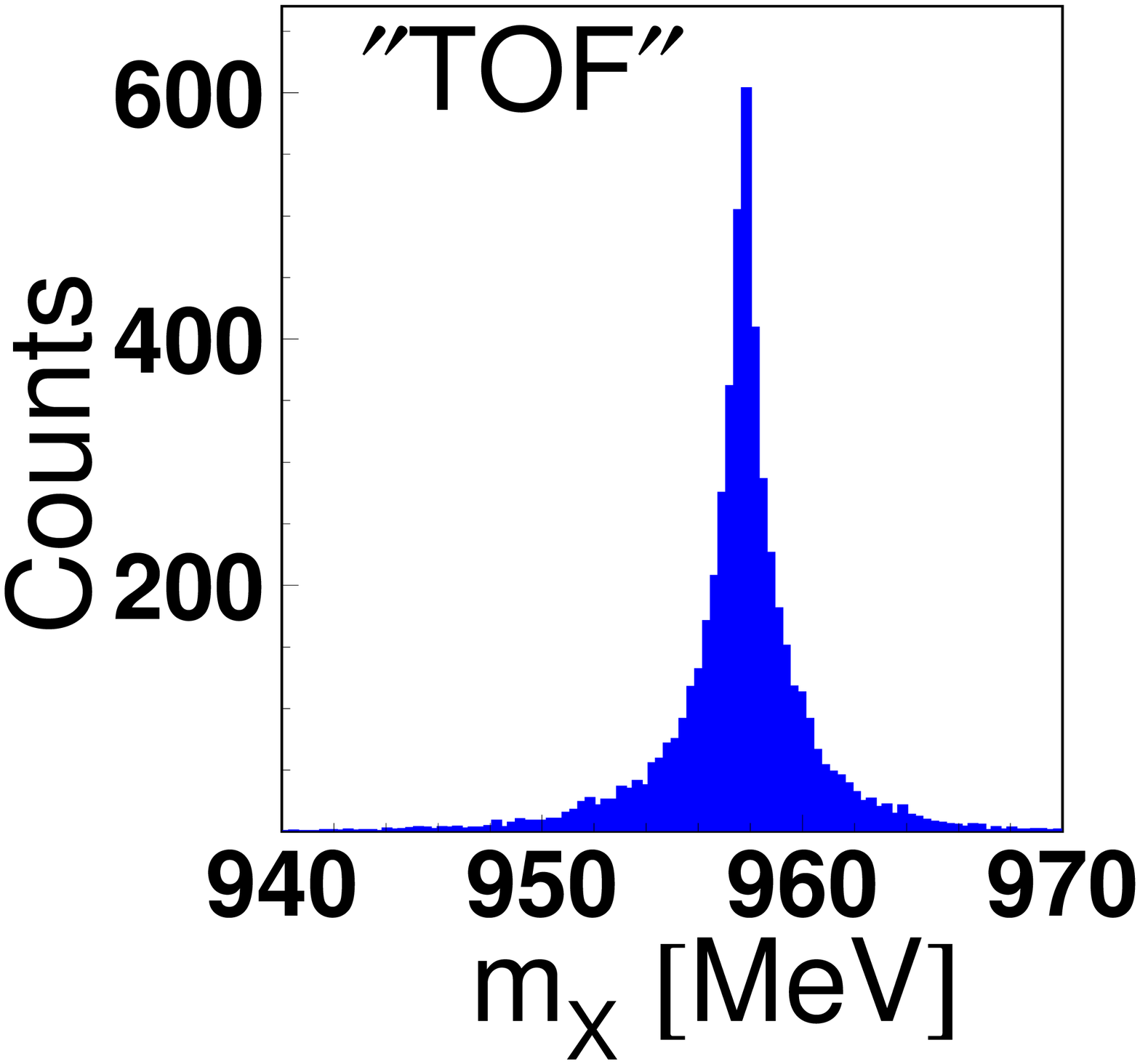,width=0.35\textwidth}}}
\caption{
Missing mass distribution to the $pp\to ppX$ reaction reconstructed under the assumption that the 
resolution of kinetic energy determination of protons will be as expected from energy loss
measurement (left plot), DIRC (middle plot), TOF with $\sigma(TOF)$~=~100~ps (right plot).  
}
\label{missmass_wykresy3}
\end{figure}

In Fig. \ref{missmass_wykresy3} the missing mass distributions for each method are shown. 
The left plot shows the result obtained using energy loss technique, middle plot corresponds 
to the DIRC technique and right to the time-of-flight method.
The simulations for the time-of-flight method and DIRC detector shows that the resolution of the missing mass 
reconstruction can be significantly improved in comparison to the results obtained when using only energy loss information. 

\section{Influence of the passive material on a precision of the energy determination}\label{sec:pasiv}
\begin{figure}[H]
\hspace{3.cm}
\parbox{0.65\textwidth}{
\centerline{\epsfig{file=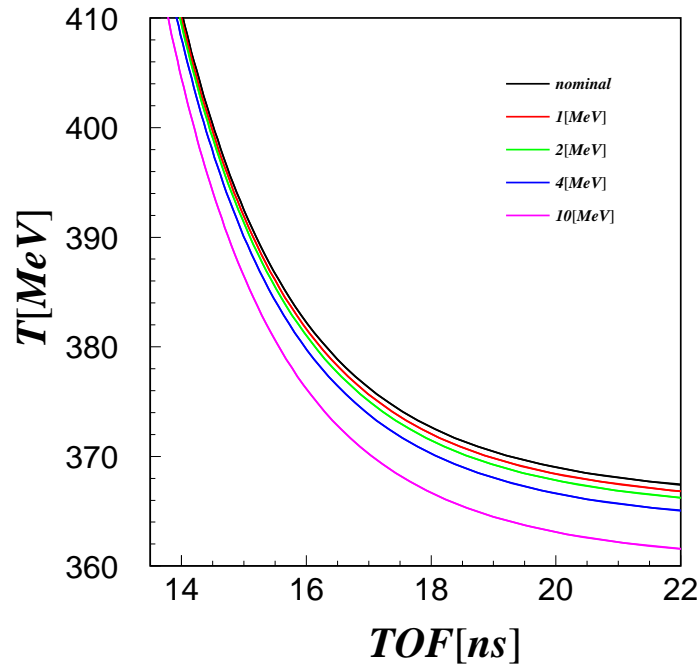,width=0.69\textwidth}}}
\caption{
The kinetic energy of protons as a function of the time-of-flight. The lines indicates different 
assumption of the error for the calculations of the energy losses in the four passive layers of plexi. 
}
\label{tof_passiv}
\end{figure}

\hspace{\parindent}
New layers of the Forward Range Hodoscope (FRH 4 and 5) are placed in thin 8 mm plexi glass shield, to keep all
scintillators modules in the correct position and to protect them against any mechanical damage. This material 
is not used in the measurement and we do not have information about the energy loss in this 
layers.  We expect that in the 32 mm (4~$\times$~8mm) of the dead material the average energy loss will be about 10~MeV, 
and that the spread of this energy loss will be in order of 1~MeV or less.
In order to estimate the scale of the effect we have conducted a simulations of a kinetic energy dependence of protons 
flying trought the Forward Detector as a function of the time-of-flight. 
In calculations we assumed a four different values for the error in calculating energy losses in the passive 
material: $\Delta E_{plexi}$ = 1~MeV, 2~MeV, 4~MeV, 10~MeV. 
The proton energy as a function of the time-of-flight for the assumed error is plotted in Fig.~\ref{tof_passiv},
where it is compared to the nominal dependence. 
We expect an error to be less then 1~MeV, however for the better visualization of the effect a much larger 
range was studied.   
One can see from Fig.~\ref{tof_passiv} that the inaccuracy of the energy loss determination 
of about 1~MeV will cause the error of kinetic energy determination in the order 
of fraction of MeV. Therefore we can neglect this influence of the passive material on energy reconstruction.
          
\section{Computational algorithm for the reconstruction of particle momenta based on
          time signals in the Forward Detector}\label{sec:algorytm}
\begin{figure}[h]
\hspace{2cm}
\parbox{0.70\textwidth}{\centerline{\epsfig{file=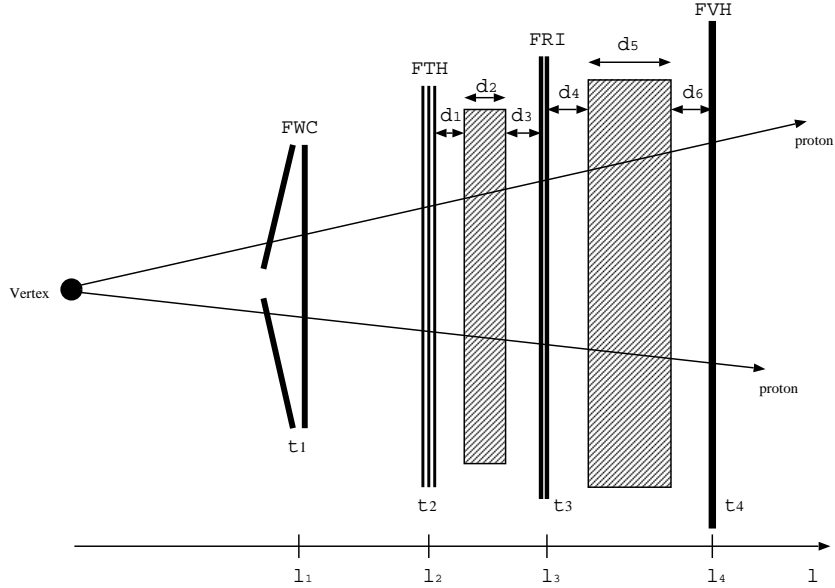,width=0.74\textwidth}}}
\caption{
Schematic view of detectors used for the explanation of the algorithm to
reconstruct protons energy based on time and energy loss information.   
}
\label{algorytm_schemat}
\end{figure}

\hspace{\parindent}
In this section we will present the scheme of the algorithm developed for the 
reconstruction of particle energy which is based on  time and energy losses measured in the Forward Detector. 
For the sake of simplicity further on we will consider only average times measured in each detector, and we will
assume for simplicity that energy loss $\Delta E$ occurs in one solid block of 
scintillating material between the FTH and FRI and another block between FRI and FVH detectors, as shown in Fig.~\ref{algorytm_schemat}. The simplification is made
having in mind that in general every detection plane can be treated as a separate detection unit,
and that the derived equations can be easily generalized for more detection planes.  

In such case each event is characterized by a set of two values of energy loss measured in the FRH  
scintillation layers, namely:
\begin{equation}
\Delta E^{exp} = (\Delta E_1^{exp}, \Delta E_2^{exp})
\end{equation}
and time information from four thin scintillators (FWC, FTH, FRI, FVH):
\begin{equation}
t_{1}^{exp},t_{2}^{exp},t_{3}^{exp},t_{4}^{exp}.
\end{equation}

The measured times can be expressed as a function of three parameters:
\begin{equation}
t_{i} = t_{i-1} + f(E_{i-1}, \Delta l, \Delta E),
\end{equation}   
where $E_i$ denotes the particle energy after i-th detector, $\Delta l$ denotes the length of the particle trajectory between the two detectors, and $\Delta E$ indicates the energy loss between (i-1)-th and i-th time counter.   

Therefore we can calculate for each event a time using following equations:  
\begin{equation}
\label{time1}
t_1^{mc} = \frac{l_1}{\beta(E_0)},
\end{equation}
\begin{equation}
t_2^{mc} = t_1^{mc} + \frac{l_2 - l_1}{\beta(E_1)} ,
\end{equation}
\begin{equation}
t_3^{mc} = t_2^{mc} + \frac{d_1}{\beta(E_2)} + \Delta t(d_2) + \frac{d_3}{\beta(E - \Delta E(d_2))},
\end{equation}
\begin{equation}
\label{time4}
t_4^{mc} = t_3^{mc}+ \frac{d_4}{\beta(E_3)} + \Delta t(d_5) + \frac{d_6}{\beta(E_3 - \Delta E(d_5))},
\end{equation}
where $\beta(E_0)$ denotes the initial velocity of the particle. 

Having $\Delta E$ measured we can calculate the time in which particles pass a thick detector $\Delta t(d)$ numerically. 
E.g. for the $\Delta t(d_2)$ the formula reads:
\begin{equation}
\label{czas_t3}
\Delta t(d) = \int\limits_{E_2}^{E_2 - \Delta E(d_2)} \frac{dE}{\beta(E)\frac{dE}{dx}(E)}.
\end{equation}  

In order to reconstruct the true particle energy $E_0$ which it had at the interaction point we compare simulated times $t_{i}^{mc}$ with the measured times $t_{i}^{exp}$. For the comparison we can construct the 
$\chi^2$ according to the method of least squares:
\begin{equation}
\label{chikwadrat}
\chi^2(E_0) = \sum\limits_{i=1}^4 \left( \frac{t_i^{exp} - t_i^{mc}(E_0)}{\sigma^t_i} \right)^2
\end{equation}
where, $\sigma^t_i$ indicates the resolution of the time mesurement in the i-th counter.
One can, treat $E_0$ as a free parameter, and determine the value of the particle energy when $\chi^2(E_0)$ 
reaches a minimum.
To optimize the algorithm we may in practice scan the value of $E_0$ in the range of uncertainty around the value 
established based on the energy loss method only, and using one of the minimalizing methods as e.g. the bisection method.   

The proposed method of particle identification using  time signals, can be generalized 
for usage with real number of scintillator layers of the WASA-at-COSY detection setup. 
         
\chapter{Summary and conclusions}
\hspace{\parindent}
This thesis aimed at the estimation of the feasibility to study the branching ratio for the 
$\eta^{\prime}\to\pi^{+}\pi^{-}\pi^{0}$ decay with the WASA-at-COSY detector.
By measuring this isospin violating decay it is possible to derive the mass difference between the 
$u$ and $d$ quarks. This in combination with the ratios of the light quarks masses, 
can be used to derive the absolute value of quark masses, which cannot be observed directly. 

In order to established the optimum beam momentum for conducting the experiments with WASA-at-COSY setup we
performed Monte-Carlo simulations taking into account  
the energy dependence of the signal and the background as well as the  missing mass resolution and the detection efficiency.
For the calculations we considered a range of values from the established upper limit of 5\% 
down to a value by one order of magnitude lower (0.5\%).
As a result we found that the best accuracy for the measurement of the BR($\eta^{\prime}\to\pi^{+}\pi^{-}\pi^{0}$)
with the WASA-at-COSY detector is achieved when the $\eta^{\prime}$ meson is produced with an excess energy in the  
range between 60 and 90 MeV corresponding to beam momenta ranging from 3.4 - 3.55~GeV/c. 

Further on we investigated the branching ratio 
uncertainty as a function of measurement time for one value of excess energy (Q = 75 MeV).
As a result  we can conclude that if the BR($\eta^{\prime}\to\pi^{+}\pi^{-}\pi^{0}$) was equal 
to 1\% as predicted based on the chiral unitary approuch~\cite{borasoy06} we would need five weeks 
of beamtime with a luminosity of 10$^{32}$~cm$^{-2}$~s$^{-1}$
in order to determine the branching ratio with an accuracy of 5\%.  
We have established also the sensitivity for the estimation of the upper limit of the BR($\eta^{\prime}\to\pi^{+}\pi^{-}\pi^{0}$)
in the case of no signal. We found that an upper limit on the branching ratio of 0.001 can be set on a confidence level of 90\% after one month of beam time.   

Because there are no available data on the $\pi^{+}\pi^{-}\pi^{0}$ production in collisions of protons near the 
kinematical threshold  for the $\eta^{\prime}$ meson production, we have used the COSY-11 data on the $pp\to ppX$ reaction and established an upper limit for the background, expected to be observed with WASA-at-COSY detector. 
Using the missing mass spectra for the $pp\to ppX$ reaction for several 
excess energies, the differential cross sections for the multimeson production was calculated. For further usage the 
excitation function was parametrized as a $\alpha\cdot Q^{\beta}$. By that means we can compute the values 
of the expected background as a function of the excess energy near the $\eta^{\prime}$ meson threshold. 

In the second part of the thesis we have proposed a possible improvement of the energy reconstruction of forward scattered 
protons using the time measured in the scintillator counters of the forward part of the WASA-at-COSY detector.

The missing mass resolution based on a time-of-flight measurement was 
investigated and it was presented that it is similar to the resolution achievable with the energy loss method. 
Thus, combining this two independent techniques we expect to improve the missing mass 
resolution by about a factor of 1.5. We also have presented the basic concept of an algorithm which will 
enable to determine the energy of particles based on the time-of-flight technique. 
In the near future we plan to implement the established 
algorithm in the WASA analysis software.    

\appendix
\chapter{$\eta^{\prime} $ as a member of the  SU(3) pseudoscalar meson nonet}\label{app:mezony}
\begin{figure}[H]
\hspace{3.4cm}
\parbox{0.55\textwidth}{
\centerline{\epsfig{file=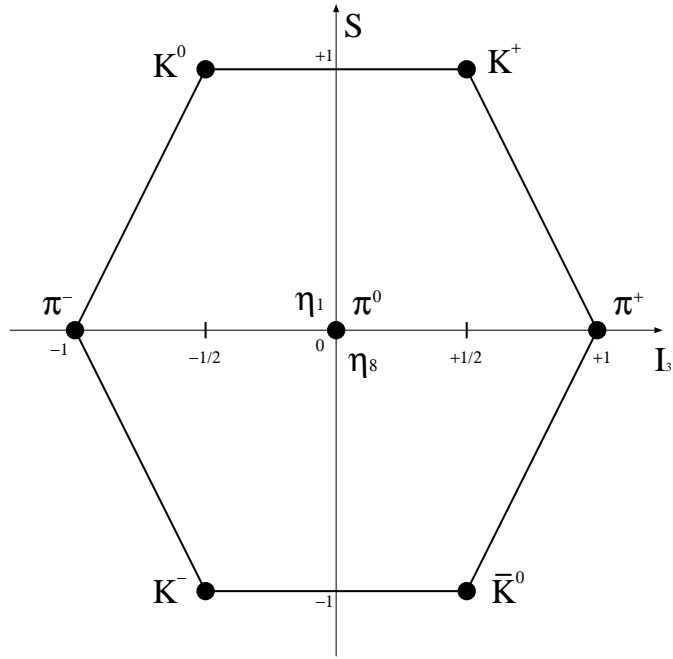,width=0.59\textwidth}}}
\caption{
Pseudoscalar meson nonet of the SU(3) symmetry group.
}
\label{su3nonet}
\end{figure}

\hspace{\parindent}
According to the quark model
each (pseudoscalar) meson is a state of a quark and antiquark system. The three lightes 
quarks $u$, $d$ and $s$, give nine possible $q\bar{q}$ combinations. 
In terms of the SU(3) symmetry we can write this as:
\begin{equation}
3 \otimes \bar{3} = 8 \oplus 1,
\end{equation}
which includes an octet and a singlet.
All possible meson states can be presented as a multiplet on a 
plot where the horizontal axis denotes the third component of the isospin -  $I_3$ and on the vertical axis the 
strangness - S is shown. Figure~\ref{su3nonet} presents a multiplet of the pseudoscalar mesons.
In this picture we have pions: $\pi^{+},\pi^{0},\pi^{-}$, kaons: $K^+$,$K^0$ and $K^-$,$\bar{K}^0$ and $\eta_1$ and 
$\eta_8$. 

The $\eta$ and $\eta^{\prime}$ meson are not  pure states of the SU(3) pseudoscalar nonet but 
a combination of a singlet $\eta_1$ and octet $\eta_8$.
The quark model give as wave functions for the pure states:
\begin{equation}
\eta_1 = \frac{1}{\sqrt{3}}(u\bar{u} + d\bar{d} + s\bar{s})
\end{equation} 
\begin{equation}
\eta_8 = \frac{1}{\sqrt{6}}(u\bar{u} + d\bar{d} - 2s\bar{s})
\end{equation} 
and from them the wave functions for physicaly observed states are constructed as:
\begin{equation}
\eta = \eta_8\cos\theta - \eta_1\sin\theta
\label{eta}
\end{equation} 
\begin{equation}
\eta^{\prime} = \eta_8\cos\theta + \eta_1\sin\theta
\label{etaprim}
\end{equation} 
where $\theta \approx -15.5^o \pm 1.3^o$ \cite{bramon99} denotes the mixing angle which 
was established experimentally. Using equation (\ref{eta}), (\ref{etaprim}) and the value of the 
mixing angle we obtain~\cite{moskal_phd}:
\begin{equation}
\eta = 0.77\frac{1}{\sqrt{2}}(u\bar{u} + d\bar{d}) - 0.63s\bar{s},.
\end{equation} 
\begin{equation}
\eta^{\prime} = 0.63\frac{1}{\sqrt{2}}(u\bar{u} + d\bar{d}) - 0.77s\bar{s}
\end{equation} 
Both wave functions above are very similar and therefore the mass of these two mesons should
in practice be also similar. 
However, the empirical mass of the $\eta^{\prime}$ meson is about two times larger than the mass of the $\eta$ meson. 

\chapter{The missing mass technique}\label{app:mm}
\hspace{\parindent}
The missing mass technique is used to identify the particles which cannot be registered
by the detectors due to short life time. 
We study the production of the $\eta^{\prime}$ meson via the $pp\to ppX$ reaction where X denotes 
the unobserved particle.
If we determine the four-momenta of all protons before and after the collision, 
the four-momentum conservation gives the relation:
\begin{equation}
\mathbb{P}_{beam} + \mathbb{P}_{target} = \mathbb{P}_{1} + \mathbb{P}_{2} + \mathbb{P}_{X},
\label{zasada4pedu}
\end{equation}
where $\mathbb{P}_{X}$ corresponds to the four-momentum of the unobserved particle. 

The mass can be calculated from (\ref{zasada4pedu}) as:
\begin{equation}
m_X^2 = \vert \mathbb{P}_{X}\vert^2 = 
\vert\mathbb{P}_{beam} + \mathbb{P}_{target} - \mathbb{P}_{1} - \mathbb{P}_{2}\vert^2 ,
\label{mm_wzor_ogolny}
\end{equation} 
with $\mathbb{P} \equiv (E,\vec{p})$ we can write:
\begin{equation}
m_X^2 = (E_{beam} + E_{target} - E_1 - E_2)^2 - (\vec{p}_{beam} + \vec{p}_{target} - \vec{p}_{1} - \vec{p}_{2})^2.
\label{missmass_3}
\end{equation} 
Using the formula above we can calculate the mass of the unobserved particle assuming that we can determine the 
four-momenta for all other particles participating in the reaction. 

In the laboratory where the proton beam collides with the hadrogen target the momentum of target is $p_{target} = 0$
and energy is equal to the mass. Therefore in the laboratory system relation (\ref{missmass_3}) can be written as:
\begin{equation}
m_X^2 = (E_{beam} + m_{target} - E_1 - E_2)^2 - (\vec{p}_{beam} - \vec{p}_{1} - \vec{p}_{2})^2.
\end{equation} 

\chapter{Relation between the fractional momentum resolution\\ and the fractional time-of-flight accuracy}\label{app:tof}
\hspace{\parindent}
To calculate the  fractional momentum resolution related to the time-of-flight fractional resolution we use the 
formula for the momentum:
\begin{equation}
p = m\gamma\beta \Longrightarrow p^2 = m^2 \gamma^2 \beta^2,
\label{wzor_1}
\end{equation}
where $\gamma$ denots the relativistic Lorentz factor
\begin{equation}
\gamma = \frac{1}{\sqrt{1 - \beta^2}}
\label{wzor_2}
\end{equation}
and $\beta$ indicates the velocity of the particle:
\begin{equation}
\beta = \frac{l}{t},
\label{wzor_3}
\end{equation}
with $l$ being the distance for the time-of-flight measurement. 
Inserting the (\ref{wzor_2}) and (\ref{wzor_3}) to (\ref{wzor_1}) we get:
\begin{equation}
p^2 = \frac{m^2 l^2}{t^2 - l^2}.
\label{wzor_4}
\end{equation}

The standard deviation of the  momentum distribution as a function of time resolution can be obtained 
by differentiating equation (\ref{wzor_4}):
\begin{equation}
\sigma(p^2) = \frac{2m^2 l^2t }{(t^2 - l^2)^2}\sigma(t).
\label{wzor_8}
\end{equation}
Applaying the relations $t = \frac{l}{\beta}$, and $\sigma(p^2) = 2p\sigma(p)$ in eq. (\ref{wzor_8}) we obtain a relation 
between the fractional momentum resolution and the fractional time-of-flight accuracy:
\begin{equation}
\frac{\sigma(p)}{p} = \frac{1}{1 - \beta^2}\frac{\sigma(t)}{t}.
\end{equation}

Further on, we will derive the relation between the fractional momentum resolution and the 
fractional energy resolution. By differentiating $p^2 = E^2 - m^2$ we obtain:
\begin{equation}
2p\sigma(p) = 2E\sigma(E),
\end{equation}
and thus we can write:
\begin{equation}
\frac{\sigma(p)}{p} = \frac{E^2}{p^2}\frac{\sigma(E)}{E} = \frac{1}{\beta^2}\frac{\sigma(E)}{E}.
\label{wzor_13}
\end{equation}

From formula (\ref{wzor_13}) we can also estimate the relation of the fractional momentum resolution 
as a function of the fractional kinetic energy resolution.
The total energy is given by $E = T + m$ and by differentiating we can write: $\sigma(E) = \sigma(T)$, 
due to constant mass.
Therefore we have:
\begin{equation}
\frac{\sigma(p)}{p} = \frac{E^2}{p^2}\frac{T}{E}\frac{\sigma(T)}{T}
\end{equation}
By further transformations we obtain:
\begin{equation}
\frac{\sigma(p)}{p} = \frac{ET}{p^2}\frac{\sigma(T)}{T} = \frac{ET}{E^2 - m^2}\frac{\sigma(T)}{T} =
                       \frac{ET}{(T + m)^2 - m^2}\frac{\sigma(T)}{T},
\end{equation}
\begin{equation}
\frac{\sigma(p)}{p} = \frac{ET}{T^2 + 2mT}\frac{\sigma(T)}{T} = \frac{E}{T + 2m}\frac{\sigma(T)}{T} .
\end{equation}
Finally the fractional momentum resolution as a function of fractional kinetic energy resolution reads:
\begin{equation}
\frac{\sigma(p)}{p} = \frac{T + m}{T + 2m}\frac{\sigma(T)}{T} .
\end{equation}

\chapter{Parametrization of the $pp\to pp\pi^{+}\pi^{-}\pi^{0}$ total cross section}\label{app:3pi}
\hspace{\parindent}
The production process of three pions is not well known. Only a few experimental data points exist
for the $pp\to pp\pi\pi\pi$ total cross section and the dynamics of the process is still 
not well understood \cite{kupsc-menu}. Therefore, for the estimation of the excitation function 
 we have used a parametrization proposed by J.~Bystricky~\cite{bystricky} which is 
based on the expansion of the total cross section in a base of the generalized Laguerre polynomials. 

One can expressed the total cross section using the effective amplitude: 
\begin{equation}
\sigma_{pp\rightarrow pp\pi^+\pi^-\pi^0} = \vert F(x) \vert^2 ,
\end{equation}
where the amplitude $F(x)$ can be expanded into series of orthonormal functions $L^{\alpha}_{n}(x)$:
\begin{equation}
F(x) = \sum_{n=0}^{\infty} a_n L^{\alpha}_{n}(x),
\end{equation}
with
\begin{equation}
L^{\alpha}_{n}(x) = e^{-\frac{x}{2}} x^{\frac{\alpha}{2}} \mathcal{L}^\alpha_n,
\end{equation}
\begin{figure}[H]
\parbox{0.48\textwidth}{\vspace{.9cm}\centerline{\epsfig{file=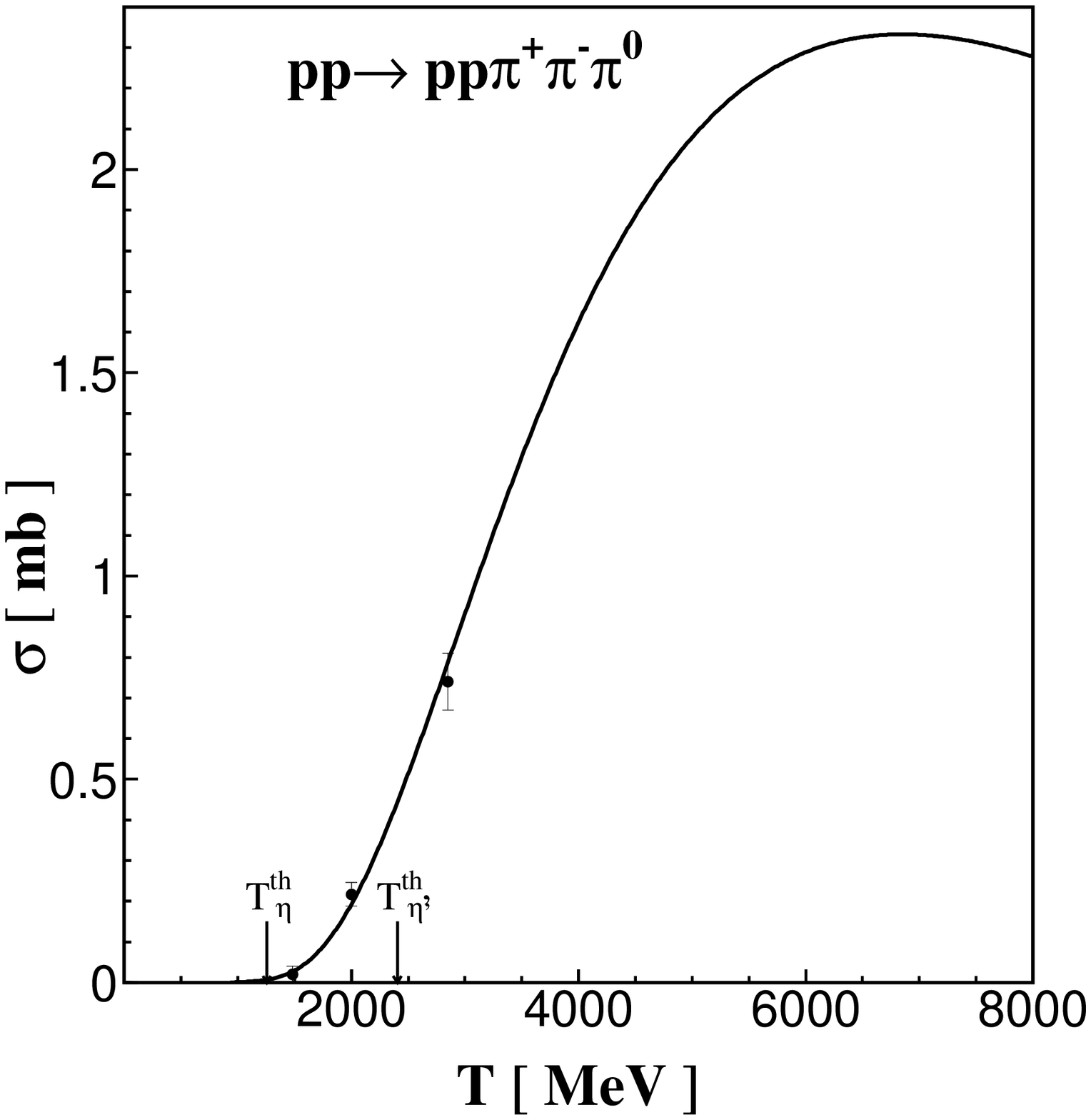,width=0.52\textwidth}}}
\parbox{0.55\textwidth}{\centerline{\epsfig{file=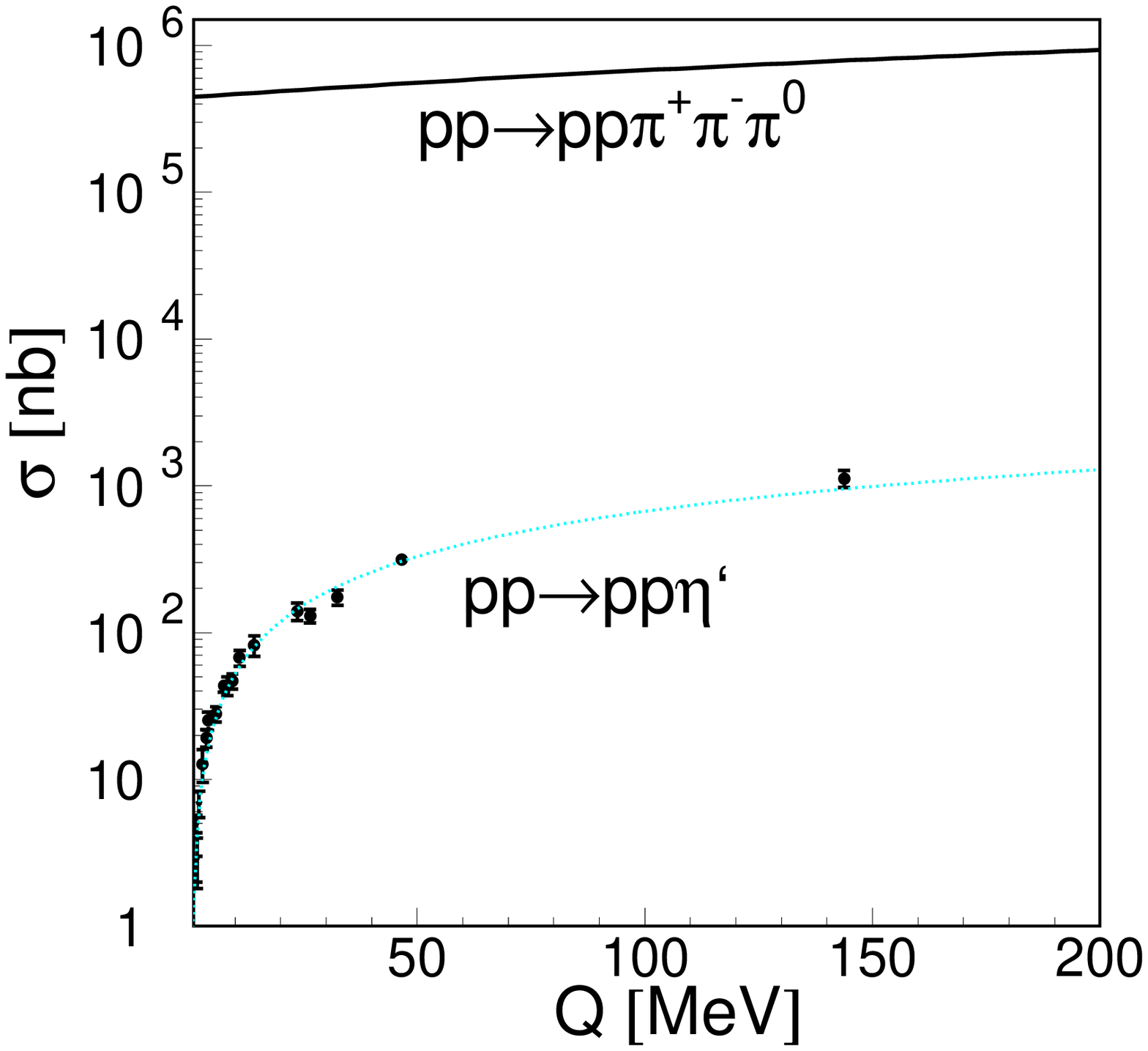,width=0.59\textwidth}}}
\caption{
{\bf (left)} The parametrization of the total cross section for $pp\rightarrow pp\pi^+\pi^-\pi^0$ 
and experimental data points from \cite{pickup,hart,eisner}.   
{\bf (right)} Cross sections of the $pp\rightarrow pp\eta^{\prime}$ reaction~\cite{cosy11-98,cosy11-00,hibou,balestra}
 compared to the total  cross section for the $pp\rightarrow pp\pi^+\pi^-\pi^0$ process (solid line).
}
\end{figure}
where $\mathcal{L}^\alpha_n$ denote the generealized Laguerre polynomials:
\begin{equation}
\mathcal{L}^\alpha_n(x) = \sum_{m=0}^{n} { n + \alpha \choose n - m } \frac{(-x)^m}{m!} . 
\end{equation}
We put 
\begin{equation}
x = c_k \ln \left( \frac{T}{T_k} \right) 
\end{equation}
where $T_k$ is the beam kinetic energy at the reaction threshold, and $c_k$ is a constant, determining the 
scale of $T$. 

One can see that the parametrization of the total cross section is in good agreement with the available 
experimental ponits for the three pion production. Considering now the production of the $\eta$, $\eta^{\prime}$ mesons
we can see that in the case of the $\eta^{\prime}$ meson the total cross section for the direct production of three pions 
increases by a factor of about 700 when going form $\eta$ to $\eta^{\prime}$ production threshold.

The direct three pion production will constitute the major ingredient of the background. We have 
plotted the total cross section of the three pion production using the introduced parametrization and compared it to the 
experimental points with a parametrization of the total cross section for the $\eta^{\prime}$ meson production. 
One can see that the cross section for the background production is three orders of magnitude larger then the 
cross section for the prodcution of the investigated meson, making the study of the $\eta^{\prime}\to\pi^{+}\pi^{-}\pi^{0}$
decay  experimentally challenging.

\chapter{Parametrization of the proton-proton Final State Interaction}\label{app:fsi}
\hspace{\parindent}
This appendix was prepared on the basis of references~\cite{swave}.
The final state interaction between two protons in the exit channel of the reaction $pp\to ppX$ is a well known 
fact, and had achieved a good theoretical description. In order to include this interaction in our calculations 
we had considered three possible models of the pp-FSI enhancement factors described in~\cite{swave}. 

The first model uses the square of the on-shell  proton-proton scattering amplitude, which for the relative angular 
momentum $l=0$ reads~\cite{wong62}:
\begin{equation}
M_{pp\to pp } = \frac{e^{i\delta_{pp}}\cdot sin(\delta_{pp})}{C\cdot p}
\end{equation}    
where $p$ indicates the momentum of both proton in the center of mass system of the colliding particles, $\delta_{pp}$
denotes the phase shift, $C$ is the Coulomb penetration factor which includes the Coulomb interaction between 
two protons. This factor can be defined as~\cite{jackson,bethe}:
\begin{equation}
C = \frac{2\pi\eta_c}{e^{2pi\eta_c}-1},
\end{equation} 
where $\eta_c=\frac{\alpha}{v}$ indicates the relativistic Coulomb parameter, with $\alpha$ being the fine 
structure constant and $v$ the proton velocity in the rest system of other proton.

The phase shift $\delta_{pp}$ is calculated using the Cini-Fubini-Stanghellini formula, including the 
Wong-Noyes Coulomb corrections~\cite{23swave}:
\begin{equation}
C^2\cdot p\cdot ctg(\delta_{pp}) + 2\cdot p\cdot \eta_c\cdot h(\eta_c) = 
-\frac{1}{a_{pp}} + \frac{1}{2}\cdot b_{pp} \cdot p^2 - \frac{P_{pp}\cdot p^4}{1+ Q_{pp}\cdot p^2},
\end{equation}
where~\cite{jackson}:
\begin{equation} 
h(\eta_c) = -ln(\eta_c) - 0.57721 + \eta_c\cdot \sum\limits_{n=1}^{\infty} \frac{1}{n\cdot(n^2+\eta_c^2)},
\end{equation}
$a_{pp} = -7.8$~fm denotes the scattering length and $b_{pp} = 2.8$~fm is the effective range. 
The parameters $P_{pp} = 0.74$~fm$^3$ and $Q_{pp} = 3.35$~fm$^{2}$
are related to the shape of the nuclear potential, and were calculated from the pion-nucleon coupling constant~\cite{22swave}.
Finally, the square of the proton-proton scattering amplitude can be expressed as: 
\begin{equation}
\vert M_{pp\to pp}\vert^2 = 
\frac{1}{
C^4\cdot p^2 + \left( -\frac{1}{a_{pp}} + \frac{1}{2}\cdot b_{pp} \cdot p^2 - \frac{P_{pp}\cdot p^4}{1+ Q_{pp}\cdot p^2} 
 - 2\cdot p\cdot \eta_c\cdot h(\eta_c) \right)^2}
\end{equation}

As a second possibility, for the pp-FSI parametrization we have approximated the enhancement factor by the 
inverse of the squared Jost function~\cite{28swave}:
\begin{equation}
M = M_0^{on} J^{-1}(-p),
\end{equation}
where for a short ranged interaction, by applaying the effective range expansion to order $p^2$ we have:
\begin{equation}
\frac{1}{J(-p)} =  \frac{b_{pp}(p^2 + \alpha^2)}{2(\frac{1}{a_{pp}} + \frac{1}{2}b_{pp}\cdot p^2 - i\cdot p)},
\label{jost}
\end{equation} 
with
\begin{equation}
\alpha = \frac{1 + \sqrt{1 + 2\frac{b_{pp}}{a_{pp}}}}{b_{pp}}.
\end{equation}
Equation \ref{jost} is true in presence of the long-ranged Coulumb force. 

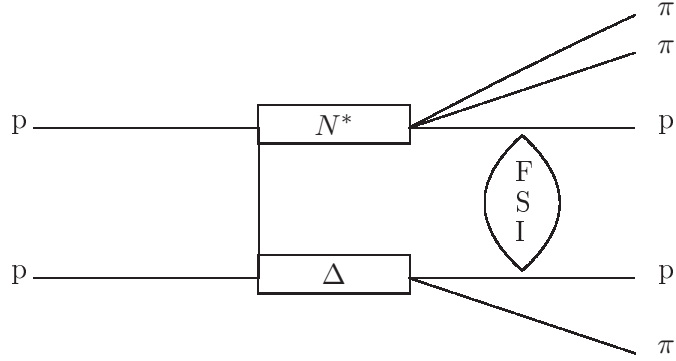
\begin{figure}[h!]
\setlength{\unitlength}{1mm}
\hspace{1.5cm}
\begin{picture}(50,50)
\put(20,35){\line(1,0){30}} 
\put(20,15){\line(1,0){30}} 
\put(50,33){\framebox(20,5)[cc]{$N^*$}} 
\put(50,13){\framebox(20,5)[cc]{$\Delta$}} 
\put(70,35){\line(1,0){30}} 
\put(70,15){\line(1,0){30}} 
\put(50,15){\line(0,1){20}} 
\put(17,35){p}
\put(17,15){p}
\put(103,35){p}
\put(103,15){p}
\put(103,45){$\pi$}
\put(103,5){$\pi$}
\put(103,50){$\pi$}
\bezier{600}(70,35)(85,40)(100,45)
\bezier{600}(70,15)(85,10)(100,5)
\bezier{600}(70,35)(85,43)(100,50)
\bezier{600}(85,34)(75,25)(85,16)
\bezier{600}(85,34)(95,25)(85,16)
\put(84,28){F}
\put(84,24){S}
\put(84,20){I}
\end{picture}
\caption{Feynman diagram of a possible production of three pion by excitation of $\Delta$ and $N^*$ resonances.}
\label{diagram}
\end{figure}

To include the possible dynamics of the three pion production along with the theoretical prediction, we had assumed 
that the production can proceed via the excitation of $\Delta$ and $N^*$ resonances according o the reaction chain:
\begin{equation}
\nonumber
pp \to  \Delta N^* \to pp \pi\pi\pi,
\end{equation} 
where $\Delta$ decays into $p\pi$ and $N^*$ into $p\pi\pi$. 
A corresponding Feynman diagram could then have a form as shown in figure \ref{diagram}.
The mechanism had been applied by using the relativistic Breite-Wigner form for the probability of the accuring of the 
$m_{\pi p}$ invariant mass:
\begin{equation}
a^2(m_{p\pi}) = \frac{m^2 \cdot \Gamma^2}{(m^2 - m_{p\pi}^2)^2 + m^2\cdot \Gamma^2}
\end{equation} 
where the $\Gamma$ and $m$ denote the width and mass of the $\Delta$ resonance, respectively, and the $m_{p\pi}$ 
indicates the invariant mass of the proton-pion subsystem. An analogous formula was used for the $m_{p\pi\pi}$ and $N^*$ 
resonance.

\chapter{Dynamics of the $\pi^{+}\pi^{-}\pi^{0}$ production in proton-proton interaction}\label{app:sym}
\begin{figure}[h!]
\hspace{2.cm}
\parbox{0.65\textwidth}{\centerline{\epsfig{file=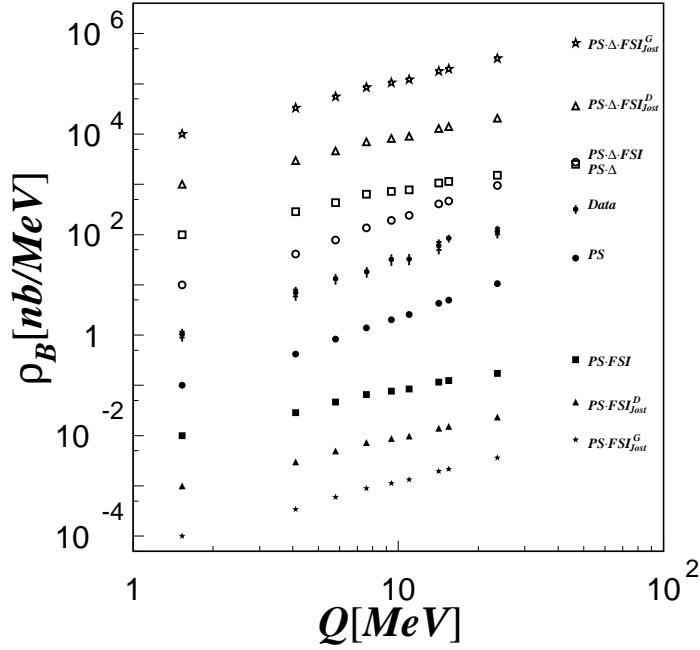,width=0.69\textwidth}}}
\caption{ Differential cross section ($\left.\frac{d\sigma_B}{dm}\right\vert_{m_{\eta^{\prime}}}$
) for the $pp\to pp 3\pi$ production
with the invariant mass of pions equal to the mass of the $\eta^{\prime}$
meson as a function of the excess energy above the $\eta^{\prime}$ threshold.
Simulations for the $pp\to pp 3\pi$ reaction are compared to the multimeson production
extracted  from the COSY-11
experiments~(see Table~\ref{tabela1} and Figure~\ref{diffcross}).
For better visalization the simulated values
have been arbirarily normalized such that the point at the lowest energy (for Q~=~1.53~MeV)
differ from the data by subsequent powers of ten.
The data points are positioned in the middle of the figure.
The results of simulations were obtained assuming
that the process is governed by
(from top to bottom):\protect\\
(i) Phase Space $\oplus$ $\Delta\Delta$ $\oplus$ $FSI_{Jost G}$ - {\it empty stars}\protect\\
(ii) Phase Space $\oplus$ $\Delta\Delta$ $\oplus$ $FSI_{Jost D}$ - {\it empty triangles}\protect\\
(iii) Phase Space $\oplus$ $\Delta\Delta$ $\oplus$ $FSI_{on-shell}$ - {\it empty squares}\protect\\
(iv) Phase Space $\oplus$ $\Delta\Delta$ - {\it empty circles}\protect\\
(v) Phase Space - {\it data points}\protect\\
(vi) Phase Space - {\it full circles}\protect\\
(vii) Phase Space $\oplus$ $FSI_{on-shell}$ - {\it full squares}\protect\\
(viii) Phase Space $\oplus$ $FSI_{Jost D}$ - {\it full triangles}\protect\\
(ix) Phase Space $\oplus$ $FSI_{Jost G}$ - {\it full stars}
\label{rysunekdynamika}
}
\end{figure}

\hspace{\parindent}
With  a  4$\pi$  facilities  such  as  WASA,  aiming  at
measurements  of  decays  of  $\eta$  and  $\eta'$  produced  in  $pp$
interactions,  the  understanding of  the  $pp \to pp 3\pi$
reaction dynamics  becomes   very  important  as  they   constitute  a  severe
background for studies of $\eta$  and $\eta'$ decays into three pions.
However, for more  than  fourty years  there  were  only  three experimental data points
available  for   the  cross  section  of
$pp\to pp \pi^+\pi^-\pi^0$  and
$pp\to pn\pi^+\pi^+\pi^-$
reactions,    all    coming    from   bubble    chamber    experiments
\cite{pickup,eisner,hart}.   Only
recently  the data  base  has  been extended  by  measurements  of the 
$pp\to pp \pi^+\pi^-\pi^0$  and
$pp\to pp\pi^0\pi^0\pi^0$
reactions  cross sections near the threshold
by  the  CELSIUS/WASA   collaboration  \cite{pauly06-celcius}.   For  the
remaining reactions there are no data in that energy region.
As a consequanece the understanding of the mechanism
of the $3\pi$ production is by far not satisfactory~\cite{kupsc-menu}.
Therefore, in order to gain more information of the dynamics of this process
we have compared the empirical $\left.\frac{d\sigma_B}{dm}\right\vert_{m_{\eta^{\prime}}}$ energy dependence
with results of simulations conducted under various assumptions as regarding
the interaction among the final state particles.
The direct  production should proceed by  an excitation of  one or two
baryon  resonances  followed     by    the    subsequent    decays
\cite{sternheimer}.  Therefore  we have tested two possibilities
assuming for the primary production a pure phase space distribution
and as a second possibility we considered the resonant production
via the
$pp\to\Delta N^*\to pp3\pi$ reaction chain.
The result of the simulations are confronted with the data
in figure~\ref{rysunekdynamika}.
Assuming the Watson-Migdal ansatz,
we factorize the proton-proton FSI and the production amplitude.
We have considered three possible parametrizations of the pp-FSI enhancement factors
which are described in detail in reference~\cite{swave} and in the appendix~\ref{app:fsi} of this thesis.
As a first possibility for the
proton-proton enhancement factor we use the square of the on-shell
proton-proton scattering amplitude calculated according
to the Cini-Fubini-Stanghellini formula including the Wong-Noyes
Coulomb corrections~\cite{22swave,23swave,24swave}.
As a second possibility, we  approximated the enhancement factor by the inverse of the
squared Jost function (Jost$_D$) derived by Druzhinin et al~\cite{28swave},
and finally we used also the inverse of the squared Jost function (Jost$_G$),
calculated according to the formulae of Goldberger and Watson~\cite{30swave}.
Figure~\ref{rysunekdynamika} indicates that the excess energy dependence of $\left.\frac{d\sigma_B}{dm}\right\vert_{m_{\eta^{\prime}}}$ can be
reproduced equally well assuming that the phase space is homegeneously populated
or assuming that the reaction  proceeds via resonances:
$pp\to \Delta N^*\to pp3\pi$.
However, it is evident from the figure that inclusion
of the FSI inhancement factors as derived for the
three body final state worsens significantly the consistency with the data.

\chapter{Description of computer programs used for simulations}\label{app:prog}
\hspace{\parindent}
Simulations were performed using the FORTRAN and C language programs with additional libraries from the 
CERNLIB package. Here only a brief description of the programs is shown without details.

All programs were using GENBOD phase space generator subroutines~\cite{genbod}, which on basis of the total 
energy in the center of mass of two colliding particles and the masses of the outgoing particles,
returns the four-momenta of all particles in the exit channel. Each generated event is weighted with the 
WT$_0$ parameter which ensures that the particles are homogeneneously distributed in the phase space.
The output four-momenta are given in the center of mass system, thus it is needed to transform them to the 
laboratory reference frame using the Lorentz transformation.
  
In most cases the input parameters (beam momentum, total energy in the CM frame, number of outgoing particles, 
mass of particles) for all programs were read from a special input file. This method prevented from compiling   
the code each time, when the values of parameters changed. 

In order to check whether the generated events fulfill the detection conditions we have implemented constrains 
of geometrical acceptance of the detector setup. 

The interaction of protons in the final state (FSI) and the production dynamics was included according to 
formulae introduced in appendix~\ref{app:fsi}. To take into account these effects we have weighted each generated 
event by the sqaure of the proton-proton elastic scattering amplitude or a square of the Breit-Wigner amplitude ($A^2$):
\begin{equation}
WT = WT_0 \cdot A^2.
\end{equation}  

As output of the programs in most cases an {\texttt{.hbook}} and an  {\texttt{ASCI}} files was produced. The first one 
contains histograms which can be directly ploted by PAW analyzing program~\cite{paw} and the second type of file can be used 
in further analysis, for example as an imput file to another program.        

\newpage
\thispagestyle{empty}
\begin{center}\end{center}
\newpage
\thispagestyle{empty}
\begin{center}\end{center}\begin{center}\end{center}\begin{center}\end{center}\begin{center}\end{center}
\begin{center}\end{center}\begin{center}\end{center}\begin{center}\end{center}\begin{center}\end{center}
\begin{center}\end{center}\begin{center}\end{center}\begin{center}\end{center}\begin{center}\end{center}
\begin{center}\end{center}\begin{center}\end{center}\begin{center}\end{center}\begin{center}\end{center}
\begin{center}\end{center}\begin{center}\end{center}\begin{center}\end{center}\begin{center}\end{center}
\begin{center}\end{center}\begin{center}\end{center}
\begin{center}
\bf ''I never did a day's work in my life. 
It was all fun.'' 
\end{center}
\begin{center}
{\it Thomas Alva Edison (1847 -- 1931)}
\end{center}

\newpage
\pagestyle{fancy}
\fancyhead{}
\fancyfoot{}
\renewcommand{\headrulewidth}{0.8pt}
\fancyhead[RO]{\textbf{\sffamily{{{\thepage}}~}}}
\fancyhead[RE]{\bf\footnotesize{\nouppercase{ }}}
\fancyhead[LE]{\textbf{\sffamily{{{\thepage}}}}}
\fancyhead[LO]{\bf\footnotesize{{\nouppercase{ }}}}
\advance\headheight by 5.3mm
\advance\headsep by 0mm
\newpage
\thispagestyle{empty}
\begin{center}\end{center}
\newpage
\thispagestyle{empty}
\begin{center}{\LARGE{\bf Acknowledgment}}\end{center}
\vspace{1.cm}
\hspace{\parindent}
I would like to take the opportunity to thank everybody who helped me during writting of this thesis.

My first words I would like to direct to Prof. dr hab. Pawe\l{} Moskal - a person without whom, this thesis would never be acomplished. He had gave me a great possibility of studing and investigating an interesting world of hadron 
and meson physics together with the COSY-11 and WASA-at-COSY groups. I would like to thank you Paweł for guiding and 
encouraging me in daily work, for an infinite patience in correcting subsequent versions of this thesis. I am also indebt 
for hours of very interesting discussions, not only on physics. And I wish to thank you for the hospitality during my stays in J\"{u}lich and for many excursions.

I am greatly indebted to Dr Andrzej Kup\'{s}\'{c} for patience in answering my questions,
for many valuable suggestions, for teaching me basics of ROOT, WASA-MC and RootSorter, and 
for perusal of the early version of this thesis. Also for hospitality during my stays in Uppsala.

I am grateful to Prof. dr hab. Walter Oelert for reading and correcting this thesis, and for support of my stays 
in J\"{u}lich and Uppsala.

I would like to thank Prof. dr hab. Bogusław Kamys for letting me work in the Nuclear Physisc Division of the 
Jagiellonian University, financial support of my stays in J\"{u}lich and for valuable remarks during our thursday meetings.

Many thanks are due to Prof. dr hab. James Ritman for support of my activities in the Research Center J\"{u}lich. 

I am also grateful to Prof. dr hab. Lucjan Jarczyk for many valuable suggestions on meson physics and 
topics considered in this thesis.

I would like to thank Dr Dieter Grzonka for proofreading of this thesis and for valuable suggestions.

I wish to thank to Dr Christian Pauly for reading a part of the manuscript, for corrections and pertinent suggestions.

I would like to thank Dr Magnus Wolke for making it possible to carry out this work within the WASA-at-COSY Collaboration.

Many thanks are due to Master of Science Jaros\l{}aw J. Zdebik for his support during the last two years, 
a lot of discussions, coffees, and for nice
atmosphere during the daily work. He has always been there when I needed him. 

I am greatly indebt to Master of Science Joanna Przerwa and Master of Science Pawe\l{} Klaja for 
their hospitality and care during my stays in J\"{u}lich, for many coffees, tasty cakes and interesting discussions. 

I would like also to thank Wojciech Krzy\.{z}yk for reading a part of the manuscript and looking into it with 
an eye of ''non-physicist'', for great support and many discussions.    

I do wish to thank: Master of Science Dagmara Rozp\k{e}dzik, Master of Science Barbara Rejdych, Master of Science Ma\l{}gorzata Hodana, 
Dr Rafa\l{} Czy\.{z}ykiewicz, Master of Science Eryk Czerwi\'{n}ski,  
Ewelina Czaicka, Jakub Bo\.{z}ek, and Micha\l{} Silarski for a nice atmosphere during the daily work in Cracow and J\"{u}lich. 

On the personal side, I would like also to thank cordially my parents and sister, they had always been entirely supportive during my 
studies.

Last but not least, I would like to thank cordially Katarzyna S\l{}oka for care and big support during last years.

\end{document}